\documentclass[a4paper]{article}

\usepackage{a4wide}
\RequirePackage{natbib}

\usepackage[pdftex,final,bookmarksnumbered,bookmarksopen=false,breaklinks,colorlinks]{hyperref}
\hypersetup{
	final,
	bookmarksnumbered=true,
	bookmarksopen=true,
	bookmarksopenlevel=0,
	unicode=false,
	pdftoolbar=true,
	pdfmenubar=true,
	pdffitwindow=false,
	pdftitle={Density estimation for compositional data using nonparametric mixtures},
	pdfdisplaydoctitle=true,
	pdftoolbar=true,
	pdfmenubar=true,
	pdflang={English},
	pdfauthor={Jiajin Xie, Yong Wang, Eduardo Garcia-Portugues},
	pdfsubject={arXiv paper},
	pdfcreator={Yong Wang},
	pdfproducer={Yong Wang},
	pdfkeywords={Simplex}{Dirichlet mixture}{Gaussian mixture}{Nonparametric maximum likelihood}{Bandwidth selection},
	pdfnewwindow=true,
	breaklinks=true,
	hidelinks,       
	linkcolor=black,
	citecolor=black,
	filecolor=magenta,
	urlcolor=cyan
}
\usepackage{url}
\usepackage[top=1.5cm,bottom=2cm,right=2.5cm,left=2.5cm]{geometry}
\usepackage{titling}

\usepackage{graphicx}%
\usepackage{multirow}%
\usepackage{amsmath,amssymb,amsfonts}%
\usepackage{amsthm}%
\usepackage{mathrsfs}%
\usepackage[title]{appendix}%
\usepackage{xcolor}%
\usepackage{textcomp}%
\usepackage{manyfoot}%
\usepackage{booktabs}%
\usepackage{algorithm}%
\usepackage{algorithmicx}%
\usepackage{algpseudocode}%
\usepackage{listings}%
\usepackage{pgffor}
\usepackage[percent]{overpic}
\usepackage{booktabs}   
\usepackage{multirow}   
\usepackage{makecell}
\usepackage{rotating}
\usepackage{caption}
\newcommand{\alphab}{\boldsymbol{\alpha}} 
\newcommand{\alphabt}{{\widetilde{\alphab}}}
\newcommand{\alphat}{{\widetilde{\alpha}}}
\newcommand{\Ab}{\boldsymbol{A}} 
\newcommand{\xb}{\boldsymbol{x}} 
\newcommand{\Xb}{\boldsymbol{X}}
\newcommand{\yb}{\boldsymbol{y}}

\newcommand{\pib}{\boldsymbol{\pi}} 
\newcommand{\mub}{\boldsymbol{\mu}}
\newcommand{\thetab}{\boldsymbol{\theta}} 

\newcommand{\gd}{\mathbf{g}}
\newcommand{\Hess}{\mathbf{H}}
\newcommand{\Rb}{\mathbb{R}} 
\newcommand{\rd}{\mathrm{d}} 
\newcommand{\Dir}{\mathrm{Dir}}
\newcommand{\Norm}{\mathrm{Norm}}

\DeclareMathOperator*{\argmax}{arg\,max}
\DeclareMathOperator*{\argmin}{arg\,min}

\title{Density estimation for compositional data using\\ nonparametric mixtures}

\author{
  Jiajin Xie\thanks{Department of Statistics, University of Auckland, New Zealand.}
  \and
  Yong Wang\thanks{Department of Statistics, University of Auckland, New Zealand. Corresponding author: \href{mailto:yongwang@auckland.ac.nz}{yongwang@auckland.ac.nz}.}
  \and
  Eduardo Garc\'ia-Portugu\'es\thanks{Department of Statistics, Universidad Carlos III de Madrid, Spain.}
}

\date{October 8, 2025}

\begin{document}

\maketitle

\begin{abstract}
Compositional data, representing proportions constrained to the simplex, arise in diverse fields such as geosciences, ecology, genomics, and microbiome research. Existing nonparametric density estimation methods often rely on transformations, which may induce substantial bias near the simplex boundary. We propose a nonparametric mixture-based framework for density estimation on compositions. Nonparametric Dirichlet mixtures are employed to naturally accommodate boundary values, thereby avoiding the transformation or zero-replacement, while also identifying components supported on the boundary, providing reliable estimates for data with zero or near-zero values. Bandwidth selection and initialization schemes are addressed. For comparison, nonparametric Gaussian mixtures, coupled with log-ratio transformations, are also considered. Extensive simulations show that the proposed estimators outperform existing approaches. Three real data applications, including GDP data analysis, handwritten digit recognition, and skin detection, demonstrate the usefulness of nonparametric Dirichlet mixtures in practice.
\end{abstract}

\noindent\textbf{Keywords:}
Bandwidth selection; Dirichlet mixture; Gaussian mixture; Nonparametric maximum likelihood; Simplex.

\section{Introduction}
\label{sec:intro}

Compositional data represent the proportions of components within a whole. These data arise in diverse fields, including geosciences \citep{Buccianti2006}, ecology \citep{Behnaz2018}, nutrition \citep{Tomova2025}, and more recently in genomics and microbiome research \citep{Matthew2016, Gloor2017, Calle2023}. The sample space of compositional data is the simplex, consisting of non-negative vectors constrained to sum to unity. Consequently, many standard statistical methods designed for the Euclidean space are spurious, motivating the development of specialized techniques for compositions.

Nonparametric density estimation reveals the underlying probability distribution of random samples without imposing a parametric form. Kernel density estimation (KDE) is the most widely used method. For compositional data, KDE was first introduced by \citet{Aitchison1985}, who proposed Dirichlet and Gaussian kernels. To render the Gaussian kernel applicable, the additive log-ratio (alr) transformation maps data from the simplex to the Euclidean space. Building on this, \citet{Chacon2011} replaced alr with the isometric log-ratio (ilr) transformation, allowing the use of a full bandwidth matrix in estimation, while \citet{Ouimet2022} studied the asymptotic behavior and boundary properties of the Dirichlet kernel.

Since neither alr nor ilr can transform compositions with zeros, various zero-replacement strategies have been proposed, such as substituting zeros with small positive values \citep{Aitchison1985, Martin2003} or applying iterative substitution algorithms \citep{Palarea2008}, as summarized in \citet[Chapter 13]{filzmoser2018}. However, these approaches may distort the data structure and yield unstable or biased estimates.

Nonparametric mixture-based density estimation (NPMDE) provides an alternative to KDE \citep{wang2012}. Its mathematical form is simpler, as the mixture typically consists of only a few components, whereas KDE records all observations. Since KDE is a convolution between the empirical mass distribution and the kernel, it often yields overly flattened densities with higher bias and slower convergence. By contrast, using de-convolution, a well-fitted mixture tends to be less biased and may achieve faster convergence, with some empirical evidence given in \citet{Wang2015}, who extended NPMDE to the multivariate setting by leveraging the fast algorithms of \citet{Wang2007}. 

The purpose of this article is to develop NPMDE for compositional data. Our main contributions are: (1) extending the NPMDE framework to both Dirichlet and Gaussian mixtures, with emphasis on Dirichlet components that naturally handle boundary data on the simplex, thereby avoiding transformation or zero-replacement, and that reliably estimate zero or near-zero values by identifying boundary-supported components; (2) proposing bandwidth-based model selection strategies and initialization schemes for Dirichlet mixtures on the simplex; (3) analyzing loss functions for density estimation, demonstrating the invariance of integrated absolute error (IAE) and Kullback--Leibler divergence (KLD) across spaces and explaining the instability of integrated squared error (ISE) on the simplex; (4) designing simulation settings by constructing a collection of Dirichlet mixtures and extending Gaussian and Dirichlet mixtures to higher dimensions; and (5) applying the proposed method to three real data density estimation and classification tasks, including GDP data analysis, handwritten digit recognition and skin color detection.

The remainder of the article is organized as follows. Section~\ref{sec:bg} reviews fundamental concepts of compositional data and associated distributions, together with existing density estimation techniques. Section~\ref{sec:nonp} presents the proposed method. Section~\ref{sec:simu} evaluates its performance against existing methods on simulated data, and Section~\ref{sec:realdata} demonstrates its application to real datasets. Section~\ref{sec:diss} concludes with a summary of findings and final remarks. The code replicating all the empirical results of the article is available at the GitHub repository \url{https://github.com/JiajinX/npmixsimplex}.

\section{Background}
\label{sec:bg}

\subsection{Compositional data and distributions}

\begin{figure*}[!h]
  \centering
  \includegraphics[page=1,width=0.325\linewidth,
  clip,trim={1.5cm 1.7cm 1.4cm 2.3cm}]{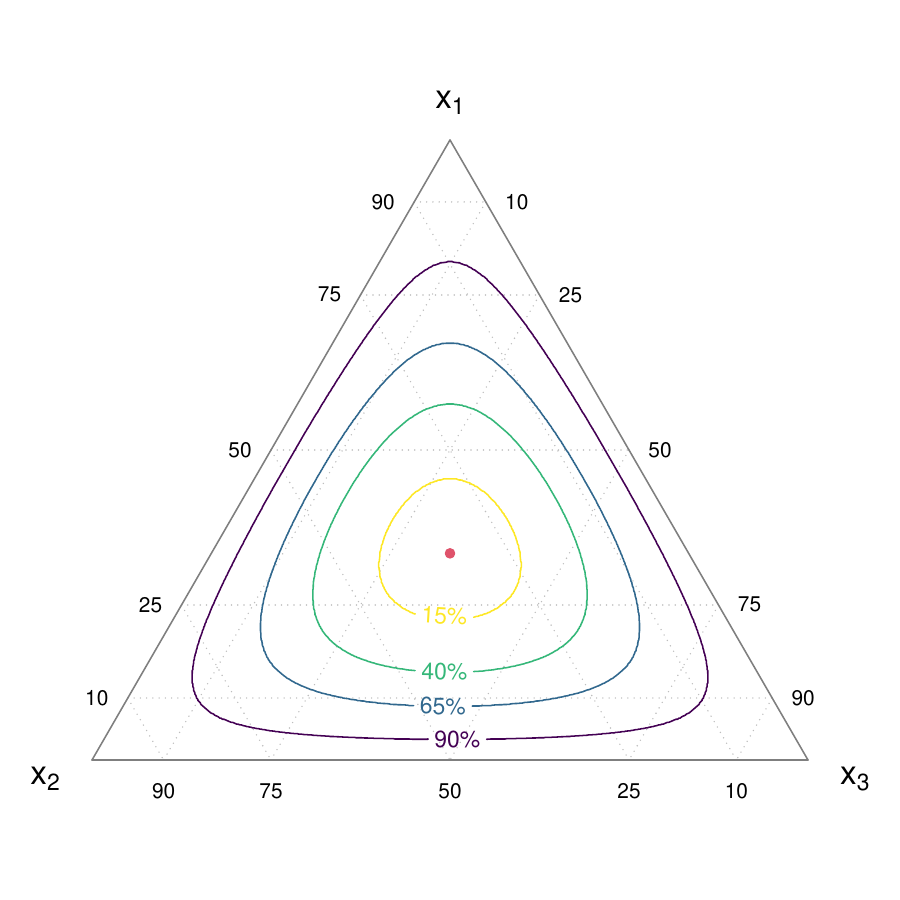}%
  \hfill
  \includegraphics[page=2,width=0.325\linewidth,
  clip,trim={1.5cm 1.7cm 1.4cm 2.3cm}]{dirichlet.pdf}%
  \hfill
  \includegraphics[page=3,width=0.325\linewidth,
  clip,trim={1.5cm 1.7cm 1.4cm 2.3cm}]{dirichlet.pdf}%

  \caption{\small Probability contours of the Dirichlet density with mode (red dot) at the center of $\Delta^2$, with $\alpha_0=5$ (left), $\alpha_0=13$ (middle), and $\alpha_0=50$ (right).}
  \label{fig:dirichlet}
\end{figure*}

The sample space of compositional data $\xb = (x_1,\ldots,x_D)^\top$ with 
$D:= d+1$ parts is the \emph{$d$-dimensional simplex}
\begin{align*}
\Delta^d := \Bigl\{\xb \in \mathbb{R}^D: x_1,\dots,x_D \geq 0, \, \sum_{i=1}^D x_i = 1 \Bigr\}.
\end{align*}
The Dirichlet distribution is the most widely used model for compositional data. Its density is
\begin{align*}
    f_{\Dir}(\xb; \alphab) := 
    \frac{\Gamma\Bigl(\sum_{i=1}^D \alpha_i\Bigr)}{\prod_{i = 1}^D\Gamma(\alpha_i)} \prod_{i=1}^D x_i^{\alpha_i - 1},~~
    \xb \in \Delta^d, \, \alphab \in \Rb^D_+. 
\end{align*}
The vector $\alphab = (\alpha_1, \dots, \alpha_D)^\top$ controls the location and shape of the distribution. When $\alphab > \mathbf{1}$ (element-wise), the distribution is unimodal and exhibits a log-concave shape. The parameter $\alpha_0 = \alphab^\top\mathbf{1}$ governs the concentration of the distribution. Figure~\ref{fig:dirichlet} illustrates three Dirichlet densities on $\Delta^2$. The label $p\%$ on each contour indicates that the distribution has $p\%$ probability mass within the enclosed region. As $\alpha_0$ increases from $D$ to $\infty$ for $\alphab = \alpha_0 \mathbf{1} / D$, the distribution transitions from the uniform to a point mass.

It is well-known Dirichlet distribution admits the following marginal property. Let $\alphab \in \Rb_+^D$, and suppose $\Xb \sim \mathrm{Dir}(\alphab)$. Partition the components of $\Xb$ into $k$ disjoint subvectors
\begin{align*}
    (X_1,\dots,X_{m_1})^\top, \dots, (X_{m_{k-1} + 1},\dots,X_D)^\top,
\end{align*}
where $1 \leq m_1 < \cdots < m_{k-1} < D$. Let $S_1, \dots, S_k$ denote the sums of the coordinates within each subvector. Then the vector of the partial sums follows a lower-dimensional Dirichlet distribution
\begin{align*}
    (S_1, \dots, S_k)^\top \sim \mathrm{Dir} \biggl(~\sum_{i=1}^{m_1} \alpha_i, \dots, \sum_{i=m_{k-1}+1}^{d+1} \alpha_i \biggr).
\end{align*}

Due to the limited flexibility of the Dirichlet distribution, compositional data on the \emph{open simplex} 
$\widetilde{\Delta}^d$, i.e., the interior of $\Delta^d$, can be modeled by Gaussian distributions after log-ratio transformations. These transformations map $\widetilde{\Delta}^d$ onto $\Rb^d$, where standard multivariate methods apply. A Gaussian distribution is then fitted in $\mathbb{R}^d$, and the resulting density is back-transformed to $\widetilde{\Delta}^d$. Formally, this procedure is expressed as
\begin{align*}
    f_{\mathrm{lr}}(\xb; \boldsymbol{\mu}, \boldsymbol{\Sigma}) :=  \bigl| \det J_{\mathrm{lr}}(\xb) \bigr| \, f_\mathrm{Norm}\bigl(\mathrm{lr}(\xb); \boldsymbol{\mu}, \boldsymbol{\Sigma}\bigr),
\end{align*}
where $\mathrm{lr}: \widetilde{\Delta}^d \to \Rb^d$ denotes a general log-ratio transformation, $J_{\mathrm{lr}}(\xb) = \partial \,\mathrm{lr}(\xb) \,/\, \partial \xb^\top$ is the corresponding Jacobian, and $\boldsymbol{\mu}$ and $\boldsymbol{\Sigma}$ represent the mean and covariance matrix of the log-ratio transformed compositions, respectively. Two common log-ratio transformations are the additive log-ratio (alr) transformation \citep{Aitchison1986} and the isometric log-ratio (ilr) transformation \citep{Egozcue2003}, defined as follows
\begin{align}
    \mathrm{alr}(\xb)_i := \log \frac{x_i}{x_D}, \qquad \mathrm{ilr}(\xb)_i := \frac{1}{\sqrt{i(i+1)}} \, \log \frac{\sum_{j = 1}^i x_j}{(x_{i+1})^i},
    \label{eq:lrt}
\end{align}
for $i = 1, \ldots, d$. Their respective Jacobian determinants are
\begin{align}
    \bigl|\det J_\mathrm{alr} (\xb) \bigr| = \left(x_1\cdots x_{D}\right)^{-1}, \qquad
    \bigl|\det J_\mathrm{ilr} (\xb) \bigr| = \bigl(\sqrt{D}~x_1 \cdots x_{D}\bigr)^{-1}.
    \label{eq:det}
\end{align}

However, as real compositional data are often sparse, direct log-ratio transformations have limited applicability. To address this, various zero-replacement strategies have been proposed \citep[Chapter 13]{filzmoser2018}. \citet{Aitchison1985} introduced a rule for compositions with $c$ zeros and $D-c$ nonzeros: each zero is replaced by $\delta (c+1)(D-c)\,/\,D^2$, and each nonzero is reduced by $\delta c(c+1)\,/,D^2$, where $\delta$ is the maximum rounding error. Additive and multiplicative replacement methods \citep{Martin2003} also substitute zeros with small positive values. The additive approach replaces zeros and then normalizes the composition, whereas the multiplicative approach proportionally adjusts only nonzero components, preserving their ratios. In contrast, \cite{Palarea2008} treats zeros as missing values and estimates them iteratively within an expectation--maximization (EM) framework, offering greater flexibility than deterministic replacements.

\subsection{Density estimation for compositional data}
\label{sec:density-estimation-compositional}

Similar to multivariate data, kernel density estimation (KDE) is widely used for compositional data. \cite{Aitchison1985} proposed using either the Dirichlet or Gaussian density as the kernel function. The KDE using the Dirichlet kernel is given by
\begin{align}
    \hat{f}_{h, \Dir}(\xb) := \frac{1}{n} \sum_{i=1}^n f_\Dir\left(\xb;\frac{\Xb_i}{h} + \mathbf{1}\right), \label{eq:kde_dir}
\end{align}
where $\{\Xb_i\}_{i=1}^n$ is an independent and identically distributed (iid) sample point. Alternatively, the Gaussian kernel is applied to alr transformed compositions, yielding
\begin{align*}
    \hat{f}_{h, \Norm}(\xb) := \frac{1}{n} \sum_{i=1}^n f_{\mathrm{lr}}\bigl(\xb; \mathrm{alr}(\Xb_i), h\mathbf{S}\bigr),
\end{align*}
where $\mathbf{S}$ is the sample covariance matrix of the transformed compositions, scaled by the bandwidth parameter $h > 0$. In both estimators, $h$ has been typically selected by the method of \cite{Habbema1974} which maximizes the pseudo-likelihood
\begin{align}
    \hat{h}_\mathrm{KL} := \argmax_{h>0} \sum_{i=1}^n \,\log \hat{f}_{h, -i}(\Xb_i),
    \label{eq:kde_kl}
\end{align}
where $\hat{f}_{h, -i}(\Xb_i)$ denotes the leave-one-out KDE at the observation $\Xb_i$.

\cite{Chacon2011} modified the Gaussian kernel method by replacing the alr with the ilr transformation, addressing asymmetry and non-isometry in alr-transformed data. They further proposed two full-bandwidth selection strategies via cross-validation and an unconstrained plug-in estimator. These modifications improved density estimation accuracy, but the methods remain tied to log-ratio transformations that inherit limitations in handling zeros.

Finite mixtures provide a flexible framework for density estimation. A Dirichlet mixture with $m$ components is given by
\begin{align*}
     f(\xb; \pib, \Ab) := \sum_{j=1}^m \pi_j \, f_\Dir(\xb; \alphab_j),~~\pi_j > 0,\, \sum_{j=1}^m \pi_j = 1, 
\end{align*}
where $\Ab = (\alphab_1, \dots, \alphab_m)^\top$ denotes the parameters of components and $\pib = (\pi_1, \dots, \pi_m)^\top$ the mixing proportions. There seems to be very little study using finite Dirichlet mixtures in the literature. This may likely be owing to the likelihood degeneracy, where components collapse onto single observations, driving the likelihood to infinity and yielding inadequate estimates. \citet{Bouguila2004} proposed an algorithm to estimate the parameters of finite Dirichlet mixtures based on maximum likelihood and Fisher scoring, but most likely it was the local likelihood maxima that were produced, as typically in the case of heteroscedastic Gaussian mixtures. As part of our study, we also implemented the EM algorithm, which confirmed our understanding above.

\section{A new approach using nonparametric mixtures}
\label{sec:nonp}

\subsection{Nonparametric Gaussian mixtures via log-ratio transformations}

A nonparametric mixture density can be expressed in the form
\begin{align*}
    f(\xb;G) := \int f(\xb; \thetab) \,\rd G(\thetab), 
\end{align*}
where $f(\cdot; \thetab)$ is the component density and $G$ a mixing distribution entirely unspecified. Maximum likelihood is a commonly used method to find the estimate of $G$. An important property is that there must exist a nonparametric maximum likelihood estimate (NPMLE) $\widehat{G}$ of $G$ that is a discrete distribution with no more support points than the number of distinct observations \citep{Laird1978, Lindsay1983}. Thus, the search for $\widehat{G}$ can be restricted to discrete distributions. For any discrete $G$ with support points $\thetab_1, \dots, \thetab_m$ and proportions $\pi_1, \dots, \pi_m$,
\begin{align*}
    G(\thetab) = \sum_{j=1}^m \pi_j \, \delta_{\thetab_j},
\end{align*}
where $\delta_{\thetab_j}$ denotes a point mass distribution at $\thetab = \thetab_j$. For an iid sample $\{\xb_i\}_{i=1}^n$, the log-likelihood is
\begin{align*}
    \ell(G) = \sum_{i=1}^n \log \Bigl\{\sum_{j=1}^m \pi_j \, f(\xb_i; \thetab_j)\Bigr\},
\end{align*}
where $m$ is the number of mixture components, which is not fixed and must also be determined from the sample.

The \emph{gradient function}, which characterizes $\widehat{G}$, is defined as
\begin{align}
    d(\thetab; G) \equiv \frac{\partial \ell \{(1-\epsilon)G+\epsilon\delta_{\thetab}\}}{\partial \epsilon} \bigg|_{\epsilon=0} = \sum_{i=1}^{n} \frac{f(\xb_i; \thetab)}{f(\xb_i; G)} - n. \label{eq:gd}
\end{align}
This function is the directional derivative of the log-likelihood in the direction from $G$ to the point mass $\delta_{\thetab}$, evaluated at $G$. When $d(\thetab; G)$ is positive, it means moving an infinitesimal step in the direction towards $\delta_{\thetab}$ increases the log-likelihood, while a negative value represents a decrease. Consequently, the NPMLE can be characterized through the gradient function, since
\begin{align*}
    \widehat{G} \text{ maximizes } \ell(G) \Longleftrightarrow \widehat{G} \text{ minimizes } \underset{\thetab}{\sup}\{d(\thetab; G)\} \Longleftrightarrow \underset{\thetab}{\sup}\{d(\thetab; \widehat{G})\} = 0.
\end{align*}

Utilizing these properties of the NPMLE, \citet{Wang2015} developed efficient algorithms for multivariate nonparametric Gaussian mixtures, which can be directly applied to density estimation of log-ratio transformed compositional data.

\subsection{Nonparametric Dirichlet mixtures}
\label{sec:NDM}

When Dirichlet distributions are used as mixture components, the density is defined as
\begin{align*}
    f_h(\xb; G) :=\; \sum_{j=1}^m \pi_j \, f_\Dir(\xb; \alphab_j)
    =\; \sum_{j=1}^m \pi_j \, f_\Dir\Bigl(\xb; \frac{\thetab_j}{h} + \mathbf{1}\Bigr),
\end{align*}
where $h > 0$ serves as the bandwidth, and $\thetab_j \in \Delta^d$ is the mode of the $j$th mixture component. We require that the component parameter satisfy $\alphab_j \geq \mathbf{1}$, which ensures each Dirichlet component is unimodal and log-concave. Observe that the parametrization of $\alphab_j$ is inspired by the KDE \eqref{eq:kde_dir}.

Maximizing the likelihood simultaneously over both $h$ and $G$ is infeasible, as $\widehat{G}$ degenerates to the empirical distribution, leading to infinite likelihood when $h \to 0$. To avoid this, $\widehat{G}_h$ is defined as the NPMLE under fixed bandwidth $h$. With the relationship 
\begin{align}
    \alpha_0 := \alphab_j^\top \mathbf{1} = h^{-1} + D, 
    \label{eq:alpha_h}
\end{align}
it yields a homoscedastic mixture structure, in the sense that all components share a common concentration parameter. The corresponding log-likelihood is
\begin{align*}
    \ell_h(G) = \sum_{i=1}^n \log\{f_h(\xb_i; G)\}.
\end{align*}
Since a discrete $G$ is fully determined by its positive masses $\pib$ and support points $\Ab = (\thetab_1, \dots, \thetab_m)$, the log-likelihood can be equivalently written as $\ell_h(G) \equiv \ell_h(\pib, \Ab)$.

\subsection{Computing \texorpdfstring{$\widehat{G}_h$}{Gh-hat}}

For multivariate observations, \citet{Wang2015} proposed a hybrid method called the CNMM to compute $\widehat{G}_h$. The method alternates between two specialized algorithms: the EM algorithm and the constrained Newton method (CNM). The EM step updates $\pib$ and $\thetab$ iteratively under fixed $m$, as in finite mixtures, while the CNM identifies and includes new support points $\thetab$ and reallocates $\pib$. Consequently, $m$ changes dynamically during the iterations. To apply CNMM in the Dirichlet mixture setting, the algorithms must be adapted to satisfy the constraints in Section~\ref{sec:NDM}: each component parameter vector obeys $\alphab_j \geq \mathbf{1}$, and all components share the same concentration parameter $\alpha_{0j} = \alpha_0$, where $\alpha_0$ is fixed given $h$.

With a constant $\alpha_0$, the EM algorithm treats the model as a homoscedastic finite mixture. At each iteration, the parameters $(\alphab_j, \pi_j)$ for the $j$th component are updated as
\begin{align*}
    p_{ij} := \frac{\pi_j f_{ij}}{\sum_{l=1}^m \pi_l f_{il}}, \qquad
    \pi_j' := \frac{1}{n} \sum_{i=1}^n p_{ij}, \qquad
    \alphab_j' := \argmax_{\alphab_j} Q_j(\alphab_j), 
\end{align*}
where $f_{ij} := f(\xb_i; \alphab_j)$, a primed variable indicates an update of it, and 
\begin{align}
    Q(\alphab_j) = \sum_{i=1}^n p_{ij} \Bigl\{- \sum_{k=1}^D \log\Gamma(\alpha_{jk}) + \sum_{k=1}^D (\alpha_{jk} - 1) \log x_{ik} \Bigr\},~~\alphab_j^\top \mathbf{1} = \alpha_0, \, \alphab_j \geq \mathbf{1}. \label{eq:obja}
\end{align}
The gradient vector $\gd_j = (g_{jk})$ and Hessian matrix $\Hess_j = (h_{jkl})$ of function $Q_j$ have the following elements, respectively,
\begin{align*}
    g_{jk} &= \frac{\partial Q_j(\alphab_j)}{\partial \alpha_{jk}} = \sum_{i=1}^n p_{ij} \left\{-\psi(\alpha_{jk}) + \log X_{ik} \right\}, \\
    h_{jkl} &=
    \begin{cases}
        \frac{\partial^2 Q_j(\alphab_j)}{\partial \alpha_{jk}^2} = - \sum_{i=1}^{n} p_{ij} \, \psi'(\alpha_{jk}), &\ell = k,  \\[0.3cm]
        \frac{\partial^2 Q_j(\alphab_j)}{\partial \alpha_{jk} \, \partial \alpha_{j\ell}} = 0, &\ell \neq k,
    \end{cases}
\end{align*}
for $k = 1, \ldots, D$ and $l = 1, \ldots, D$, where $\psi$ and $\psi'$ denote the digamma and trigamma functions. Since $\Hess_j$ is diagonal and negative-definite, $Q_j$ is strictly concave with a unique global maximizer. However, the maximizer has no closed-form expression due to the log-gamma term. We approximate the objective function \eqref{eq:obja} with a second-order Taylor series expansion about the current iterate $\alphab_j$, giving 
\begin{align*}
Q_\lambda(\alphabt_j) \approx \; \gd_j^\top(\alphabt_j - \alphab_j)
    + \tfrac{1}{2}(\alphabt_j - \alphab_j)^\top 
     \Hess_j (\alphabt_j - \alphab_j)
      + \lambda (\alphabt_j^\top \mathbf{1} - \alpha_0),~~\alphabt_j \geq \mathbf{1},
\end{align*}
where the equality constraint $\alphabt_j^\top \mathbf{1} = \alpha_0$ is enforced using a Lagrange multiplier. The stationary point of this quadratic approximation satisfies
\begin{align}
\lambda = \frac{\mathbf{1}^\top \Hess_j^{-1} \gd_j}{\mathbf{1}^\top \Hess_j^{-1} \mathbf{1}}, \qquad
\alphabt_j = \alphab_j - \Hess_j^{-1} \gd_j + \lambda \Hess_j^{-1} \mathbf{1}. \label{eq:alphat}
\end{align}
This provides the formulae for updating $\alphab_j$, which is essentially a single constrained Newton step. Note that with a diagonal $\Hess_j$,
\eqref{eq:alphat} simplifies to  
\begin{align}
\lambda = \frac {\sum_{k=1}^D g_{jk} h_{jkk}^{-1}} {\sum_{k=1}^D h_{jkk}^{-1}}, \qquad
\alphat_{jk} = \alpha_{jk} - \frac {g_{jk} - \lambda} {h_{jkk}},~~k = 1, \dots, D. \label{eq:alphat2}
\end{align}

To enforce $\alphab_j \geq \mathbf{1}$ throughout the optimization, we resort to the active set method \citep{Wolfe1959} and partition the elements of $\alphab_j$ into two sets: the set of the active restrictions $S_{=1} = \{k: \alpha_{jk} = 1\}$ and the inactive set $S_{>1} = \{k: \alpha_{jk} > 1\}$. It is then decided if any element in $S_{=1}$ can be included in the next update by \eqref{eq:alphat2}. Consider a $k \in S_{=1}$. Since 
$h_{jk}$ is negative, we have an update $\alphat_{jk}> 1$, thus valid, if and only if $g_{jk} - \lambda > 0$. It is thus safe to move $k$ from $S_{=1}$ to $S_{>1}$, if $g_{jk} - \lambda_{k} > 0$,
where 
\begin{align*}
\lambda_{k} = \frac {\sum_{l \in S_{>1} \cup \{k\}} g_{jl} h_{jll}^{-1}} {\sum_{l \in S_{>1} \cup \{k\}} h_{jll}^{-1}}. 
\end{align*}
Out of all such feasible $k$s, we include the one that gives the largest positive value of $g_{jk} - \lambda_{k}$ when updating with \eqref{eq:alphat2}, along with those in $S_{>1}$, where the corresponding $\lambda_k$ is used in place of $\lambda$. The other elements in $S_{=1}$ will remain unchanged, i.e., being $1$.
The resulting update, however, may potentially make some elements in $S_{>1}$ violate the restriction $\ge 1$. If this occurs, we pull the updating vector back to the point on the boundary and then move possibly one element from $S_{>1}$ to $S_{=1}$. This thus allows for free movement, if beneficial, of any element between $S_{=1}$ and $S_{>1}$. In aid of a line search criterion, the objective function is ensured to be monotonically increasing and eventually optimized at the unique global maximizer. 

After updating $\pib$ and $\Ab$ via the EM algorithm, the CNMM method expands the support set $\Ab$ with the local maxima of the gradient function \eqref{eq:gd}, due to their importance for characterizing the NPMLE. However, in multivariate cases, direct optimization to locate multiple local maxima can be computationally very expensive. To alleviate this, the CNMM makes a novel use of the property of the gradient function. It resorts to a random grid, i.e., a random sample drawn from the mixture density that is proportional to 
\begin{align*}
    f(\alphab) = \sum_{i=1}^n w_i\,f_\Dir(\xb_i; \alphab), 
\end{align*}
where $w_i = f(\xb_i; G)^{-1}$. This function is the gradient function \eqref{eq:gd} without the additive constant ``$-n$''. Notably, $\alphab \mapsto f_i(\alphab) = c_i f_\Dir(\xb_i; \alphab)$ defines a density in $\alphab$ with $\xb_i$ as a parameter, where $c_i = c(\xb_i)$ is the normalizing constant. It can be viewed as a continuous version of the multinomial probability mass function, up to a normalizing constant. However, for a multinomial distribution, $\alphab$ can only take on integer values. To obtain a well-dispersed sample, simulated integers are jittered with random perturbations drawn from a uniform distribution $\mathrm{Unif}(-0.5, 0.5)$. This random grid approach tends to produce more points in the neighborhood of the maxima of the gradient function. Using each of these points as an initial value $\alphab$, the modal EM algorithm \citep{Li2007} is employed to iteratively update $\alphab$ to locate the local maximum nearby, using the iteration formulae
\begin{align}
    p_i &= \frac{w_i f_\Dir(\xb_i; \alphab)}{f(\alphab)}, ~~ i = 1, \dots, n,  \nonumber\\
    \alphab' &= \argmax_{\alphab} ~ \sum_{i=1}^n p_i \log f_\Dir(\xb_i; \alphab),~~\alphab^\top \mathbf{1} = \alpha_0, ~~ \alphab \geq \mathbf{1}. \label{eq:mem}
\end{align}

Note that although the normalizing constant $c_i$ of the ``continuous'' multinomial distribution has no closed-form expression, not knowing it does not hinder the application of the modal EM algorithm here. This is because all of the unknown normalizing constants cancel out nicely in the iteration formulae. The objective function in \eqref{eq:mem} for updating $\alphab$ retains the same structure and constraints as \eqref{eq:obja}, so the same optimization strategy applies. 

The addition of new support points induces a reallocation of the mixing proportions. For a fixed $\Ab$, $\pib$ is updated by maximizing the second-order Taylor expansion
\begin{align*}
    \ell_h(\pib',\Ab) \approx \ell_h(\pib, \Ab) - \frac{1}{2} \Vert\mathbf{S} \pib'-\mathbf{2}\Vert^2 + \frac{n}{2},
\end{align*}
where $\mathbf{s}_i = \partial \log \{f_h(\xb_i; G)\} / \partial \pib$, $\mathbf{S} = \left(\mathbf{s}_1, \dots, \mathbf{s}_n \right)^\top$, and $\Vert\cdot\Vert$ denotes the $\ell_2$-norm \citep{Wang2007}. Maximizing the quadratic approximation above is equivalent to solving
\begin{align*}
    \min \limits_{\pib'} \Vert\mathbf{S} \pib' - \mathbf{2}\Vert^2,~~\pib'\mathbf{1} = \mathbf{1}, \, \pib' \geq \mathbf{0}.
\end{align*}
During this optimization, some components may be assigned a mixing proportion of exactly zero. Since a support point with null weight contributes nothing to either the mixture density or the likelihood, it can be discarded. This serves as a shrinkage step which, together with the expansion step, dynamically adjusts the number of mixture components.

\begin{figure*}[!ht]
    \centering
    \includegraphics[page=1, width=0.3\linewidth,
        clip,trim={1cm 0.8cm 1cm 0.7cm}]{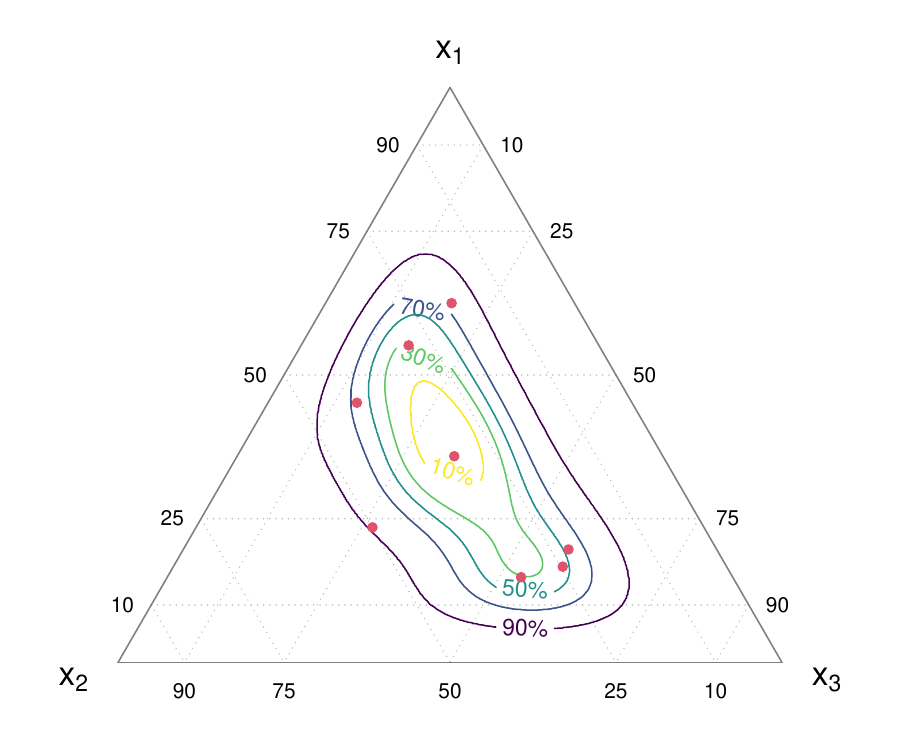}%
    \hspace{0.1cm}
    \includegraphics[page=2, width=0.3\linewidth,
        clip,trim={0.9cm 0.8cm 0.9cm 0.7cm}]{gradient.pdf}%
    \hspace{0.1cm}
    \includegraphics[page=3, width=0.3\linewidth,
        clip,trim={1cm 0.8cm 1cm 0.7cm}]{gradient.pdf}%
    
    \caption{\small Process of one CNMM iteration: 
    mixture density after EM step (left); 
    modal EM iteration paths from random grid points (gray dots) to modes (blue dots) of the gradient function within the feasible region $\alphab \geq \mathbf{1}$ (blue dashed triangle) (middle); 
    mixture density after CNM step (right).}
    \label{fig:cnmm}
\end{figure*}

To enhance convergence speed, alternating five EM iterations with one CNM iteration within each CNMM iteration was found most efficient. Figure~\ref{fig:cnmm} illustrates a single CNMM iteration applied to simulated data from the mixture in Figure~\ref{fig:chacon}(4). The left and right panels display the density contours and support points on the simplex of the estimated mixture after the EM and CNM steps, respectively. Some red support points in the right panel are retained from the left, while others are removed during the shrinkage step. The two blue points denote new support points expanded from the modal EM, with the iteration process shown in the middle as paths from random grid points (gray dots) to local maxima (blue dots) of the gradient function in the left panel. Note that the simplex in the middle represents the parameter space of $\alphab$, analogous to $\Delta^2$ since $\alphab$ is positive, has the same dimension as $\xb$, and sums to a fixed $\alpha_0$.

\subsection{Loss functions}
\label{sec:LossFun}

In density estimation for compositional data, three widely used criteria are IAE, ISE, and KLD, which are defined as
\begin{align*}
    \mathrm{IAE}(f, \hat{f}) &= \int \bigl|f(\xb) - \hat{f}(\xb)\bigr| \,\rd \xb, \nonumber\\
    \mathrm{ISE}(f, \hat{f}) &= \int \bigl[f(\xb) - \hat{f}(\xb)\bigr]^2 \,\rd \xb, \nonumber\\
    \mathrm{KLD}(f, \hat{f}) &= \int f(\xb)\,
    \log \biggr(\frac{f(\xb)}{\hat{f}(\xb)}\biggl) \,\rd \xb,
\end{align*}
where $\hat{f}$ is an estimate of the true density $f$. The integrals are rarely available in closed form, but they can be approximated using Monte Carlo methods. For IAE and ISE, importance sampling is particularly convenient. Let $\{\xb_i\}_{i=1}^n$ be iid samples from a proposal density $q$ supported on the simplex. Expressing the integrals as expectations with respect to $q$ gives
\begin{align}
    \mathrm{ISE}(f, \hat{f}) = \frac{1}{n} \sum_{i=1}^n 
   \frac{\bigl[f(\xb_i) - \hat{f}(\xb_i) \bigr]^2}{q(\xb_i)}, \qquad
    \mathrm{IAE}(f, \hat{f}) = \frac{1}{n} \sum_{i=1}^n 
   \frac{\bigl| f(\xb_i) - \hat{f}(\xb_i) \bigr|}{q(\xb_i)}.
   \label{eq:imp_samp}
\end{align}
A proposal $q$ well aligned with the target improves the accuracy of these approximations. For KLD, if a sample is available directly from $f$, the approximation can be computed using
\begin{align}
    \mathrm{KLD}(f, \hat{f}) =  \frac{1}{n} \sum_{i=1}^n \log f(\xb_i) - \frac{1}{n} \sum_{i=1}^n \log \hat{f}(\xb_i).
    \label{eq:kld} 
\end{align}

In almost all practical scenarios, $f$ is unknown. Consequently, the IAE cannot be used, as it requires direct evaluation of $f$, making an accurate approximation infeasible. By contrast, the first term of KLD in \eqref{eq:kld} is constant with respect to $\hat{f}$ and can therefore be omitted when comparing estimators. A similar observation applies to the ISE. Let $\{\xb_i\}_{i=1}^n$ and $\{\yb_j\}_{j=1}^m$ be iid draws from $f$ and $\hat{f}$, respectively. The expanded ISE can be approximated as
\begin{align}
    \mathrm{ISE}(f, \hat{f}) 
    &= \int f(\xb)^2 \,\rd\xb - 2\int f(\xb)\hat{f}(\xb) \,\rd\xb ~~ + \int \hat{f}(\xb)^2 \,\rd\xb, \label{eq:ISExp}\\
    & \approx C - \frac{2}{n} \sum_{i=1}^n \hat{f}(\xb_i) + \frac{1}{m}\sum_{j=1}^m \hat{f}(\yb_j),
    \label{eq:ISE}
\end{align}
where $C$ is a constant independent of $\hat{f}$ that does not affect estimator comparison. Since samples cannot be generated when $f$ is unknown, $\{\xb_i\}_{i=1}^n$ typically correspond to observed real-world data.

Since the log-ratio transformation and its inverse provide a mapping between the simplex and Euclidean spaces, both compositions and densities can be consistently transferred across the two domains. Consequently, loss functions can be evaluated in either space. Specifically, IAE and KLD are invariant under a one-to-one transformation of the density, as exemplified next by the ilr transformation. Let $\yb = \mathrm{ilr}(\xb)$ and $f$ be defined on $\Delta^d$. The induced density on $\Rb^d$ is 
\begin{align*}
    g(\yb) = f\left(\mathrm{ilr}^{-1}(\yb)\right)\, \bigl|\det J_{\mathrm{ilr}^{-1}}(\yb)\bigr|,
\end{align*}
where $|\det J_{\mathrm{ilr}^{-1}}(\yb)| = |\det \partial \, \mathrm{ilr}^{-1}(\yb) \,/\,\partial \yb^\top|=|\det J_{\mathrm{ilr}}(\mathrm{ilr}^{-1}(\yb))|^{-1}$. By change of variables,
\begin{align*}
    \mathrm{IAE}_{\Delta^d} (f, \hat{f})
    & = \int_{\Delta^d} \bigl|f(\xb) - \hat{f}(\xb) \bigr| \,\rd \xb \\
    &= \int_{\Rb^d} \bigl|f(\mathrm{ilr}^{-1}(\yb)) - \hat{f}(\mathrm{ilr}^{-1}(\yb))\bigr| \, \bigl|\det J_{\mathrm{ilr}^{-1}}(\yb)\bigr| \,\rd \yb = \mathrm{IAE}_{\Rb^d}(g, \hat{g}); \\[0.2cm]
    \mathrm{KLD}_{\Delta^d} (f, \hat{f})
    & = \int_{\Delta^d} f(\xb) \log \frac{f(\xb)}{\hat{f}(\xb)} \,\rd \xb \\
    & = \int_{\Rb^d} f\left(\mathrm{ilr}^{-1}(\yb)\right)\, \bigl|\det J_{\mathrm{ilr}^{-1}}(\yb)\bigr| \, \log \frac{g(\mathrm{ilr}^{-1}(\yb)) \,/\, \bigl|\det J_{\mathrm{ilr}^{-1}}(\yb)\bigr|}{\hat{g}(\mathrm{ilr}^{-1}(\yb)) \,/\, \bigl|\det J_{\mathrm{ilr}^{-1}}(\yb)\bigr|} \,\rd \yb = \mathrm{KLD}_{\Rb^d}(g, \hat{g}).
\end{align*}
However, $\mathrm{ISE}_{\Delta^d} (f, \hat{f}) \neq \mathrm{ISE}_{\Rb^d} (g, \hat{g})$, i.e., the performance of an estimator $\hat{f}$ is different from its equivalent version $\hat{g}$, under the expected squared error loss.

\subsection{Issues of squared loss}
\label{sec:ISE_Issues}

For the ISE, numerical challenges arise when applying Gaussian-based methods (with Gaussian kernels or mixture components) directly on the simplex. Consider a two-part composition $\xb = (x, 1-x)^\top \in \Delta^1$. According to \eqref{eq:lrt} and \eqref{eq:det}, we have 
\begin{align*}
    y = \mathrm{ilr}(\xb) = \frac{\mathrm{logit}(x)}{\sqrt{2}}, \qquad\bigl|\det J_\mathrm{ilr}(\xb) \bigr| = \bigl[\sqrt{2}x(1-x)\bigr]^{-1}.
\end{align*}
Suppose $Y \sim \mathcal{N}(\mu, \sigma^2)$. By change of variables, the induced density of $\Xb = \mathrm{ilr}^{-1}(Y)$ on $\Delta^1$ is
\begin{align}
    f(\xb; \mu, \sigma) 
    = \bigl|\det J_\mathrm{ilr}(\xb) \bigr| \, f_{\mathrm{Norm}}(\mathrm{ilr}(\xb); \mu, \sigma) = \frac{\exp \biggl[
    -\frac{1}{2\sigma^{2}}
    \Bigl(\,\frac{\mathrm{logit}(x)}{\sqrt{2}}-\mu \Bigr)^{2}
    \biggr]}{2\sigma\sqrt{\pi}\,x(1-x)}.
    \label{eq:simplexone}
\end{align}
For $\mu = 0$ and $\sigma = 1$, the density is symmetric with mode $x = 0.5$ and modal value $1.13$. Increasing $\mu = 3$ shifts the mode to $x \approx 0.998$ with density $53.6$; while increasing $\sigma = 3$ (with $\mu=0$) yields a bimodal distribution with modes approaching the two boundaries, both reaching a density of $107.8$. Based on \eqref{eq:simplexone}, the integral of the squared density on $\Delta^1$ is
\begin{align*}
     \int_{\Delta^1} f(\xb; \mu, \sigma)^2\, \rd \xb = \frac{\sqrt{2}}{2\sigma \sqrt{\pi}} \biggl[\exp\Bigl(\frac{\sigma^2}{2}\cosh(\sqrt{2}\mu) +1\Bigr)\biggr],
\end{align*}
which constitutes one term of \eqref{eq:ISExp} and grows exponentially with respect to both $\mu$ and $\sigma$. This demonstrates that even moderate changes in $\mu$ or $\sigma$ can result in highly peaked densities near the simplex boundaries and an exponential growth of the squared density integral after inverse log-ratio transformation. These two effects explain the numerical instability of Gaussian-based methods for ISE computation on the simplex: the sharp concentration near the boundaries is too localized to be captured by standard quadrature or Monte Carlo integration, while the rapidly increasing squared integral amplifies numerical errors and undermines computational stability. 

The numerical instability motivates evaluating ISE in the Euclidean space, where Gaussian-based methods can be applied without transformation issues. However, this requires expressing Dirichlet-based methods (with Dirichlet kernels or mixture components) under the log-ratio map. For $X \sim \mathrm{Beta}(\alpha, \beta)$, or $\Xb = (X, 1 - X)^\top \sim \Dir((\alpha,\beta)^\top)$ on $\Delta^1$, the induced density of $Y = \mathrm{ilr}(\Xb)$ is
\begin{align*}
    f(y;\alpha, \beta) = \bigl|\det J_\mathrm{ilr}(y)\bigr|^{-1} \, f_\mathrm{Beta}\bigl(\mathrm{ilr^{-1}}(y); \alpha,\beta\bigr) = \frac{2 \exp\bigl(2\sqrt{2}\alpha y\bigr)}{\mathrm{B}(\alpha, \beta)^{2} \Bigl(1+\exp\bigl(\sqrt{2}y\bigr)\Bigr)^{2(\alpha+\beta)}}.
\end{align*}
Closed forms are available for the modal density value and the squared-density integral:
\begin{align*}
     & f_\mathrm{mode} = 2\mathrm{B}(\alpha, \beta)^{-2} \alpha^{2\alpha} \beta^{2\beta} (\alpha + \beta)^{-2 (\alpha + \beta)}, \\
     & \int_{\Rb^1} f(y; \alpha, \beta)^2 \,\rd y
     = \frac{\sqrt{2}\mathrm{B}(2\alpha, 2\beta)}{\mathrm{B}(\alpha, \beta)^2}.
\end{align*}
The density function $f(\cdot; \alpha, \beta)$ is smooth without sharp peaks, and both modal density and squared density integral vary smoothly in $(\alpha,\beta)$. Thus, comparing ISE across methods in the Euclidean space is numerically stable. An important limitation is that the log-ratio transform is undefined for compositions containing zeros, so such observations have to be excluded or relocated.

To deal with the non-invariance of ISE on $\Delta^d$ and $\mathbb{R}^d$, \citet{Chacon2011} considered the measure $\lambda_a$ with density with respect to the Lebesgue measure $\lambda$ given by the Radon--Nikodym derivative $[\rd\lambda_a/\rd \lambda](\xb)=|\det J_{\mathrm{ilr}}(\xb)|=(\sqrt{D}x_1\cdots x_D)^{-1}$ on $\Delta^d$, defining $f_a$ and $\hat{f}_a$ as densities with respect to this measure, and hence making $\smash{\mathrm{ISE}_{\Delta^d,\lambda_a} (f_a, \hat{f}_a)}$ and $\mathrm{ISE}_{\mathbb{R}^d} (g, \hat{g})$ equal. This has the advantage of transferring integrals on $\Delta^d$ to more manageable integrals on $\mathbb{R}^d$, but at the expense of introducing a measure that assigns unbounded weights to the boundaries of the simplex. Note that considering $\lambda$ or $\lambda_a$ within the IAE and KLD metrics makes no difference: $\smash{\mathrm{IAE}_{\Delta^d,\lambda_a} (f_a, \hat{f}_a)=\mathrm{IAE}_{\Delta^d} (f, \hat{f})}$ and $\smash{\mathrm{KLD}_{\Delta^d,\lambda_a} (f_a, \hat{f}_a)=\mathrm{KLD}_{\Delta^d} (f, \hat{f})}$.

\subsection{Bandwidth selection and initialization}
\label{sec:BandSelect}

The bandwidth $h$ controls the concentration of mixture components and thus the smoothness of the estimated density. For each fixed $h$, the resulting mixture is a density estimator. Hence, bandwidth selection is a model selection problem, and we address this using deterministic criteria or $k$-fold cross-validation.

For a deterministic criterion, one can adopt, e.g., the Akaike information criterion (AIC), as used by \cite{wang2012} and \cite{Wang2015}, which is given by
\begin{align*}
    \mathrm{AIC}(h) = - 2\widetilde{\ell}(h) + 2p
\end{align*}
where $p$ is the number of free parameters. The optimal bandwidth with this criterion is then
\begin{align}
    \hat{h}_\mathrm{AIC} = \argmin_{h > 0} \, \mathrm{AIC}(h). \label{eq:haic}
\end{align}
We search for $\hat{h}_\mathrm{AIC}$ by gradually decreasing $h$ from a initial value. The AIC typically decreases at first and then increases. The search stops once the AIC begins to rise and exceeds a specified threshold. Note that the asymptotic normality of maximum likelihood does not hold for mixture models \citep{Lindsay1995}, and thus AIC may lack reliability in this context.

For $k$-fold cross-validation, the data are randomly partitioned into $K$ subsets $P_1, \dots, P_K$ of approximately equal size. Each subset is used once as validation, while the remaining $K-1$ subsets serve as the training set. Averaging over folds yields the validation loss. We define two loss functions based on \eqref{eq:ISE} and \eqref{eq:kld} with constant term omitted, given by
\begin{align*}
\mathrm{CVISE}(h) &=  \frac{1}{K} \sum_{j=1}^K 
     \int_{\Delta^d} f_h^{(j)}(\xb)^2 \,\rd\xb - \frac{2}{K} \sum_{j=1}^K \frac{1}{n_j} \sum_{\xb_i \in P_j} f_h^{(j)}(\xb_i), \\
\mathrm{CVKLD}(h) & = -\frac{1}{K} \sum_{j=1}^K \frac{1}{n_j} 
     \sum_{\xb_i \in P_j} \log f_h^{(j)} (\xb_i).
\end{align*}
where $f_h^{(j)}$ is the Dirichlet mixture density estimated on the training data excluding fold $P_j$, and $n_j$ is the number of observations in $P_j$. The optimal bandwidths are
\begin{align}
    \hat{h}_\mathrm{CVISE} = \argmin_{h>0} \, \mathrm{CVISE}(h), \qquad
    \hat{h}_\mathrm{CVKLD} = \argmin_{h>0} \, \mathrm{CVKLD}(h).
    \label{eq:hcv}
\end{align}
In most cases, zero-containing compositions are included in model fitting and bandwidth selection. The only exception occurs when a validation point $\xb_i$ with zeros lies on the simplex boundary where $\smash{f_h^{(j)}}$ has no support, leading to $\smash{f_h^{(j)}(\xb_i)=0}$ and causing $\mathrm{CVKLD}(h)$ to diverge to $-\infty$. To avoid this, such validation points with zero densities are excluded from the computation of $\hat{h}_\mathrm{CVKLD}$.

To select $h$ with \eqref{eq:haic} and \eqref{eq:hcv}, a grid search is conducted over a data-adaptive range. A single Dirichlet distribution is first fitted to the data, and the initial bandwidth $\hat{h}_0$ is computed based on the relationship of \eqref{eq:alpha_h}. The value $\hat{h}_0$ corresponds to the smoothest admissible mixture component and is taken as the upper bound of the search grid. To explore mixture components with higher concentration, we scale down the standard deviation of the single Dirichlet distribution by a factor $\eta \in (0, 1)$. Assuming $\Xb \sim \Dir(\alphab)$, the standard deviation of an element $X_j$ of $\Xb$ is
\begin{align}
    \mathsf{sd}(X_j) = \sqrt{\frac{\mu_j(1-\mu_j)}{\alpha_0+1}}, \label{eq:Dir_sd}
\end{align}
where $\mub = \alphab\,/\,\alpha_0$ is the mean of the distribution. The standard deviation is inversely proportional to $\sqrt{\alpha_0 + 1}$. Consequently, reducing the standard deviation by a factor of $\eta$ increases $\alpha_0 + 1$ by a factor of $\eta^{-2}$. Using that $\hat{\alpha}_0 = \hat{h}_0^{-1} + D$ (see \eqref{eq:alpha_h}), the grid of explored bandwidths is then given explicitly by
\begin{align*}
    \hat{h}_\eta = \{(\hat{h}_0^{-1} + D + 1)\,\eta^{-2} - (1 + D)\}^{-1},~~\eta \in (0, 1).
\end{align*}

Once $h$ is fixed, a Dirichlet mixture is initialized by a clustering-based procedure. The first cluster center is defined as the sample mean, with all points within radius $r$ forming the initial cluster. From the remaining points, the one nearest to the sample mean is shifted outward to reduce overlap, and a new cluster is formed from points within radius $r$ of this location. The procedure is repeated until all points are assigned. Each cluster center serves as the mean $\mub$ of a mixture component. Data are reassigned to their nearest center, and the initial $\pib$ are proportional to cluster sizes.

However, directly applying the above procedure to the simplex is not ideal. With $\alpha_0$ fixed, the variance of the Dirichlet distribution is not spatially homogeneous across the simplex: distributions near the boundary are more concentrated than those at the center. Consequently, boundary clusters should occupy smaller regions than central ones. Since a fixed $r$ is used to form clusters, the initial mixture fails to capture the data structure faithfully.

To mitigate this, a variance-stabilizing transformation is applied to the compositions before clustering. It decorrelates the local variance from the mean, ensuring fixed $r$ operates uniformly across the simplex. Formally, the transformation is defined as
\begin{align*}
   y_j = f(x_j) \propto \int^{x_j}_0 v(\mu_j)^{-1/2} \, \rd\mu_j,
\end{align*}
where $v(\mu_j) = \mathsf{var}(X_j)$ for mean $\mu_j$. For the Dirichlet distribution, using the square of \eqref{eq:Dir_sd}, the explicit form becomes
\begin{align*}
y_j = f(x_j) \propto \int_0^{x_j} \left\{\frac{\mu_j(1 - \mu_j)}{\alpha_0 + 1}\right\}^{-1/2} \, \rd\mu_j = 2\sqrt{\alpha_0 + 1} \,\arcsin(\sqrt {x_j}).
\end{align*}
In this transformed space, the local variability is more homogeneous, which enables clustering-based procedure. The resulting cluster centers are then mapped back to the simplex and used to initialize the Dirichlet mixture.

\section{Simulation studies}
\label{sec:simu}

\subsection{Evaluation setup}

We evaluate the numerical performance of the proposed nonparametric mixture-based estimator (NPMDE) against KDE. The finite mixture estimator described in Section~\ref{sec:density-estimation-compositional} is excluded owing to its likelihood degeneracy phenomenon. Three KDE and four NPMDE are considered, with different bandwidth selection strategies. For the Gaussian kernel, two selectors are used: cross-validation (CV) and the unconstrained plug-in (PI), both introduced in \citet{Chacon2011} and implemented in the \texttt{ks} package \citep{package:ks}. For the Gaussian mixture of \citet{Wang2015}, we use the implementation in the \texttt{npden} package \citep{package:npden}. For the Dirichlet KDE, the bandwidth $\hat{h}_\mathrm{KL}$ in \eqref{eq:kde_kl} is adopted. For the Dirichlet NPMDE, three selectors are considered: $\hat{h}_\mathrm{AIC}$, $\hat{h}_\mathrm{CVISE}$, and $\hat{h}_\mathrm{CVKLD}$.

We employ only the ilr transformation rather than alr, since the two are linearly related and yield equivalent results in the MDE framework. In KDE, the bandwidth selector for alr-transformed data cannot estimate a full bandwidth matrix \citep[Section 3.2]{Chacon2011}, making it less flexible than the ilr-based approach.

Three loss functions from Section~\ref{sec:LossFun} are employed to evaluate performance. Loss values are approximated via Monte Carlo with an iid sample of size $10,000$ to ensure numerical accuracy. The IAE and KLD are computed on the simplex, as both are invariant under transformations between the simplex and Euclidean space. In contrast, the ISE is computed in the Euclidean space to avoid numerical instability on the simplex. The ISE and IAE are computed via \eqref{eq:imp_samp}, with proposal density $q$ chosen as a smoothed version of the true density $f$, while the KLD is computed via \eqref{eq:kld}.

The methods are evaluated under two mixture families: Gaussian mixtures and Dirichlet mixtures. Experiments are conducted on $\Delta^2$ and $\Delta^5$. For each family and dimension, twelve distinct mixtures are constructed as test cases. From each mixture, independent samples of sizes $100$ and $500$ are generated for density estimation.

\subsection{Gaussian mixtures}
\label{sec:GaussianMix}

\begin{figure*}[!ht]
  \centering
  \foreach \p [count=\i from 1] in {1,2,3,4,5,6,7,8,9,10,11,12}{
    \begin{overpic}[page=\p,width=0.15\linewidth,
      clip,trim={1.9cm 0.8cm 1.8cm 1.4cm}]{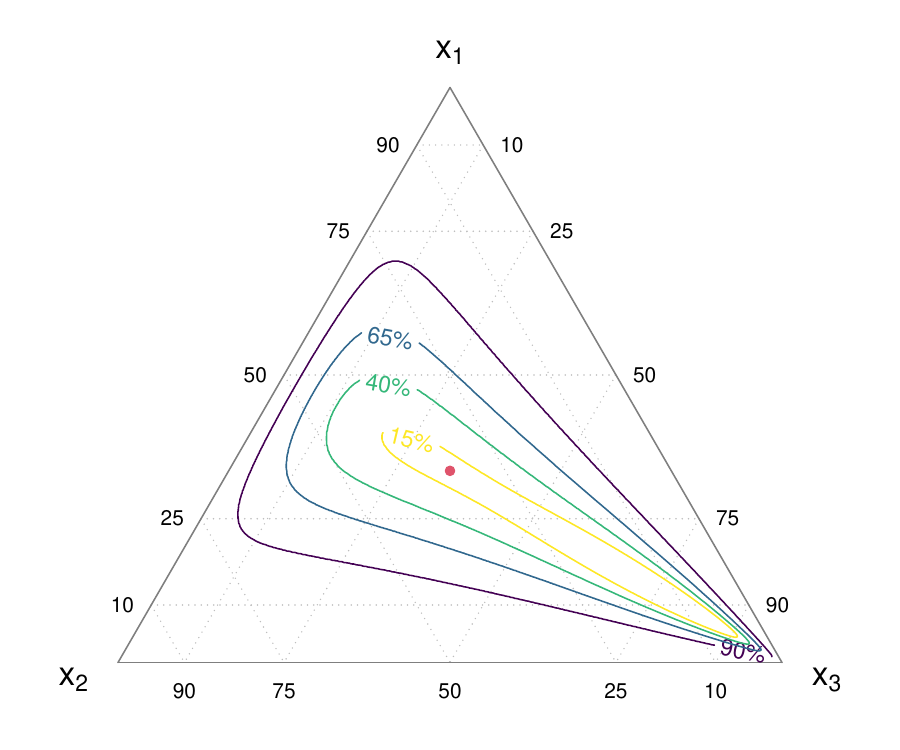}
      \put(42,-12){\footnotesize (\i)}%
    \end{overpic}%
    \ifnum\i=6\par\vspace{15pt}\fi
  }%
  \vspace{0.5cm}
  \caption{\small Probability contour plots for the ilr-Gaussian mixtures on $\Delta^2$.}
  \label{fig:chacon}
\end{figure*}

\begin{figure*}[!ht]
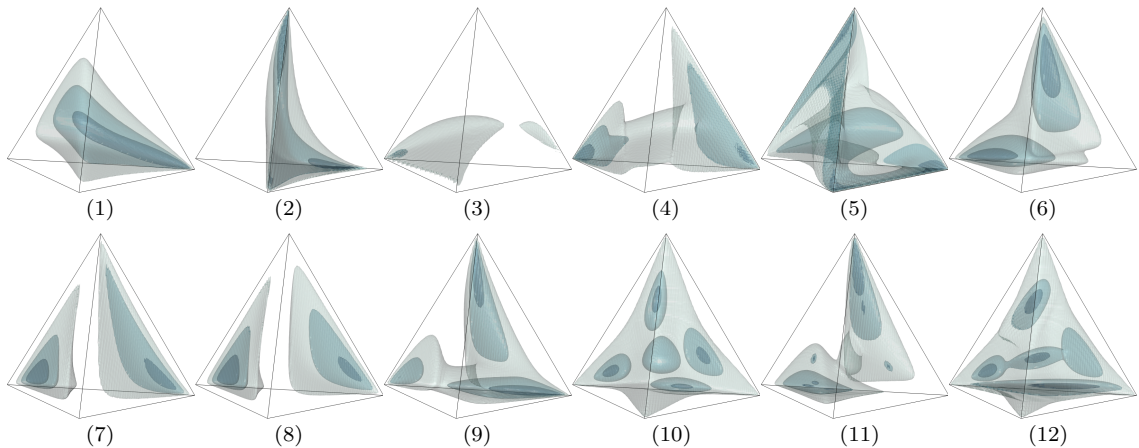

  \centering
  \foreach \i in {1,...,12}{%
    \begin{overpic}[width=0.155\linewidth,
      clip,trim={17.3cm 8.6cm 14.9cm 4.1cm}]{chacon\i.png}
      \put(42,-12){\footnotesize (\i)}
    \end{overpic}%
    \ifnum\i=6\par\vspace{15pt}\fi
  }%
  \vspace{0.5cm}
  \caption{\small Probability contour plots for the ilr-Gaussian mixtures on $\Delta^3$.}
  \label{fig:chacon3D}
\end{figure*}

Samples of Gaussian mixtures are first constructed in $\Rb^d$ and then mapped to $\Delta^d$ via the inverse ilr transformation. For $d=2$, the mixture parameters are given in \citet[Table 1]{Chacon2009}, and the corresponding visualizations on $\Delta^2$ are shown in Figure~\ref{fig:chacon}, where the red dots denote the component means after inverse ilr transformation. These mixtures are extended to higher dimensions with minor adjustments: zeros are appended to the mean vectors to preserve relative positions, and the covariance matrices are expanded by inserting diagonal elements from an arithmetic sequence spanning the original two eigenvalues, thereby defining variance in the additional dimensions. Off-diagonal elements for these dimensions are set to zero, maintaining orthogonality. For illustration, Figure~\ref{fig:chacon3D} displays the extended mixtures in $\Delta^3$.

Tables~\ref{tab:chacon_MIAE}--\ref{tab:chacon_KLD} report the mean IAE (MIAE) and mean KLD (MKLD), respectively, along with their standard errors (subscripts), obtained from $1,000$ Monte Carlo replications. These values are scaled and rounded to two decimal places for presentation purposes. In each scenario, the minimum mean loss is highlighted in bold, with some other values also highlighted if a pairwise $t$-test shows no significant difference from the minimum at the $5\%$ level.

The relative performance of the methods using MIAE or MKLD is fairly similar. Overall, ilr-MDE has the best performance. This is somewhat reasonable, as this estimator uses the same distribution family as for data generation, although ilr-MDE only uses a covariance matrix common to all mixture components. As the sample size increases, the outperformance of the ilr-MDE over the others is more remarkable. The Dirichlet-based estimators outperform ilr-PI and ilr-CV in most cases, despite the use of Gaussian kernels by the latter two. It is more so in the higher-dimensional case with $d=5$, showing a higher level of robustness in such cases. Occasionally, as in scenario (3) where the true distribution is highly skewed with a sharp peak near a corner of the simplex, the Dirichlet-based estimators may perform poorly for their relative flatness near the boundary, as compared with inverse ilr transformed Gaussian densities, as discussed in Section~\ref{sec:ISE_Issues}. 

\begin{sidewaystable*}[p]
    \centering
    \caption{\small MIAE with standard errors (subscripts) based on Monte Carlo replications in the Gaussian mixture simulations; bold highlights the minimum and those not significantly different in each scenario.}
    \setlength{\tabcolsep}{3pt}
    \small
    \begin{tabular}{llcccccccccccc}
    \toprule
    Settings & Estimators & (1) & (2) & (3) & (4) & (5) & (6) & (7) & (8) & (9) & (10) & (11) & (12) \\
    \midrule
    \multirow{7}{*}{\centering\makecell[l]{$d=2$, \\ $n=100$ \\[3pt] ($\times 10$)}}
    
    & ilr-PI & $2.72_{0.02}$ & $2.69_{0.02}$ & $\boldsymbol{3.95}_{0.01}$ & $\boldsymbol{3.98}_{0.01}$ & $3.80_{0.01}$ & $2.98_{0.01}$ & $4.20_{0.02}$ & $4.34_{0.02}$ & $3.47_{0.02}$ & $4.79_{0.01}$ & $3.78_{0.02}$ & $5.20_{0.01}$ \\
    & ilr-CV & $3.38_{0.03}$ & $3.29_{0.03}$ & $4.66_{0.03}$ & $4.48_{0.03}$ & $4.36_{0.03}$ & $3.64_{0.03}$ & $3.96_{0.03}$ & $4.06_{0.03}$ & $4.04_{0.03}$ & $5.06_{0.02}$ & $4.05_{0.03}$ & $5.27_{0.03}$ \\
    & ilr-MDE & $\boldsymbol{1.99}_{0.02}$ & $\boldsymbol{2.03}_{0.02}$ & $4.07_{0.02}$ & $4.24_{0.02}$ & $3.69_{0.02}$ & $\boldsymbol{2.63}_{0.02}$ & $\boldsymbol{2.81}_{0.04}$ & $\boldsymbol{3.22}_{0.02}$ & $\boldsymbol{3.23}_{0.02}$ & $4.93_{0.02}$ & $\boldsymbol{2.93}_{0.03}$ & $\boldsymbol{5.10}_{0.02}$ \\
    & Dir-KDE & $2.46_{0.01}$ & $3.48_{0.01}$ & $8.86_{0.02}$ & $4.20_{0.02}$ & $\boldsymbol{3.54}_{0.01}$ & $2.88_{0.02}$ & $3.93_{0.02}$ & $3.74_{0.02}$ & $3.80_{0.01}$ & $5.37_{0.03}$ & $4.02_{0.02}$ & $6.11_{0.02}$ \\
    & Dir-AIC & $2.71_{0.02}$ & $3.84_{0.02}$ & $8.44_{0.01}$ & $4.31_{0.02}$ & $3.81_{0.02}$ & $2.93_{0.02}$ & $3.98_{0.02}$ & $3.68_{0.02}$ & $4.12_{0.02}$ & $5.00_{0.02}$ & $4.29_{0.02}$ & $6.10_{0.01}$ \\
    & Dir-CVISE & $2.77_{0.02}$ & $3.87_{0.02}$ & $8.47_{0.01}$ & $4.35_{0.02}$ & $4.30_{0.04}$ & $2.93_{0.02}$ & $3.98_{0.02}$ & $3.60_{0.02}$ & $4.08_{0.02}$ & $\boldsymbol{4.67}_{0.02}$ & $4.24_{0.02}$ & $5.70_{0.02}$ \\
    & Dir-CVKLD & $2.70_{0.02}$ & $3.77_{0.02}$ & $9.60_{0.02}$ & $4.25_{0.02}$ & $3.72_{0.02}$ & $2.88_{0.02}$ & $3.89_{0.01}$ & $3.57_{0.02}$ & $3.97_{0.02}$ & $4.78_{0.02}$ & $4.14_{0.01}$ & $5.83_{0.01}$ \\

    \midrule

        \multirow{7}{*}{\centering\makecell[l]{$d=2$, \\ $n=500$ \\[3pt] ($\times 10$)}}
    & ilr-PI & $1.58_{0.01}$ & $1.59_{0.01}$ & $2.48_{0.01}$ & $2.44_{0.01}$ & $2.43_{0.01}$ & $1.79_{0.01}$ & $2.27_{0.01}$ & $2.39_{0.01}$ & $2.08_{0.01}$ & $3.04_{0.01}$ & $2.22_{0.01}$ & $3.52_{0.01}$ \\
    & ilr-CV & $1.79_{0.01}$ & $1.80_{0.01}$ & $2.75_{0.01}$ & $2.54_{0.01}$ & $2.53_{0.01}$ & $1.95_{0.01}$ & $2.16_{0.01}$ & $2.25_{0.01}$ & $2.22_{0.01}$ & $3.05_{0.01}$ & $2.27_{0.01}$ & $3.24_{0.01}$ \\
    & ilr-MDE & $\boldsymbol{0.92}_{0.01}$ & $\boldsymbol{0.93}_{0.01}$ & $\boldsymbol{2.35}_{0.01}$ & $\boldsymbol{2.25}_{0.01}$ & $\boldsymbol{1.96}_{0.01}$ & $\boldsymbol{1.30}_{0.01}$ & $\boldsymbol{1.36}_{0.02}$ & $\boldsymbol{1.63}_{0.01}$ & $\boldsymbol{1.64}_{0.01}$ & $\boldsymbol{2.70}_{0.01}$ & $\boldsymbol{1.57}_{0.02}$ & $\boldsymbol{3.16}_{0.01}$ \\
    & Dir-KDE & $1.50_{0.01}$ & $2.18_{0.01}$ & $6.44_{0.01}$ & $2.61_{0.01}$ & $2.33_{0.01}$ & $1.72_{0.01}$ & $2.37_{0.01}$ & $2.27_{0.01}$ & $2.31_{0.01}$ & $3.19_{0.03}$ & $2.52_{0.01}$ & $4.50_{0.04}$ \\
    & Dir-AIC & $1.60_{0.01}$ & $2.35_{0.01}$ & $6.69_{0.01}$ & $2.56_{0.01}$ & $2.38_{0.01}$ & $1.68_{0.01}$ & $2.27_{0.01}$ & $2.11_{0.01}$ & $2.36_{0.01}$ & $3.11_{0.01}$ & $2.71_{0.02}$ & $4.15_{0.01}$ \\
    & Dir-CVISE & $1.62_{0.01}$ & $2.39_{0.01}$ & $6.70_{0.01}$ & $2.53_{0.01}$ & $3.43_{0.03}$ & $1.70_{0.01}$ & $2.21_{0.01}$ & $2.18_{0.01}$ & $2.38_{0.01}$ & $2.93_{0.01}$ & $2.72_{0.01}$ & $3.86_{0.01}$ \\
    & Dir-CVKLD & $1.61_{0.01}$ & $2.34_{0.01}$ & $6.69_{0.01}$ & $2.44_{0.01}$ & $2.41_{0.01}$ & $1.69_{0.01}$ & $2.16_{0.01}$ & $2.15_{0.01}$ & $2.37_{0.01}$ & $3.00_{0.01}$ & $2.63_{0.01}$ & $4.01_{0.01}$ \\
    
    \midrule
    
    \multirow{7}{*}{\centering\makecell[l]{$d=5$, \\ $n=100$ \\[3pt] ($\times 10$)}}

    & ilr-PI & $10.90_{0.02}$ & $10.22_{0.03}$ & $10.92_{0.04}$ & $10.66_{0.03}$ & $11.27_{0.02}$ & $10.93_{0.03}$ & $10.98_{0.02}$ & $10.92_{0.02}$ & $10.62_{0.02}$ & $12.02_{0.02}$ & $11.03_{0.02}$ & $12.12_{0.02}$ \\
    & ilr-CV & $9.36_{0.04}$ & $8.98_{0.03}$ & $12.49_{0.11}$ & $10.06_{0.05}$ & $11.34_{0.06}$ & $9.41_{0.04}$ & $9.99_{0.06}$ & $10.13_{0.04}$ & $9.82_{0.04}$ & $13.48_{0.10}$ & $10.01_{0.04}$ & $12.47_{0.05}$ \\
    & ilr-MDE & $\boldsymbol{4.20}_{0.02}$ & $\boldsymbol{4.22}_{0.02}$ & $\boldsymbol{8.75}_{0.02}$ & $\boldsymbol{7.06}_{0.02}$ & $\boldsymbol{7.19}_{0.01}$ & $\boldsymbol{5.04}_{0.02}$ & $8.42_{0.06}$ & $8.03_{0.04}$ & $\boldsymbol{6.47}_{0.02}$ & $10.53_{0.02}$ & $7.69_{0.04}$ & $9.38_{0.01}$ \\
    & Dir-KDE & $5.12_{0.01}$ & $7.67_{0.01}$ & $13.45_{0.06}$ & $7.69_{0.01}$ & $7.54_{0.02}$ & $5.54_{0.01}$ & $7.80_{0.01}$ & $7.29_{0.01}$ & $6.88_{0.02}$ & $9.71_{0.01}$ & $\boldsymbol{7.47}_{0.01}$ & $\boldsymbol{9.14}_{0.01}$ \\
    & Dir-AIC & $6.06_{0.02}$ & $10.31_{0.02}$ & $15.57_{0.10}$ & $9.09_{0.02}$ & $8.06_{0.02}$ & $6.55_{0.02}$ & $9.35_{0.02}$ & $8.68_{0.03}$ & $8.10_{0.03}$ & $\boldsymbol{9.64}_{0.01}$ & $8.94_{0.04}$ & $\boldsymbol{9.14}_{0.01}$ \\
    & Dir-CVISE & $5.79_{0.02}$ & $9.02_{0.02}$ & $14.74_{0.07}$ & $8.05_{0.04}$ & $8.32_{0.06}$ & $5.46_{0.02}$ & $7.76_{0.02}$ & $7.25_{0.03}$ & $6.95_{0.02}$ & $10.53_{0.04}$ & $7.53_{0.02}$ & $9.94_{0.03}$ \\
    & Dir-CVKLD & $5.55_{0.01}$ & $8.58_{0.01}$ & $14.59_{0.13}$ & $7.67_{0.02}$ & $7.27_{0.01}$ & $5.55_{0.02}$ & $\boldsymbol{7.66}_{0.02}$ & $\boldsymbol{7.07}_{0.02}$ & $6.94_{0.02}$ & $9.78_{0.01}$ & $\boldsymbol{7.50}_{0.03}$ & $9.17_{0.01}$ \\
        
    \midrule

    \multirow{7}{*}{\centering\makecell[l]{$d=5$, \\ $n=500$ \\[3pt] ($\times 10$)}}
    
    & ilr-PI & $7.47_{0.01}$ & $6.87_{0.01}$ & $9.26_{0.07}$ & $7.50_{0.01}$ & $8.67_{0.03}$ & $7.49_{0.01}$ & $7.77_{0.01}$ & $7.70_{0.01}$ & $7.33_{0.01}$ & $10.15_{0.02}$ & $7.81_{0.01}$ & $9.82_{0.01}$ \\
    & ilr-CV & $5.29_{0.01}$ & $5.36_{0.01}$ & $9.43_{0.04}$ & $6.38_{0.01}$ & $7.87_{0.02}$ & $5.49_{0.01}$ & $5.87_{0.05}$ & $5.92_{0.01}$ & $5.83_{0.01}$ & $10.63_{0.02}$ & $6.05_{0.01}$ & $9.49_{0.02}$ \\
    & ilr-MDE & $\boldsymbol{1.81}_{0.01}$ & $\boldsymbol{1.82}_{0.01}$ & $\boldsymbol{6.38}_{0.01}$ & $\boldsymbol{4.78}_{0.02}$ & $\boldsymbol{5.69}_{0.01}$ & $\boldsymbol{2.56}_{0.02}$ & $\boldsymbol{2.88}_{0.06}$ & $\boldsymbol{3.36}_{0.01}$ & $\boldsymbol{3.18}_{0.01}$ & $9.35_{0.01}$ & $\boldsymbol{3.89}_{0.06}$ & $8.42_{0.01}$ \\
    & Dir-KDE & $3.86_{0.01}$ & $6.00_{0.01}$ & $11.76_{0.02}$ & $6.08_{0.01}$ & $6.04_{0.00}$ & $4.07_{0.01}$ & $5.97_{0.01}$ & $5.52_{0.01}$ & $5.29_{0.01}$ & $8.62_{0.00}$ & $5.85_{0.01}$ & $\boldsymbol{8.20}_{0.00}$ \\
    & Dir-AIC & $4.78_{0.02}$ & $8.64_{0.02}$ & $13.71_{0.05}$ & $7.50_{0.03}$ & $6.77_{0.01}$ & $4.58_{0.01}$ & $7.34_{0.02}$ & $6.65_{0.02}$ & $6.25_{0.01}$ & $9.23_{0.01}$ & $7.13_{0.06}$ & $8.58_{0.01}$ \\
    & Dir-CVISE & $3.90_{0.01}$ & $6.85_{0.02}$ & $13.90_{0.05}$ & $5.98_{0.02}$ & $8.25_{0.08}$ & $3.62_{0.01}$ & $5.71_{0.02}$ & $5.06_{0.01}$ & $5.09_{0.01}$ & $8.79_{0.01}$ & $5.70_{0.01}$ & $8.83_{0.02}$ \\
    & Dir-CVKLD & $3.83_{0.01}$ & $6.58_{0.01}$ & $11.93_{0.03}$ & $5.98_{0.01}$ & $5.72_{0.01}$ & $3.61_{0.01}$ & $5.58_{0.01}$ & $5.13_{0.01}$ & $5.09_{0.01}$ & $\boldsymbol{8.28}_{0.01}$ & $5.62_{0.02}$ & $8.28_{0.01}$ \\
    
    \bottomrule
    \end{tabular}
    \label{tab:chacon_MIAE}
\end{sidewaystable*}

\begin{sidewaystable*}[p]
    \centering
    \caption{\small MKLD with standard errors (subscripts) based on Monte Carlo replications in the Gaussian mixture simulations; bold highlights the minimum and those not significantly different in each scenario.}
    \setlength{\tabcolsep}{3pt}
    \small
    \begin{tabular}{llcccccccccccc}
    \toprule
    Settings & Estimators & (1) & (2) & (3) & (4) & (5) & (6) & (7) & (8) & (9) & (10) & (11) & (12) \\
    \midrule
    \multirow{7}{*}{\centering\makecell[l]{$d=2$, \\ $n=100$ \\[3pt] ($\times 100$)}}
    
    & ilr-PI & $9.01_{0.10}$ & $9.78_{0.10}$ & $39.93_{0.55}$ & $\boldsymbol{17.85}_{0.14}$ & $20.73_{0.26}$ & $9.53_{0.09}$ & $15.75_{0.11}$ & $16.80_{0.11}$ & $11.55_{0.10}$ & $23.59_{0.12}$ & $14.55_{0.12}$ & $24.22_{0.10}$ \\
    & ilr-CV & $21.36_{1.31}$ & $19.35_{1.01}$ & $104.75_{3.66}$ & $41.95_{1.91}$ & $53.83_{2.34}$ & $19.97_{0.91}$ & $24.41_{1.21}$ & $23.25_{0.84}$ & $23.35_{1.00}$ & $56.42_{1.25}$ & $23.14_{0.83}$ & $65.34_{1.84}$ \\
    & ilr-MDE & $\boldsymbol{4.75}_{0.11}$ & $\boldsymbol{5.02}_{0.13}$ & $\boldsymbol{20.08}_{0.19}$ & $18.69_{0.16}$ & $\boldsymbol{13.23}_{0.15}$ & $\boldsymbol{6.94}_{0.12}$ & $\boldsymbol{8.08}_{0.25}$ & $\boldsymbol{10.40}_{0.16}$ & $\boldsymbol{10.16}_{0.12}$ & $22.59_{0.16}$ & $\boldsymbol{8.67}_{0.18}$ & $\boldsymbol{23.09}_{0.18}$ \\
    & Dir-KDE & $7.54_{0.08}$ & $13.89_{0.12}$ & $86.47_{0.80}$ & $18.51_{0.14}$ & $17.77_{0.16}$ & $8.26_{0.09}$ & $16.38_{0.14}$ & $13.80_{0.12}$ & $14.29_{0.11}$ & $26.31_{0.22}$ & $16.74_{0.12}$ & $30.47_{0.16}$ \\
    & Dir-AIC & $8.93_{0.09}$ & $15.92_{0.12}$ & $114.42_{1.33}$ & $18.75_{0.15}$ & $17.51_{0.14}$ & $9.01_{0.08}$ & $17.36_{0.22}$ & $14.05_{0.11}$ & $17.01_{0.13}$ & $22.26_{0.19}$ & $19.93_{0.14}$ & $30.01_{0.17}$ \\
    & Dir-CVISE & $9.13_{0.10}$ & $17.00_{0.22}$ & $118.07_{1.42}$ & $20.57_{0.38}$ & $27.27_{1.14}$ & $9.14_{0.12}$ & $17.17_{0.15}$ & $13.93_{0.14}$ & $16.67_{0.14}$ & $22.42_{0.32}$ & $19.77_{0.26}$ & $29.44_{0.27}$ \\
    & Dir-CVKLD & $8.70_{0.10}$ & $15.43_{0.12}$ & $89.59_{0.85}$ & $18.23_{0.13}$ & $16.73_{0.13}$ & $8.72_{0.08}$ & $16.68_{0.14}$ & $13.45_{0.12}$ & $16.08_{0.13}$ & $\boldsymbol{20.50}_{0.15}$ & $18.78_{0.12}$ & $28.03_{0.14}$ \\
    
    \midrule
    
    \multirow{7}{*}{\centering\makecell[l]{$d=2$, \\ $n=500$ \\[3pt] ($\times 100$)}}
    
    & ilr-PI & $3.34_{0.02}$ & $3.77_{0.03}$ & $19.96_{0.17}$ & $7.46_{0.04}$ & $8.52_{0.06}$ & $3.69_{0.03}$ & $5.27_{0.03}$ & $5.75_{0.03}$ & $4.42_{0.03}$ & $13.08_{0.08}$ & $5.74_{0.04}$ & $14.11_{0.06}$ \\
    & ilr-CV & $4.83_{0.13}$ & $5.48_{0.14}$ & $35.30_{0.68}$ & $11.32_{0.21}$ & $14.85_{0.24}$ & $4.69_{0.08}$ & $6.04_{0.11}$ & $6.39_{0.10}$ & $5.95_{0.09}$ & $25.50_{0.23}$ & $6.48_{0.10}$ & $28.05_{0.31}$ \\
    & ilr-MDE & $\boldsymbol{1.10}_{0.03}$ & $\boldsymbol{1.12}_{0.03}$ & $\boldsymbol{7.56}_{0.06}$ & $\boldsymbol{5.51}_{0.04}$ & $\boldsymbol{4.09}_{0.04}$ & $\boldsymbol{1.73}_{0.03}$ & $\boldsymbol{2.20}_{0.11}$ & $\boldsymbol{2.79}_{0.03}$ & $\boldsymbol{2.78}_{0.03}$ & $8.23_{0.05}$ & $\boldsymbol{3.15}_{0.11}$ & $\boldsymbol{10.36}_{0.06}$ \\
    & Dir-KDE & $2.84_{0.02}$ & $5.53_{0.03}$ & $45.64_{0.15}$ & $7.32_{0.04}$ & $7.36_{0.04}$ & $3.08_{0.02}$ & $6.09_{0.03}$ & $5.34_{0.03}$ & $5.32_{0.03}$ & $9.67_{0.21}$ & $6.99_{0.04}$ & $18.21_{0.33}$ \\
    & Dir-AIC & $3.41_{0.02}$ & $6.89_{0.04}$ & $46.88_{0.13}$ & $6.75_{0.04}$ & $7.88_{0.04}$ & $3.24_{0.02}$ & $5.76_{0.04}$ & $4.97_{0.03}$ & $6.01_{0.04}$ & $8.36_{0.04}$ & $7.92_{0.04}$ & $14.07_{0.05}$ \\
    & Dir-CVISE & $3.50_{0.03}$ & $6.70_{0.04}$ & $47.11_{0.13}$ & $6.84_{0.06}$ & $14.38_{0.34}$ & $3.28_{0.03}$ & $5.46_{0.04}$ & $5.11_{0.03}$ & $5.97_{0.03}$ & $8.59_{0.07}$ & $8.26_{0.05}$ & $14.21_{0.06}$ \\
    & Dir-CVKLD & $3.24_{0.02}$ & $6.44_{0.03}$ & $46.87_{0.13}$ & $6.25_{0.04}$ & $7.49_{0.03}$ & $3.14_{0.02}$ & $5.25_{0.03}$ & $5.09_{0.03}$ & $5.89_{0.03}$ & $\boldsymbol{8.12}_{0.04}$ & $7.51_{0.04}$ & $13.51_{0.05}$ \\

    \midrule
    
    \multirow{7}{*}{\centering\makecell[l]{$d=5$, \\ $n=100$ \\[3pt] ($\times 10$)}}
    & ilr-PI & $17.06_{0.06}$ & $14.12_{0.06}$ & $26.16_{0.15}$ & $16.52_{0.06}$ & $23.43_{0.11}$ & $16.66_{0.06}$ & $16.80_{0.06}$ & $16.26_{0.06}$ & $15.88_{0.06}$ & $29.42_{0.15}$ & $17.88_{0.07}$ & $24.59_{0.11}$ \\
    & ilr-CV & $11.52_{0.18}$ & $9.22_{0.13}$ & $32.07_{0.43}$ & $12.95_{0.18}$ & $28.28_{0.44}$ & $11.45_{0.18}$ & $12.41_{0.19}$ & $12.87_{0.19}$ & $12.53_{0.19}$ & $42.77_{0.54}$ & $13.48_{0.20}$ & $31.86_{0.47}$ \\
    & ilr-MDE & $\boldsymbol{1.74}_{0.01}$ & $\boldsymbol{1.75}_{0.01}$ & $\boldsymbol{8.45}_{0.03}$ & $\boldsymbol{4.31}_{0.01}$ & $\boldsymbol{4.86}_{0.02}$ & $\boldsymbol{2.36}_{0.01}$ & $5.92_{0.07}$ & $5.41_{0.05}$ & $\boldsymbol{3.54}_{0.02}$ & $11.83_{0.02}$ & $\boldsymbol{5.54}_{0.04}$ & $8.75_{0.02}$ \\
    & Dir-KDE & $2.78_{0.01}$ & $6.37_{0.02}$ & $21.47_{0.04}$ & $5.61_{0.02}$ & $8.28_{0.04}$ & $2.86_{0.01}$ & $5.41_{0.01}$ & $4.67_{0.01}$ & $4.43_{0.01}$ & $10.33_{0.02}$ & $5.77_{0.01}$ & $\boldsymbol{8.40}_{0.01}$ \\
    & Dir-AIC & $3.48_{0.01}$ & $8.98_{0.03}$ & $182.43_{2.36}$ & $7.32_{0.03}$ & $7.79_{0.04}$ & $3.57_{0.01}$ & $7.25_{0.03}$ & $6.09_{0.03}$ & $5.86_{0.03}$ & $12.39_{0.03}$ & $7.33_{0.03}$ & $9.49_{0.01}$ \\
    & Dir-CVISE & $3.20_{0.01}$ & $7.49_{0.12}$ & $98.53_{2.54}$ & $6.11_{0.12}$ & $11.37_{0.45}$ & $2.88_{0.01}$ & $5.50_{0.02}$ & $4.53_{0.02}$ & $4.61_{0.02}$ & $16.30_{0.17}$ & $6.07_{0.06}$ & $10.46_{0.09}$ \\
    & Dir-CVKLD & $3.10_{0.01}$ & $6.91_{0.02}$ & $20.85_{0.04}$ & $5.39_{0.01}$ & $6.09_{0.02}$ & $2.85_{0.01}$ & $\boldsymbol{5.39}_{0.01}$ & $\boldsymbol{4.46}_{0.01}$ & $4.51_{0.01}$ & $\boldsymbol{10.22}_{0.02}$ & $5.71_{0.01}$ & $8.59_{0.02}$ \\
    
    \midrule
    
    \multirow{7}{*}{\centering\makecell[l]{$d=5$, \\ $n=500$ \\[3pt] ($\times 10$)}}
    & ilr-PI & $7.22_{0.01}$ & $5.68_{0.01}$ & $16.17_{0.05}$ & $6.96_{0.01}$ & $11.89_{0.03}$ & $6.96_{0.01}$ & $6.89_{0.01}$ & $6.83_{0.01}$ & $6.42_{0.01}$ & $18.75_{0.05}$ & $7.77_{0.01}$ & $14.25_{0.03}$ \\
    & ilr-CV & $2.56_{0.01}$ & $2.46_{0.01}$ & $14.57_{0.09}$ & $3.66_{0.02}$ & $8.53_{0.05}$ & $2.69_{0.01}$ & $2.89_{0.01}$ & $3.02_{0.01}$ & $2.98_{0.01}$ & $18.89_{0.08}$ & $3.87_{0.02}$ & $12.58_{0.06}$ \\
    & ilr-MDE & $\boldsymbol{0.39}_{0.00}$ & $\boldsymbol{0.40}_{0.00}$ & $\boldsymbol{4.46}_{0.01}$ & $\boldsymbol{2.06}_{0.02}$ & $\boldsymbol{2.93}_{0.01}$ & $\boldsymbol{0.66}_{0.01}$ & $\boldsymbol{0.98}_{0.05}$ & $\boldsymbol{1.03}_{0.01}$ & $\boldsymbol{0.94}_{0.01}$ & $8.41_{0.01}$ & $\boldsymbol{2.20}_{0.04}$ & $6.57_{0.01}$ \\
    & Dir-KDE & $1.63_{0.00}$ & $3.83_{0.01}$ & $15.60_{0.02}$ & $3.39_{0.01}$ & $4.89_{0.01}$ & $1.67_{0.00}$ & $3.19_{0.01}$ & $2.72_{0.01}$ & $2.60_{0.00}$ & $7.01_{0.01}$ & $3.70_{0.01}$ & $6.14_{0.01}$ \\
    & Dir-AIC & $2.24_{0.01}$ & $6.12_{0.02}$ & $84.21_{0.73}$ & $4.53_{0.02}$ & $5.12_{0.01}$ & $2.05_{0.01}$ & $4.51_{0.02}$ & $3.62_{0.01}$ & $3.52_{0.01}$ & $10.38_{0.02}$ & $4.78_{0.01}$ & $7.91_{0.01}$ \\
    & Dir-CVISE & $1.64_{0.00}$ & $4.43_{0.03}$ & $91.73_{0.15}$ & $3.51_{0.12}$ & $12.50_{0.73}$ & $1.41_{0.00}$ & $3.02_{0.01}$ & $2.47_{0.01}$ & $2.51_{0.01}$ & $12.83_{0.06}$ & $3.81_{0.02}$ & $8.20_{0.08}$ \\
    & Dir-CVKLD & $1.57_{0.00}$ & $4.13_{0.01}$ & $14.80_{0.01}$ & $3.12_{0.00}$ & $3.53_{0.01}$ & $1.39_{0.00}$ & $2.98_{0.01}$ & $2.46_{0.00}$ & $2.48_{0.00}$ & $\boldsymbol{6.01}_{0.01}$ & $3.54_{0.01}$ & $\boldsymbol{5.99}_{0.01}$ \\
    \bottomrule
    \end{tabular}
    \label{tab:chacon_KLD}
\end{sidewaystable*}

\subsection{Dirichlet mixtures}

\renewcommand{\arraystretch}{1.6}
\begin{table*}[!htb]
    \centering
    \caption{\small Parameters of the Dirichlet mixtures.}
    \small
    \begin{tabular}{l
        >{\raggedright\hangindent=1em\arraybackslash}p{15cm}}
        \toprule
        & $\pi_1 \Dir(\thetab_1, h_1) + \cdots + \pi_m \Dir(\thetab_m, h_m)$ \\
        \midrule
        1 & $\Dir\bigl((\frac{1}{3}, \frac{1}{3}, \frac{1}{3}), \, 0.1\bigr)$ \\
        2 & $\Dir\bigl((0, 0.3, 0.7), \, 0.05\bigr)$ \\
        3 & $\Dir\bigl((0, 0, 1), \, 0.5\bigr)$ \\
        4 & $\frac{1}{2} \,\Dir\bigl((0.5, 0.3, 0.2), \, 0.05 \bigr) +  \frac{1}{2} \, \Dir\bigl((0.2, 0.3, 0.5), \, 0.05\bigr)$ \\
        5 & $\frac{1}{18} \, \Dir\bigl((0.02, 0.8, 0.18), \, 0.03\bigr) + \frac{5}{18} \, \Dir\bigl((0.65, 0, 0.35), \, 0.03 \bigr) + \frac{2}{3} \, \Dir\bigl((0, 0, 1), \, 0.03\bigr)$ \\
        6 & $ \frac{1}{6} \, \sum_{i=1}^6 \Dir\Bigl(\bigl(\frac{i}{6}, \frac{i}{6} (1 - \frac{i}{6}), (1 - \frac{i}{6})^2\bigr), \, 0.1\Bigr) $ \\
        7 & $\frac{3}{10} \, \Dir\bigl((0.8, 0, 0.2), \, 0.02\bigr) + \frac{7}{10} \, \Dir\bigl((0.6, 0.4, 0), \, 0.1\bigr)$ \\   
        8 & $\frac{1}{5} \, \Dir\bigl((\frac{1}{3}, \frac{31}{75}, \frac{19}{75}), \, 0.1 \bigr) + \frac{7}{15} \, \Dir\bigl((\frac{1}{3}, 0.1, \frac{17}{30}), \, 0.01\bigr) + \frac{4}{15} \, \Dir\left((\frac{2}{15}, \frac{77}{150}, \frac{53}{150}), \, 0.004\right) + \frac{1}{15} \, \Dir\bigl((0.44, 0.42, 0.14), \, 0.001\bigr)$ \\
        9 & $\frac{2}{15} \, \Dir\bigl((1,0,0), \, 0.3\bigr) + \frac{2}{15} \, \Dir\bigl((0.01,0.98,0.01), \, 0.1\bigr) + \frac{2}{15} \, \Dir\bigl((0,0.01,0.99), \, 0.01\bigr) + \frac{1}{5} \, \Dir\bigl((0,0.5,0.5), \, 0.05\bigr) + \frac{1}{5} \, \Dir\bigl((0.3,0,0.7), \, 0.3\bigr) + \frac{1}{5} \, \Dir\bigl((0.6,0.4,0), \, 0.01\bigr)$ \\
        10 & $\frac{1}{12} \, \sum_{i=1}^6 \Dir\left(\bigl(\frac{2i}{15}, \frac{1}{8}(1 - \frac{2i}{15}), \frac{7}{8}(1 - \frac{2i}{15})\bigr), \, \frac{2(7-i)}{75}\right) +$ $\frac{1}{12} \, \sum_{i=1}^6 \Dir\Bigl(\bigl(\frac{i}{12}, \frac{9}{10}(1 - \frac{i}{12}), \frac{1}{10}(1 - \frac{i}{12})\bigr), \, \frac{7-i}{30}\Bigr)$ \\     
        11 & $\frac{3}{40} \, \sum_{i=1}^6 \Dir\Bigl(\bigl(\frac{2i}{15}, \frac{2i}{15} (1 - \frac{2i}{15}), (1 - \frac{2i}{15})^2\bigr), \, \frac{1}{10}\sqrt{\frac{2i}{15}}\Bigr)$ + $\frac{11}{120} \, \sum_{i=1}^6 \Dir\Bigl(\bigl(\frac{2i}{15}(1 - \frac{2i}{15}) + \frac{4}{225}, \frac{2i}{15} - \frac{4}{225}, (1 - \frac{2i}{15})^2\bigr), \, \frac{1}{10}\sqrt{\frac{14-2i}{15}}\Bigr)$ \\
        12 & $\frac{3}{70} \, \sum_{i=1}^5 \Dir\left((0, \frac{i-1}{6}, \frac{7-i}{6}), \, 16-3i\right) +
        \frac{3}{40} \, \sum_{i=1}^6 \Dir\Bigl(\bigl(\frac{i}{6}, \frac{i}{6} (1 - \frac{i}{6}), (1 - \frac{i}{6})^2\bigr), \, \frac{58-8i}{5}\Bigr) + 
        \frac{3}{70} \, \Dir \bigl((0, \frac{5}{6}, \frac{1}{6}), \, 5 \bigr) +
        \frac{3}{70} \, \Dir \bigl((0, 1, 0), \, 9 \bigr) + 
        \frac{1}{20} \, \Dir \bigl((0.25,0.75,0), \, 0.1 \bigr) + \frac{1}{20} \, \Dir \bigl((0.75,0.25,0), \, 0.05 \bigr) + \frac{1}{20} \, \Dir \bigl((0.9,0,0.1), \, 0.05 \bigr) + \frac{1}{20} \, \Dir \bigl((0.25,0.5,0.25), \, 0.02 \bigr) + \frac{1}{20} \, \Dir \bigl((0.39,0.05,0.56), \, 0.005 \bigr)$ \\
        \bottomrule
    \end{tabular}
    \label{tab:Dir_mix}
\end{table*}
\renewcommand{\arraystretch}{1}

\begin{figure*}[!htb]
  \centering
  \foreach \p [count=\i from 1] in {1,2,3,4,5,6,7,8,9,10,11,12}{
    \begin{overpic}[page=\p,width=0.15\linewidth,
      clip,trim={1.9cm 0.8cm 1.8cm 1.4cm}]{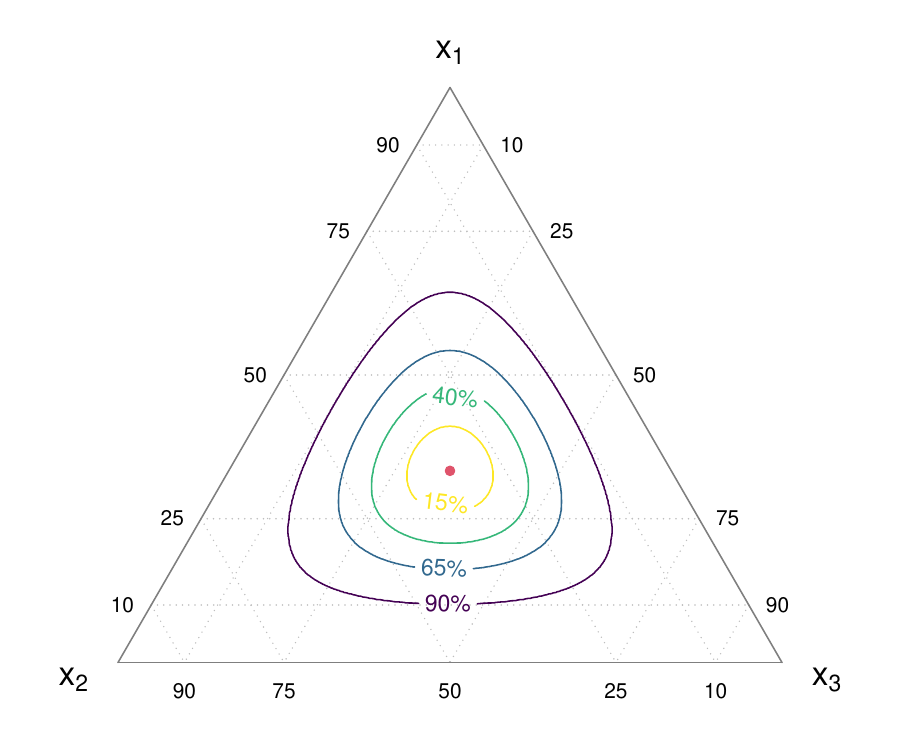}
      \put(42,-12){\footnotesize (\i)}%
    \end{overpic}%
    \ifnum\i=6\par\vspace{15pt}\fi
  }%
  \vspace{0.5cm}
  \caption{\small Probability contour plots for the ilr-Gaussian mixtures on $\Delta^2$.}
  \label{fig:dirmix}
\end{figure*}

\begin{figure*}[!htb]
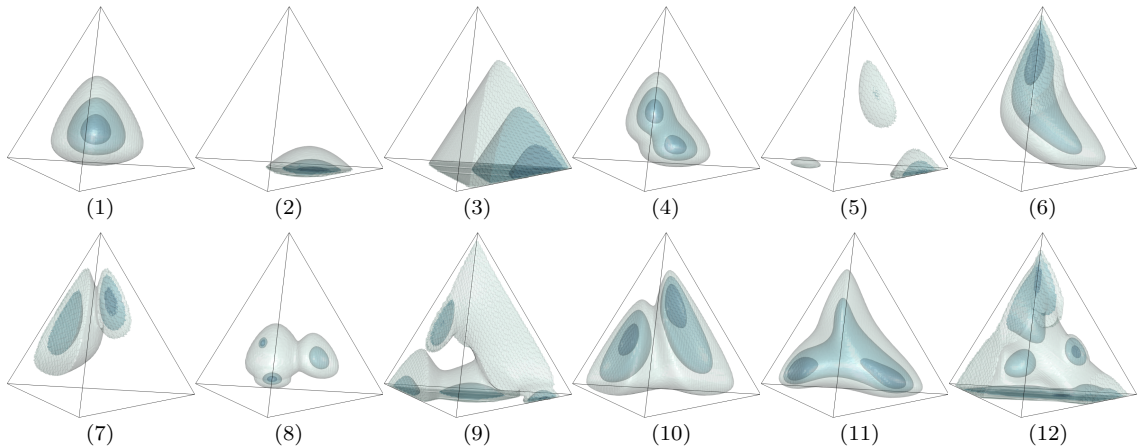

  \centering
  \foreach \i in {1,...,12}{%
    \begin{overpic}[width=0.155\linewidth,
      clip,trim={17.3cm 8.6cm 14.9cm 4.1cm}]{dirichlet\i.png}
      \put(42,-12){\footnotesize (\i)}
    \end{overpic}%
    \ifnum\i=6\par\vspace{15pt}\fi
  }%
  \vspace{0.5cm}
  \caption{\small Probability contour plots for the ilr-Gaussian mixtures on $\Delta^3$.}
  \label{fig:dirmix3D}
\end{figure*}

Twelve Dirichlet mixtures are constructed in each dimension: the first six are homogeneous, whereas the latter six are heterogeneous. For $\Delta^2$, component modes $\thetab$ and bandwidths $h$ are listed in Table~\ref{tab:Dir_mix}, and their visualizations are shown in Figure~\ref{fig:dirmix}, with $\thetab$ indicated by red dots. To extend mixtures to higher dimensions, $h$ remains unchanged so as to be independent of dimension. Additional modes are set to $1/D$, while the original modes are scaled by $3/D$ to ensure $\thetab^\top \mathbf{1} = 1$. Figure~\ref{fig:dirmix3D} illustrates the extended mixtures in $\Delta^3$. Tables~\ref{tab:dirichlet_MIAE}--\ref{tab:dirichlet_KLD} report MIAE and MKLD in the same format as the Gaussian mixture tables.

The results show that Dirichlet-based estimators naturally outperform Gaussian-based estimators, as they share the Dirichlet family used for data generation. Among them, Dir-CVKLD generally achieves the best performance across both measures, particularly at $d=5$. Dir-AIC performs satisfactorily at $d=3$ but is less so at $d=5$, since AIC is not a reliable bandwidth selector for mixture models. The three Dirichlet mixture estimators show greater advantage on homogeneous mixtures, owing to their use of a fixed bandwidth across components. Among the Gaussian-based methods, ilr-MDE outperforms the two KDE methods overall, consistently with the results from the Gaussian mixture simulation. Furthermore, the outperformance of the Dir-CVKLD over the others increases with the sample size. 

In summary, NPMDE estimators outperform the corresponding KDE within the same distribution family, achieving the best performance in both Gaussian and Dirichlet mixture simulations, with the advantage particularly evident at the higher dimension $d=5$ and the larger sample size $n=500$. Among the three bandwidth selectors of the Dirichlet mixture, $\hat{h}_\mathrm{CVKLD}$ delivers better and more stable performance than the other two.

\begin{sidewaystable*}[p]
    \centering
    \caption{\small MIAE with standard errors (subscripts) based on Monte Carlo replications in the Dirichlet mixture simulations; bold highlights the minimum and those not significantly different in each scenario.}
    \setlength{\tabcolsep}{3pt}
    \small
    \begin{tabular}{llcccccccccccc}
    \toprule
    Settings & Estimators & (1) & (2) & (3) & (4) & (5) & (6) & (7) & (8) & (9) & (10) & (11) & (12) \\
    \midrule
    \multirow{7}{*}{\centering\makecell[l]{$d=2$, \\ $n=100$ \\[3pt] ($\times 10$)}}
    & ilr-PI & $2.78_{0.02}$ & $2.94_{0.02}$ & $2.98_{0.02}$ & $3.05_{0.01}$ & $5.05_{0.02}$ & $3.10_{0.02}$ & $4.64_{0.02}$ & $6.11_{0.02}$ & $6.67_{0.01}$ & $3.72_{0.01}$ & $3.07_{0.02}$ & $\boldsymbol{4.51}_{0.01}$ \\
    & ilr-CV & $3.39_{0.03}$ & $3.65_{0.03}$ & $3.62_{0.03}$ & $3.61_{0.03}$ & $5.01_{0.03}$ & $3.76_{0.03}$ & $4.52_{0.03}$ & $5.82_{0.03}$ & $6.56_{0.02}$ & $4.13_{0.03}$ & $3.73_{0.03}$ & $5.29_{0.03}$ \\
    & ilr-MDE & $2.24_{0.02}$ & $2.53_{0.02}$ & $2.66_{0.02}$ & $2.89_{0.02}$ & $4.62_{0.02}$ & $2.79_{0.02}$ & $4.00_{0.02}$ & $5.33_{0.02}$ & $7.01_{0.02}$ & $3.66_{0.02}$ & $2.99_{0.02}$ & $4.53_{0.01}$ \\
    & Dir-KDE & $2.19_{0.02}$ & $2.83_{0.02}$ & $2.81_{0.02}$ & $2.69_{0.02}$ & $4.43_{0.02}$ & $2.55_{0.01}$ & $3.84_{0.01}$ & $5.56_{0.02}$ & $6.35_{0.01}$ & $3.40_{0.03}$ & $\boldsymbol{2.73}_{0.01}$ & $4.77_{0.02}$ \\
    & Dir-AIC & $1.63_{0.02}$ & $\boldsymbol{1.48}_{0.02}$ & $1.29_{0.02}$ & $2.22_{0.02}$ & $\boldsymbol{2.29}_{0.02}$ & $2.46_{0.02}$ & $3.37_{0.02}$ & $4.82_{0.02}$ & $6.32_{0.04}$ & $3.26_{0.02}$ & $3.01_{0.02}$ & $4.73_{0.01}$ \\
    & Dir-CVISE & $1.73_{0.03}$ & $1.84_{0.02}$ & $1.40_{0.03}$ & $2.26_{0.02}$ & $3.05_{0.03}$ & $2.66_{0.02}$ & $3.37_{0.02}$ & $\boldsymbol{4.63}_{0.02}$ & $\boldsymbol{5.96}_{0.03}$ & $3.23_{0.02}$ & $3.06_{0.02}$ & $4.97_{0.03}$ \\
    & Dir-CVKLD & $\boldsymbol{1.54}_{0.02}$ & $1.71_{0.01}$ & $\boldsymbol{1.16}_{0.03}$ & $\boldsymbol{2.16}_{0.02}$ & $2.85_{0.03}$ & $\boldsymbol{2.41}_{0.02}$ & $\boldsymbol{3.33}_{0.02}$ & $4.78_{0.02}$ & $\boldsymbol{5.97}_{0.01}$ & $\boldsymbol{3.21}_{0.02}$ & $2.97_{0.02}$ & $4.64_{0.01}$ \\
    
    \midrule
    
    \multirow{7}{*}{\centering\makecell[l]{$d=2$, \\ $n=500$ \\[3pt] ($\times 10$)}}
    & ilr-PI & $1.66_{0.01}$ & $1.75_{0.01}$ & $1.81_{0.01}$ & $1.87_{0.01}$ & $3.11_{0.01}$ & $1.88_{0.01}$ & $2.72_{0.01}$ & $3.90_{0.01}$ & $4.50_{0.01}$ & $2.27_{0.01}$ & $1.84_{0.01}$ & $3.34_{0.01}$ \\
    & ilr-CV & $1.86_{0.01}$ & $1.95_{0.01}$ & $2.03_{0.01}$ & $2.02_{0.01}$ & $2.92_{0.01}$ & $2.10_{0.01}$ & $2.62_{0.01}$ & $3.66_{0.01}$ & $4.18_{0.01}$ & $2.37_{0.01}$ & $2.04_{0.01}$ & $3.79_{0.01}$ \\
    & ilr-MDE & $1.22_{0.01}$ & $1.37_{0.01}$ & $1.45_{0.01}$ & $1.50_{0.01}$ & $2.74_{0.01}$ & $1.66_{0.01}$ & $2.05_{0.01}$ & $2.95_{0.01}$ & $4.33_{0.01}$ & $2.11_{0.01}$ & $1.71_{0.01}$ & $3.52_{0.01}$ \\
    & Dir-KDE & $1.31_{0.01}$ & $1.73_{0.01}$ & $1.75_{0.01}$ & $1.62_{0.01}$ & $2.70_{0.01}$ & $1.56_{0.01}$ & $2.38_{0.01}$ & $3.64_{0.01}$ & $4.04_{0.01}$ & $1.97_{0.01}$ & $1.63_{0.01}$ & $3.16_{0.01}$ \\
    & Dir-AIC & $0.75_{0.01}$ & $\boldsymbol{0.68}_{0.01}$ & $0.58_{0.01}$ & $1.07_{0.01}$ & $\boldsymbol{1.00}_{0.01}$ & $1.21_{0.01}$ & $1.83_{0.01}$ & $2.98_{0.01}$ & $3.55_{0.01}$ & $1.88_{0.01}$ & $1.62_{0.01}$ & $3.13_{0.01}$ \\
    & Dir-CVISE & $0.76_{0.01}$ & $1.09_{0.01}$ & $0.63_{0.01}$ & $1.05_{0.01}$ & $1.39_{0.02}$ & $1.33_{0.01}$ & $1.73_{0.01}$ & $\boldsymbol{2.92}_{0.01}$ & $\boldsymbol{3.30}_{0.01}$ & $1.88_{0.01}$ & $1.63_{0.01}$ & $3.20_{0.01}$ \\
    & Dir-CVKLD & $\boldsymbol{0.70}_{0.01}$ & $1.04_{0.01}$ & $\boldsymbol{0.55}_{0.01}$ & $\boldsymbol{1.02}_{0.01}$ & $1.24_{0.01}$ & $\boldsymbol{1.18}_{0.01}$ & $\boldsymbol{1.67}_{0.01}$ & $\boldsymbol{2.90}_{0.01}$ & $3.45_{0.01}$ & $\boldsymbol{1.85}_{0.01}$ & $\boldsymbol{1.61}_{0.01}$ & $\boldsymbol{3.07}_{0.01}$ \\
    
    \midrule
    
    \multirow{7}{*}{\centering\makecell[l]{$d=5$, \\ $n=100$ \\[3pt] ($\times 10$)}}
    & ilr-PI & $10.58_{0.02}$ & $10.16_{0.02}$ & $10.33_{0.03}$ & $10.62_{0.02}$ & $10.42_{0.03}$ & $10.47_{0.11}$ & $9.92_{0.02}$ & $11.14_{0.05}$ & $11.70_{0.06}$ & $10.28_{0.03}$ & $10.44_{0.02}$ & $10.57_{0.07}$ \\
    & ilr-CV & $9.40_{0.03}$ & $9.21_{0.03}$ & $9.55_{0.04}$ & $9.40_{0.03}$ & $10.41_{0.04}$ & $9.81_{0.06}$ & $10.53_{0.04}$ & $13.66_{0.32}$ & $13.08_{0.15}$ & $10.16_{0.05}$ & $9.74_{0.05}$ & $11.24_{0.34}$ \\
    & ilr-MDE & $5.08_{0.02}$ & $5.04_{0.02}$ & $5.98_{0.02}$ & $5.10_{0.02}$ & $8.49_{0.02}$ & $5.99_{0.02}$ & $8.94_{0.03}$ & $11.91_{0.06}$ & $11.19_{0.28}$ & $6.97_{0.04}$ & $5.88_{0.03}$ & $8.15_{0.02}$ \\
    & Dir-KDE & $3.69_{0.01}$ & $5.47_{0.02}$ & $3.89_{0.01}$ & $4.79_{0.01}$ & $7.40_{0.02}$ & $3.94_{0.01}$ & $6.59_{0.01}$ & $10.69_{0.02}$ & $9.24_{0.01}$ & $4.69_{0.01}$ & $4.56_{0.01}$ & $6.19_{0.01}$ \\
    & Dir-AIC & $2.82_{0.02}$ & $3.89_{0.03}$ & $2.15_{0.01}$ & $\boldsymbol{3.54}_{0.02}$ & $5.82_{0.04}$ & $3.71_{0.02}$ & $6.28_{0.02}$ & $\boldsymbol{8.93}_{0.08}$ & $9.08_{0.04}$ & $4.93_{0.01}$ & $4.84_{0.02}$ & $7.02_{0.02}$ \\
    & Dir-CVISE & $2.89_{0.02}$ & $3.01_{0.03}$ & $2.06_{0.02}$ & $3.73_{0.02}$ & $4.33_{0.04}$ & $3.67_{0.02}$ & $6.38_{0.03}$ & $9.24_{0.05}$ & $10.46_{0.03}$ & $4.66_{0.02}$ & $4.34_{0.02}$ & $6.85_{0.04}$ \\
    & Dir-CVKLD & $\boldsymbol{2.73}_{0.02}$ & $\boldsymbol{2.78}_{0.02}$ & $\boldsymbol{2.02}_{0.02}$ & $3.62_{0.02}$ & $\boldsymbol{3.99}_{0.04}$ & $\boldsymbol{3.58}_{0.02}$ & $\boldsymbol{5.73}_{0.01}$ & $\boldsymbol{8.92}_{0.02}$ & $\boldsymbol{8.84}_{0.01}$ & $\boldsymbol{4.48}_{0.01}$ & $\boldsymbol{4.26}_{0.02}$ & $\boldsymbol{5.87}_{0.01}$ \\
    
    \midrule
    
    \multirow{7}{*}{\centering\makecell[l]{$d=5$, \\ $n=500$ \\[3pt] ($\times 10$)}}
    & ilr-PI & $7.47_{0.19}$ & $6.73_{0.02}$ & $6.95_{0.01}$ & $7.20_{0.01}$ & $7.54_{0.01}$ & $6.92_{0.02}$ & $7.11_{0.03}$ & $9.11_{0.02}$ & $9.35_{0.02}$ & $7.07_{0.01}$ & $7.11_{0.02}$ & $7.73_{0.04}$ \\
    & ilr-CV & $5.61_{0.03}$ & $5.51_{0.02}$ & $5.85_{0.01}$ & $5.59_{0.02}$ & $7.22_{0.02}$ & $5.84_{0.02}$ & $7.04_{0.02}$ & $10.16_{0.04}$ & $10.24_{0.03}$ & $6.25_{0.02}$ & $5.79_{0.01}$ & $7.73_{0.03}$ \\
    & ilr-MDE & $3.07_{0.01}$ & $3.00_{0.01}$ & $3.75_{0.01}$ & $3.19_{0.01}$ & $4.98_{0.01}$ & $3.69_{0.01}$ & $5.59_{0.01}$ & $7.31_{0.02}$ & $8.68_{0.03}$ & $4.61_{0.01}$ & $3.69_{0.01}$ & $5.95_{0.02}$ \\
    & Dir-KDE & $2.58_{0.01}$ & $3.99_{0.01}$ & $3.33_{0.01}$ & $3.46_{0.01}$ & $5.47_{0.01}$ & $2.81_{0.01}$ & $5.09_{0.01}$ & $9.11_{0.01}$ & $7.83_{0.01}$ & $3.57_{0.01}$ & $3.31_{0.01}$ & $4.92_{0.01}$ \\
    & Dir-AIC & $1.31_{0.01}$ & $2.08_{0.02}$ & $1.07_{0.01}$ & $2.21_{0.01}$ & $3.17_{0.02}$ & $2.50_{0.01}$ & $4.79_{0.01}$ & $\boldsymbol{6.47}_{0.01}$ & $7.70_{0.01}$ & $3.76_{0.01}$ & $3.31_{0.02}$ & $4.95_{0.01}$ \\
    & Dir-CVISE & $1.31_{0.01}$ & $1.35_{0.01}$ & $0.95_{0.01}$ & $1.72_{0.01}$ & $1.92_{0.02}$ & $1.87_{0.01}$ & $4.15_{0.01}$ & $8.89_{0.07}$ & $7.28_{0.03}$ & $3.21_{0.01}$ & $2.42_{0.01}$ & $6.00_{0.03}$ \\
    & Dir-CVKLD & $\boldsymbol{1.23}_{0.01}$ & $\boldsymbol{1.29}_{0.01}$ & $\boldsymbol{0.92}_{0.01}$ & $\boldsymbol{1.65}_{0.01}$ & $\boldsymbol{1.78}_{0.01}$ & $\boldsymbol{1.82}_{0.01}$ & $\boldsymbol{3.90}_{0.01}$ & $6.50_{0.01}$ & $\boldsymbol{6.98}_{0.01}$ & $\boldsymbol{2.91}_{0.01}$ & $\boldsymbol{2.33}_{0.01}$ & $\boldsymbol{4.28}_{0.01}$ \\
    \bottomrule
    \end{tabular}
    \label{tab:dirichlet_MIAE}
\end{sidewaystable*}

\begin{sidewaystable*}[p]
    \centering
    \caption{\small MKLD with standard errors (subscripts) based on Monte Carlo replications in the Dirichlet mixture simulations; bold highlights the minimum and those not significantly different in each scenario.}
    \setlength{\tabcolsep}{3pt}
    \small
    \begin{tabular}{llcccccccccccc}
    \toprule
    Settings & Estimators & (1) & (2) & (3) & (4) & (5) & (6) & (7) & (8) & (9) & (10) & (11) & (12) \\
    \midrule
    \multirow{7}{*}{\centering\makecell[l]{$d=2$, \\ $n=100$ \\[3pt] ($\times 100$)}}
    & ilr-PI & $11.73_{0.13}$ & $17.94_{0.26}$ & $19.86_{0.24}$ & $10.18_{0.10}$ & $35.07_{0.38}$ & $22.14_{0.27}$ & $28.85_{0.26}$ & $39.82_{0.17}$ & $44.55_{0.17}$ & $16.28_{0.12}$ & $13.42_{0.12}$ & $28.01_{0.20}$ \\
    & ilr-CV & $21.48_{0.83}$ & $35.56_{1.54}$ & $35.39_{1.82}$ & $20.43_{1.03}$ & $77.61_{3.52}$ & $38.45_{1.28}$ & $46.96_{1.29}$ & $117.10_{2.53}$ & $107.46_{1.84}$ & $29.83_{0.77}$ & $27.91_{1.47}$ & $66.54_{1.89}$ \\
    & ilr-MDE & $6.23_{0.12}$ & $8.79_{0.14}$ & $9.87_{0.17}$ & $8.69_{0.14}$ & $30.42_{0.28}$ & $11.86_{0.18}$ & $23.92_{0.28}$ & $33.89_{0.24}$ & $48.11_{0.21}$ & $14.24_{0.12}$ & $10.13_{0.13}$ & $\boldsymbol{23.13}_{0.20}$ \\
    & Dir-KDE & $5.04_{0.06}$ & $9.61_{0.10}$ & $9.46_{0.10}$ & $6.92_{0.08}$ & $25.44_{1.41}$ & $7.99_{0.09}$ & $16.63_{0.13}$ & $32.93_{0.19}$ & $43.13_{0.24}$ & $11.40_{0.16}$ & $\boldsymbol{7.91}_{0.08}$ & $27.34_{0.25}$ \\
    & Dir-AIC & $3.14_{0.07}$ & $\boldsymbol{2.45}_{0.06}$ & $2.17_{0.09}$ & $\boldsymbol{4.64}_{0.08}$ & $\boldsymbol{7.44}_{0.47}$ & $6.05_{0.10}$ & $11.97_{0.14}$ & $38.56_{0.85}$ & $64.74_{2.17}$ & $10.23_{0.09}$ & $8.87_{0.10}$ & $24.47_{0.18}$ \\
    & Dir-CVISE & $3.76_{0.14}$ & $3.97_{0.17}$ & $2.92_{0.13}$ & $5.08_{0.13}$ & $13.12_{0.71}$ & $7.66_{0.24}$ & $13.47_{0.32}$ & $37.86_{0.81}$ & $60.18_{1.58}$ & $10.36_{0.12}$ & $9.32_{0.13}$ & $30.47_{0.85}$ \\
    & Dir-CVKLD & $\boldsymbol{2.81}_{0.07}$ & $3.01_{0.07}$ & $\boldsymbol{2.03}_{0.09}$ & $\boldsymbol{4.64}_{0.10}$ & $10.02_{0.58}$ & $\boldsymbol{5.77}_{0.08}$ & $\boldsymbol{11.74}_{0.11}$ & $\boldsymbol{27.62}_{0.20}$ & $\boldsymbol{35.11}_{0.22}$ & $\boldsymbol{10.07}_{0.10}$ & $8.64_{0.10}$ & $23.15_{0.17}$ \\
    
    \midrule
    
    \multirow{7}{*}{\centering\makecell[l]{$d=2$, \\ $n=500$ \\[3pt] ($\times 100$)}}
    & ilr-PI & $4.78_{0.04}$ & $7.09_{0.08}$ & $8.37_{0.07}$ & $4.33_{0.03}$ & $15.88_{0.08}$ & $8.90_{0.08}$ & $11.16_{0.09}$ & $22.03_{0.10}$ & $24.07_{0.08}$ & $8.00_{0.06}$ & $5.77_{0.04}$ & $16.73_{0.07}$ \\
    & ilr-CV & $6.49_{0.13}$ & $9.57_{0.21}$ & $11.29_{0.22}$ & $5.88_{0.09}$ & $22.00_{0.27}$ & $11.99_{0.30}$ & $14.55_{0.17}$ & $50.51_{0.47}$ & $44.97_{0.31}$ & $11.64_{0.15}$ & $7.82_{0.13}$ & $41.75_{0.45}$ \\
    & ilr-MDE & $1.94_{0.04}$ & $2.81_{0.05}$ & $3.16_{0.05}$ & $2.53_{0.03}$ & $11.77_{0.10}$ & $4.28_{0.05}$ & $6.70_{0.07}$ & $14.28_{0.07}$ & $22.96_{0.10}$ & $5.44_{0.05}$ & $3.72_{0.04}$ & $15.28_{0.06}$ \\
    & Dir-KDE & $1.88_{0.02}$ & $4.25_{0.03}$ & $4.20_{0.03}$ & $2.69_{0.02}$ & $9.04_{0.05}$ & $3.15_{0.03}$ & $7.13_{0.04}$ & $15.69_{0.07}$ & $16.79_{0.06}$ & $3.94_{0.03}$ & $2.86_{0.02}$ & $10.95_{0.05}$ \\
    & Dir-AIC & $0.72_{0.01}$ & $\boldsymbol{0.60}_{0.02}$ & $\boldsymbol{0.43}_{0.01}$ & $1.12_{0.02}$ & $\boldsymbol{1.35}_{0.03}$ & $1.48_{0.02}$ & $3.47_{0.03}$ & $\boldsymbol{12.83}_{0.07}$ & $\boldsymbol{11.89}_{0.06}$ & $3.56_{0.03}$ & $2.72_{0.02}$ & $9.98_{0.06}$ \\
    & Dir-CVISE & $0.75_{0.02}$ & $1.23_{0.02}$ & $0.57_{0.02}$ & $1.18_{0.02}$ & $2.70_{0.09}$ & $1.82_{0.03}$ & $3.38_{0.05}$ & $14.92_{0.27}$ & $13.63_{0.20}$ & $3.53_{0.03}$ & $2.65_{0.03}$ & $10.25_{0.15}$ \\
    & Dir-CVKLD & $\boldsymbol{0.62}_{0.02}$ & $1.12_{0.02}$ & $\boldsymbol{0.42}_{0.02}$ & $\boldsymbol{1.05}_{0.02}$ & $1.93_{0.02}$ & $\boldsymbol{1.42}_{0.02}$ & $\boldsymbol{3.16}_{0.03}$ & $\boldsymbol{12.88}_{0.07}$ & $13.44_{0.07}$ & $\boldsymbol{3.41}_{0.03}$ & $\boldsymbol{2.61}_{0.02}$ & $\boldsymbol{9.26}_{0.05}$ \\
    
    \midrule
    
    \multirow{7}{*}{\centering\makecell[l]{$d=5$, \\ $n=100$ \\[3pt] ($\times 10$)}}
    & ilr-PI & $17.10_{0.08}$ & $15.28_{0.08}$ & $17.22_{0.09}$ & $16.79_{0.07}$ & $16.32_{0.08}$ & $16.94_{0.09}$ & $15.08_{0.08}$ & $32.44_{0.20}$ & $20.83_{0.10}$ & $17.27_{0.09}$ & $17.34_{0.08}$ & $17.95_{0.10}$ \\
    & ilr-CV & $12.67_{0.18}$ & $11.80_{0.18}$ & $15.35_{0.23}$ & $13.19_{0.22}$ & $15.76_{0.21}$ & $15.55_{0.26}$ & $18.29_{0.28}$ & $58.91_{0.72}$ & $37.36_{0.49}$ & $17.23_{0.27}$ & $14.78_{0.23}$ & $22.47_{0.34}$ \\
    & ilr-MDE & $2.51_{0.01}$ & $2.51_{0.01}$ & $3.53_{0.02}$ & $2.61_{0.01}$ & $6.95_{0.02}$ & $3.47_{0.01}$ & $7.68_{0.02}$ & $16.15_{0.05}$ & $14.60_{0.03}$ & $4.53_{0.02}$ & $3.27_{0.01}$ & $7.62_{0.02}$ \\
    & Dir-KDE & $1.18_{0.01}$ & $2.72_{0.01}$ & $1.37_{0.01}$ & $1.95_{0.01}$ & $5.25_{0.04}$ & $1.54_{0.01}$ & $4.28_{0.01}$ & $12.27_{0.02}$ & $9.89_{0.02}$ & $2.10_{0.01}$ & $1.92_{0.01}$ & $4.75_{0.02}$ \\
    & Dir-AIC & $0.82_{0.01}$ & $1.31_{0.02}$ & $0.45_{0.01}$ & $\boldsymbol{1.16}_{0.01}$ & $2.80_{0.03}$ & $1.18_{0.01}$ & $4.80_{0.02}$ & $26.60_{2.34}$ & $18.58_{1.59}$ & $2.26_{0.01}$ & $2.01_{0.01}$ & $5.75_{0.03}$ \\
    & Dir-CVISE & $0.86_{0.01}$ & $0.94_{0.02}$ & $0.42_{0.01}$ & $1.31_{0.01}$ & $2.24_{0.06}$ & $1.25_{0.01}$ & $4.35_{0.07}$ & $26.96_{0.53}$ & $21.47_{0.32}$ & $2.06_{0.02}$ & $1.75_{0.01}$ & $5.59_{0.10}$ \\
    & Dir-CVKLD & $\boldsymbol{0.77}_{0.01}$ & $\boldsymbol{0.78}_{0.01}$ & $\boldsymbol{0.40}_{0.01}$ & $1.23_{0.01}$ & $\boldsymbol{1.76}_{0.02}$ & $\boldsymbol{1.16}_{0.01}$ & $\boldsymbol{3.07}_{0.01}$ & $\boldsymbol{10.07}_{0.02}$ & $\boldsymbol{8.37}_{0.02}$ & $\boldsymbol{1.87}_{0.01}$ & $\boldsymbol{1.67}_{0.01}$ & $\boldsymbol{3.97}_{0.01}$ \\
    
    \midrule
    
    \multirow{7}{*}{\centering\makecell[l]{$d=5$, \\ $n=500$ \\[3pt] ($\times 10$)}}
    & ilr-PI & $7.37_{0.02}$ & $6.21_{0.02}$ & $7.31_{0.02}$ & $6.98_{0.02}$ & $6.90_{0.02}$ & $7.11_{0.02}$ & $6.47_{0.02}$ & $21.59_{0.07}$ & $11.79_{0.03}$ & $7.58_{0.02}$ & $7.45_{0.02}$ & $8.59_{0.02}$ \\
    & ilr-CV & $3.10_{0.02}$ & $2.92_{0.01}$ & $3.64_{0.02}$ & $2.94_{0.02}$ & $4.86_{0.02}$ & $3.61_{0.02}$ & $5.20_{0.02}$ & $26.96_{0.15}$ & $14.55_{0.07}$ & $4.23_{0.02}$ & $3.51_{0.02}$ & $7.84_{0.04}$ \\
    & ilr-MDE & $1.01_{0.00}$ & $0.99_{0.00}$ & $1.48_{0.01}$ & $1.05_{0.00}$ & $2.64_{0.01}$ & $1.45_{0.01}$ & $3.07_{0.01}$ & $8.03_{0.02}$ & $8.38_{0.02}$ & $2.07_{0.01}$ & $1.47_{0.01}$ & $4.15_{0.01}$ \\
    & Dir-KDE & $0.61_{0.00}$ & $1.55_{0.00}$ & $1.07_{0.00}$ & $1.05_{0.00}$ & $2.88_{0.01}$ & $0.84_{0.00}$ & $2.62_{0.01}$ & $8.84_{0.01}$ & $6.70_{0.01}$ & $1.25_{0.00}$ & $1.03_{0.00}$ & $3.02_{0.01}$ \\
    & Dir-AIC & $0.22_{0.00}$ & $0.39_{0.01}$ & $0.12_{0.00}$ & $0.43_{0.00}$ & $0.86_{0.01}$ & $0.52_{0.00}$ & $3.22_{0.02}$ & $7.13_{0.03}$ & $8.67_{0.03}$ & $1.46_{0.01}$ & $0.98_{0.01}$ & $3.32_{0.01}$ \\
    & Dir-CVISE & $0.22_{0.00}$ & $0.25_{0.00}$ & $0.10_{0.00}$ & $0.35_{0.00}$ & $0.50_{0.01}$ & $0.35_{0.00}$ & $1.87_{0.02}$ & $33.16_{0.33}$ & $8.95_{0.13}$ & $1.03_{0.01}$ & $0.61_{0.00}$ & $4.16_{0.06}$ \\
    & Dir-CVKLD & $\boldsymbol{0.19}_{0.00}$ & $\boldsymbol{0.22}_{0.00}$ & $\boldsymbol{0.09}_{0.00}$ & $\boldsymbol{0.32}_{0.00}$ & $\boldsymbol{0.44}_{0.00}$ & $\boldsymbol{0.32}_{0.00}$ & $\boldsymbol{1.39}_{0.00}$ & $\boldsymbol{6.41}_{0.01}$ & $\boldsymbol{4.74}_{0.01}$ & $\boldsymbol{0.85}_{0.00}$ & $\boldsymbol{0.55}_{0.00}$ & $\boldsymbol{2.25}_{0.00}$ \\
    \bottomrule
    \end{tabular}
    \label{tab:dirichlet_KLD}
\end{sidewaystable*}

\subsection{Computational cost}

\begin{figure}[!ht]
    \centering
    \includegraphics[width=0.45\textwidth,clip, trim = {0cm 0.6cm 1cm 2cm}]{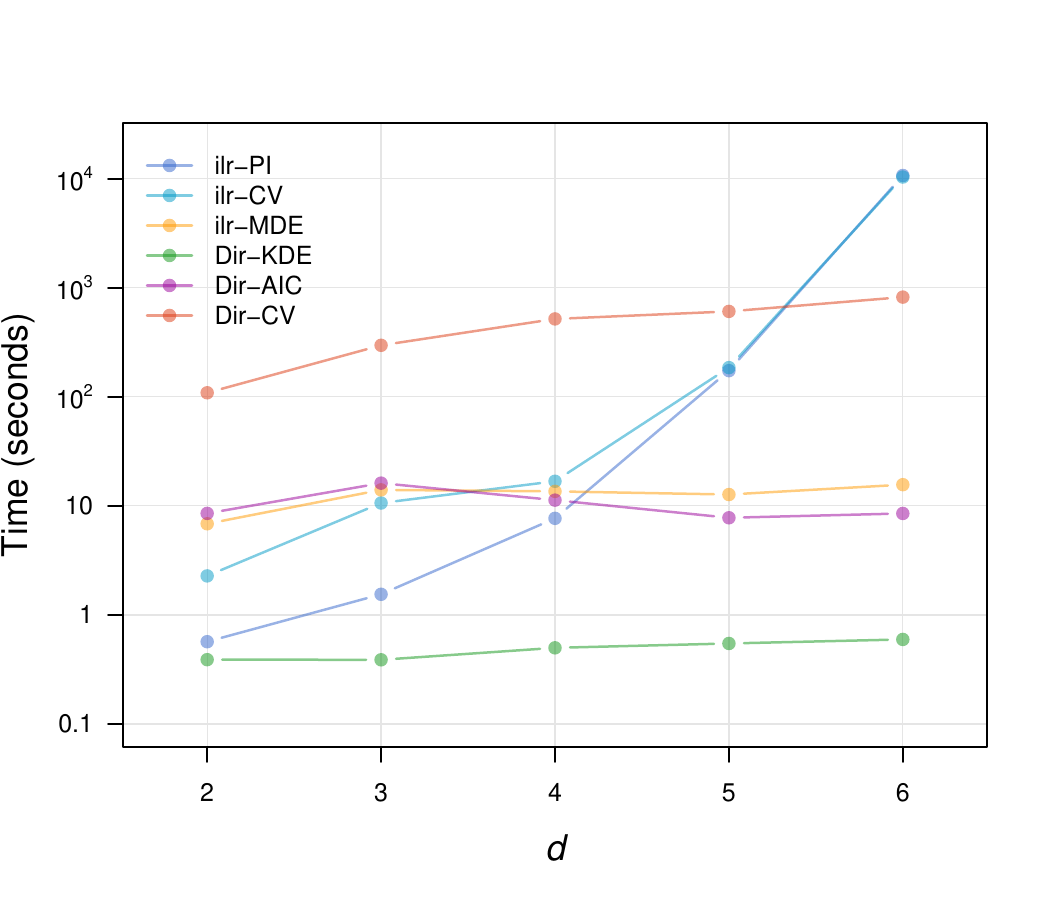}
    \caption{\small Computation time ($\log_{10}$-scaled) of different estimators for dimensions two to six.}
    \label{fig:time}
\end{figure}

We also compared the computational cost of the estimators on a computer with a processor speed of 2.40 GHz. The mixture in Figure~\ref{fig:chacon}(10) and its higher-dimensional extensions were used as reference models, from which 500 samples were generated for density estimation and runtimes recorded. Since Dir-CVISE and Dir-CVKLD share the same cross-validation partitions and are evaluated jointly, resulting in identical running times, they are labeled collectively as Dir-CV. Figure~\ref{fig:time} reports the average runtime over five repetitions for $\Delta^2$ through $\Delta^6$. Dir-KDE is fastest, owing to a single pseudo-likelihood-tuned parameter and weak dependence on $d$. Dir-AIC and ilr-MDE exhibit similar costs with also weak $d$-dependence. Ilr-PI and ilr-CV perform well for $d\le 4$ but scale poorly due to bandwidth-matrix estimation. Dir-CV has the largest overall run time yet grows smoothly with $d$; for higher dimensions ($d\ge 6$), it can be faster than the two ilr-KDE methods.

\section{Real data applications}
\label{sec:realdata}

\subsection{Density estimation for national GDP dataset}

\begin{figure*}[!htb]
    \centering
    \includegraphics[width=0.4\textwidth,
        clip,trim={0.8cm 0.8cm 1cm 0.6cm}]{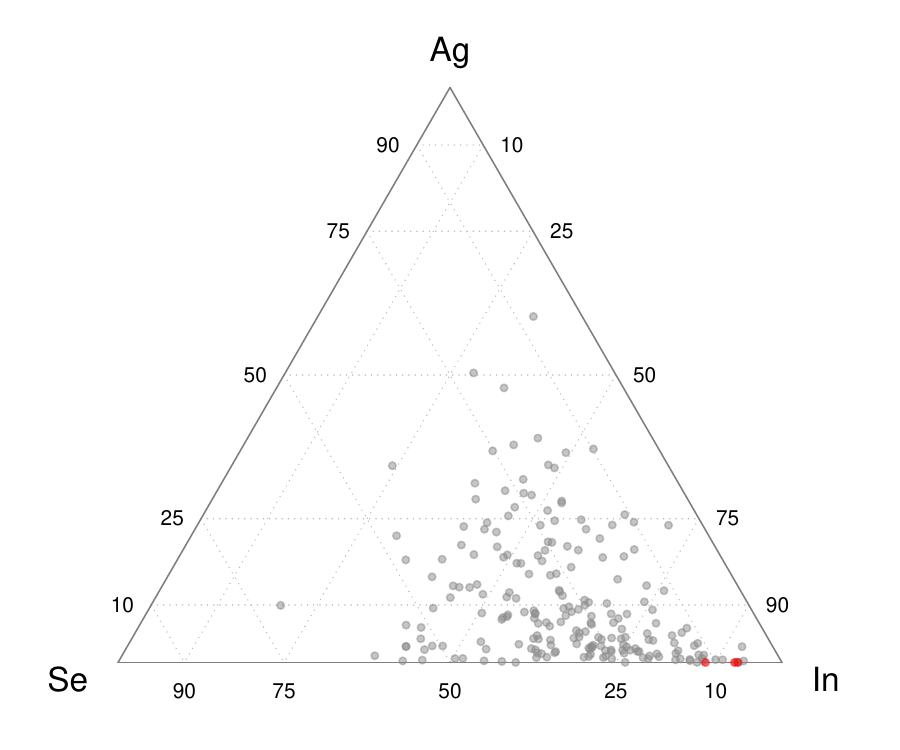}
    \hspace{1.5cm}
    \includegraphics[width=0.4\textwidth,
        clip,trim={0.8cm 0.8cm 1cm 0.6cm}]{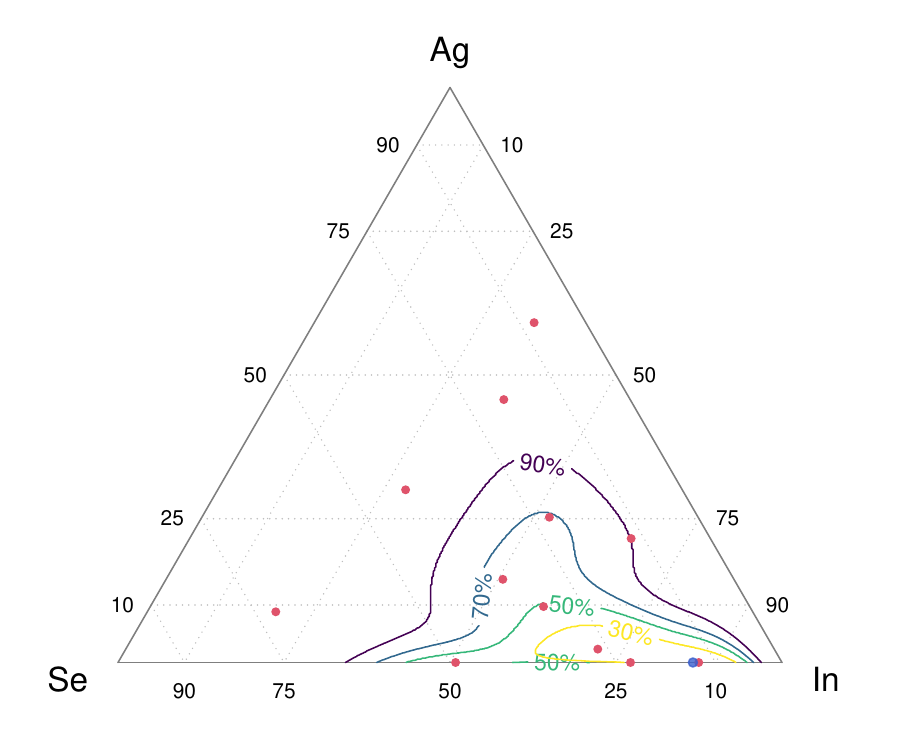}
    \caption{\small Plots of three-part GDP compositions: scatter plot (left); probability contour plot of Dir-CVKLD estimated from data after excluding zero-containing observations (right).}
    \label{fig:gdpest}
\end{figure*}

The National GDP Data by \citet{GDP2025} reports global GDP compositions by economic activity at current prices. It covers 212 countries from 1970 to 2023. We use the most recent data from 2023, comprising six sectors classified under the International Standard Industrial Classification (ISIC). These sectors are further aggregated into three broad categories: agriculture, industry, and services, yielding a six-part and a three-part compositional dataset. Table~\ref{tab:gdp_group} summarizes the component mapping, and the left panel of Figure~\ref{fig:gdpest} presents the scatter plot of the three-part dataset. Many observations lie along the lower edge of $\Delta^2$, indicating that agriculture contributes only a small proportion in many countries. In particular, three cases have zero agriculture, marked in red.

\begin{table*}[!htb]
    \centering
    \caption{\small Mapping of ISIC codes to economic sectors.}
    \small
    \begin{tabular}{lll}
    \toprule
    \textbf{Sector ($d = 2$)} & \textbf{ISIC ($d = 5$)} & \textbf{Included Activities} \\
    \midrule
    Agriculture (Ag) & A--B & Agriculture, hunting, forestry, fishing \\
    Industry (In)   & C--E & Mining, utilities, manufacturing \\
                & F    & Construction \\
    Services (Se)   & G--H & Wholesale, retail trade, restaurants and hotels \\
                & I    & Transport, storage and communication \\
                & J--P & Other activities \\
    \bottomrule
    \end{tabular}
    \label{tab:gdp_group}
\end{table*}

The right panel of Figure~\ref{fig:gdpest} presents density contours for Dir-CVKLD on the three-part GDP compositions after excluding zero-containing observations. The method successfully identifies three support points (red dots) on the boundary nonparametrically, and the contours truncate at the boundary, with density transitioning smoothly into the interior (even with the three zero-containing cases removed). Under the grid search, the maximum density value is found at the blue point on the boundary, consistent with the distribution observed in the left panel of Figure~\ref{fig:gdpest}. By contrast, except for Dirichlet-based mixture methods, the estimated density is zero on the boundary for other estimators: Gaussian-based methods operate only on $\widetilde{\Delta}^d$, while Dir-KDE places kernels only at the observations. Since zero-containing cases are removed, no kernels lie on the boundary, resulting in zero density there.

Because the data contains zeros, it is difficult to fairly compare Dirichlet-based models on the original data with Gaussian-based models on the zero-replaced data. To address this, estimator performance is evaluated via a parametric bootstrap. A reference density $\hat{f}$ is first constructed from the GDP compositions using one estimator. Then we generate a synthetic sample from $\hat{f}$ with the same size as the original dataset. To simulate real data, zeros are introduced to the samples. Specifically, given a proportion $x_j$, zeros are generated using the following process:
\begin{align*}
    x_j' =
    \begin{cases}
        0, & x_j < \tau, \\
        2\tau, & \tau \leq x_j < 3\tau, \\
        x_j, & x_j \geq 3\tau,
    \end{cases}
\end{align*}
where $\tau$ is a small positive threshold. Unlike a simple threshold, which sets all values below $\tau$ to zero, this scheme avoids an excessive concentration of exact zeros. Here, we set $\tau = 0.01$ so that roughly $10\%$ of samples contain zeros. The resulting vector $\xb'$ is then normalized to the simplex, denoted $\xb^{(0)}$, with density estimates $\hat{f}^{(0)}$. We further estimate densities on $\xb^{(0)}$ after applying three zero-replacement strategies: Aitchison's method, multiplicative replacement, and log-ratio EM. We denote the resulting estimate after this zero-replacement as $\hat{f}^{(*)}$. We evaluate the performance of $\hat{f}^{(0)}$ and $\hat{f}^{(*)}$ against $\hat{f}$ using the loss functions. As both are measured relative to the same $\hat{f}$, $\hat{f}^{(0)}$ and $\hat{f}^{(*)}$ are directly comparable. 

To avoid favoring estimators that share the distribution family of a single reference model, we employ two reference densities $\hat{f}$: ilr-PI from the Gaussian family and Dir-CVKLD from the Dirichlet family. Zero-containing observations are excluded when ilr-PI is used as $\hat{f}$, since it can only be constructed on $\widetilde{\Delta}^d$. Table~\ref{tab:GDP_simu} reports the MIAE and MKLD, along with their standard errors (subscripts), under the two reference densities, based on $1,000$ Monte Carlo replications. For each fixed $d$, the minimum value in each scenario (no replacement or any of the three zero-replacement strategies), along with statistically indistinguishable values at the 5\% level, is highlighted in bold.

\begin{table*}[!ht]
    \centering
    \caption{\small Comparison of density estimators with zero-replacement strategies (Achson, Mult, lrEM) and without (zero), for different estimators. Mean loss values with standard errors (subscripts) based on Monte Carlo replications under the parametric bootstrap of GDP compositions are reported. Bold highlights the minimum and those not significantly different from it in each replacement method.}
    
    \text{(a) With ilr-PI as the reference density.}\\[0.5em]
    \setlength{\tabcolsep}{3pt}
    \small
    \begin{tabular}{clcccccccc}
        \toprule
        & Method & 
        \multicolumn{4}{c}{MIAE ($\times$10)} & \multicolumn{4}{c}{MKLD ($\times$10)} \\
        \midrule
        & & Zero & Achson & Mult & lrEM & Zero & Achson & Mult & lrEM  \\
        \cmidrule(lr){3-6} \cmidrule(lr){7-10}
        \multirow{7}{*}{\makecell[l]{$d = 2$}}
        & ilr-PI  & -- & $\boldsymbol{2.88}_{0.01}$ & $3.54_{0.01}$ & $\boldsymbol{2.67}_{0.01}$ & -- & $4.16_{0.02}$ & $9.89_{0.04}$ & $2.81_{0.03}$\\
        & ilr-CV  & -- & $10.02_{0.12}$ & $11.68_{0.05}$ & $10.45_{0.09}$ & -- & $321.67_{5.62}$ & $293.03_{2.66}$ & $290.58_{4.13}$\\
        & ilr-MDE & -- & $4.87_{0.06}$ & $5.45_{0.04}$ & $4.86_{0.04}$ & -- & $40.06_{1.66}$ & $59.11_{2.39}$ & $15.35_{0.38}$\\
        & Dir-KDE & $3.27_{0.02}$ & $3.23_{0.01}$ & $3.49_{0.01}$ & $3.14_{0.01}$ & $1.69_{0.03}$ & $1.62_{0.01}$ & $2.22_{0.02}$ & $1.44_{0.01}$\\
        & Dir-AIC  & $\boldsymbol{2.94}_{0.01}$ & $3.00_{0.01}$ & $\boldsymbol{3.23}_{0.02}$ & $2.98_{0.01}$ & $\boldsymbol{1.17}_{0.02}$ & $\boldsymbol{1.17}_{0.01}$ & $\boldsymbol{1.40}_{0.02}$ & $1.15_{0.01}$ \\
        & Dir-CVISE  & $6.67_{0.01}$ & $4.16_{0.04}$ & $5.69_{0.05}$ & $4.29_{0.05}$ & $7.76_{0.04}$ & $3.00_{0.09}$ & $7.28_{0.16}$ & $3.57_{0.11}$ \\
        & Dir-CVKLD  & $\boldsymbol{2.92}_{0.01}$ & $3.00_{0.01}$ & $3.28_{0.02}$ & $2.96_{0.01}$ & $\boldsymbol{1.14}_{0.01}$ & $\boldsymbol{1.16}_{0.01}$ & $1.46_{0.02}$ & $\boldsymbol{1.13}_{0.01}$ \\
        \midrule
        \multirow{7}{*}{\makecell[l]{$d = 5$}}
        & ilr-PI & -- & $7.95_{0.01}$ & $8.63_{0.01}$ & $8.40_{0.01}$ & -- & $14.11_{0.06}$ & $18.85_{0.06}$ & $16.61_{0.06}$\\
        & ilr-CV & -- & $7.77_{0.01}$ & $8.02_{0.01}$ & $7.87_{0.02}$ & -- & $14.12_{0.09}$ & $17.48_{0.20}$ & $16.05_{0.15}$\\
        & ilr-MDE & -- & $\boldsymbol{7.24}_{0.01}$ & $\boldsymbol{7.27}_{0.01}$ & $\boldsymbol{7.22}_{0.01}$ & -- & $\boldsymbol{11.56}_{0.03}$ & $12.36_{0.03}$ & $12.20_{0.02}$\\
        & Dir-KDE & $7.42_{0.01}$ & $7.41_{0.01}$ & $7.46_{0.01}$ & $7.43_{0.01}$ & $12.01_{0.02}$ & $12.12_{0.02}$ & $12.83_{0.02}$ & $12.40_{0.02}$\\
        & Dir-AIC & $8.14_{0.01}$ & $8.20_{0.01}$ & $8.08_{0.01}$ & $8.11_{0.01}$ & $14.27_{0.03}$ & $14.32_{0.03}$ & $14.15_{0.03}$ & $14.13_{0.03}$\\
        & Dir-CVISE & $9.04_{0.07}$ & $7.89_{0.04}$ & $8.29_{0.03}$ & $8.15_{0.04}$ & $20.69_{0.70}$ & $13.89_{0.32}$ & $14.93_{0.15}$ & $15.03_{0.46}$\\
        & Dir-CVKLD & $\boldsymbol{7.37}_{0.01}$ & $7.31_{0.01}$ & $7.34_{0.01}$ & $7.32_{0.01}$ & $\boldsymbol{11.91}_{0.02}$ & $11.67_{0.02}$ & $\boldsymbol{11.92}_{0.02}$ & $\boldsymbol{11.72}_{0.02}$\\
        \bottomrule
    \end{tabular}

    \bigskip

    \text{(b) With Dir-CVKLD as the reference density.}\\[0.5em]
    \setlength{\tabcolsep}{3pt}
    \small
    \begin{tabular}{clcccccccc}
        \toprule
        & Method & 
        \multicolumn{4}{c}{MIAE ($\times$10)} & \multicolumn{4}{c}{MKLD ($\times$10)} \\
        \midrule
        & & Zero & Achson & Mult & lrEM & Zero & Achson & Mult & lrEM  \\
        \cmidrule(lr){3-6} \cmidrule(lr){7-10}
        \multirow{7}{*}{\makecell[l]{$d = 2$}}
        & ilr-PI  & -- & $2.83_{0.01}$ & $3.30_{0.01}$ & $2.86_{0.01}$ & -- & $5.37_{0.02}$ & $9.53_{0.03}$ & $5.67_{0.03}$\\
        & ilr-CV  & -- & $9.39_{0.10}$ & $10.51_{0.06}$ & $9.74_{0.10}$ & -- & $273.14_{4.71}$ & $233.88_{2.73}$ & $226.54_{4.06}$\\
        & ilr-MDE & -- & $4.25_{0.06}$ & $4.81_{0.04}$ & $4.44_{0.04}$ & -- & $46.69_{2.24}$ & $65.50_{2.90}$ & $31.70_{1.01}$\\
        & Dir-KDE & $2.61_{0.02}$ & $2.67_{0.01}$ & $2.85_{0.01}$ & $2.66_{0.01}$ & $1.11_{0.03}$ & $1.11_{0.01}$ & $1.56_{0.02}$ & $1.10_{0.01}$\\
        & Dir-AIC  & $2.20_{0.01}$ & $\boldsymbol{2.21}_{0.01}$ & $\boldsymbol{2.49}_{0.02}$ & $\boldsymbol{2.20}_{0.01}$ & $0.53_{0.01}$ & $\boldsymbol{0.60}_{0.01}$ & $\boldsymbol{0.82}_{0.01}$ & $\boldsymbol{0.59}_{0.01}$ \\
        & Dir-CVISE  & $6.11_{0.03}$ & $2.88_{0.03}$ & $4.19_{0.05}$ & $2.82_{0.03}$ & $6.93_{0.06}$ & $1.40_{0.06}$ & $4.20_{0.14}$ & $1.42_{0.07}$ \\
        & Dir-CVKLD  & $\boldsymbol{2.10}_{0.01}$ & $\boldsymbol{2.21}_{0.01}$ & $\boldsymbol{2.51}_{0.02}$ & $\boldsymbol{2.18}_{0.01}$ & $\boldsymbol{0.49}_{0.01}$ & $\boldsymbol{0.61}_{0.01}$ & $0.85_{0.02}$ & $\boldsymbol{0.60}_{0.01}$ \\
        \midrule
        \multirow{7}{*}{\makecell[l]{$d = 5$}}
        & ilr-PI & -- & $8.99_{0.12}$ & $9.38_{0.01}$ & $9.31_{0.02}$ & -- & $14.19_{0.05}$ & $20.02_{0.05}$ & $18.81_{0.06}$\\
        & ilr-CV & -- & $8.41_{0.19}$ & $8.21_{0.02}$ & $8.12_{0.02}$ & -- & $10.21_{0.13}$ & $14.49_{0.30}$ & $13.43_{0.23}$\\
        & ilr-MDE & -- & $5.65_{0.01}$ & $5.58_{0.01}$ & $5.52_{0.01}$ & -- & $3.46_{0.01}$ & $4.25_{0.03}$ & $3.99_{0.01}$\\
        & Dir-KDE & $5.95_{0.01}$ & $5.76_{0.01}$ & $5.78_{0.01}$ & $5.74_{0.01}$ & $3.29_{0.01}$ & $3.31_{0.01}$ & $3.65_{0.01}$ & $3.51_{0.01}$\\
        & Dir-AIC & $5.92_{0.04}$ & $5.85_{0.03}$ & $5.69_{0.02}$ & $5.65_{0.02}$ & $2.99_{0.03}$ & $2.90_{0.02}$ & $2.79_{0.02}$ & $2.76_{0.02}$\\
        & Dir-CVISE & $\boldsymbol{4.48}_{0.03}$ & $4.31_{0.01}$ & $4.53_{0.02}$ & $4.38_{0.02}$ & $2.68_{0.31}$ & $2.00_{0.01}$ & $2.35_{0.02}$ & $2.17_{0.02}$\\
        & Dir-CVKLD & $4.71_{0.02}$ & $\boldsymbol{4.28}_{0.01}$ & $\boldsymbol{4.32}_{0.01}$ & $\boldsymbol{4.24}_{0.01}$ & $\boldsymbol{2.17}_{0.01}$ & $\boldsymbol{1.91}_{0.01}$ & $\boldsymbol{2.03}_{0.01}$ & $\boldsymbol{1.95}_{0.01}$\\
        \bottomrule
    \end{tabular}
    \label{tab:GDP_simu}
\end{table*}

Table~\ref{tab:GDP_simu}(a) presents the case with ilr-PI as the reference density. In terms of MIAE, ilr-PI and ilr-MDE achieve the best performance at $d=2$ and $d=5$, respectively. This is unsurprising, as both estimators belong to the Gaussian distributional family of the reference density. Dir-AIC and Dir-CVKLD, though based on the Dirichlet family, also yield competitive results. For MKLD, Dir-CVKLD generally achieves the best performance, as its bandwidth is selected by minimizing MKLD, while Dir-AIC performs comparably to Dir-CVKLD at $d=2$. Table~\ref{tab:GDP_simu}(b) reports the results with Dir-CVKLD as the reference density. In this setting, the relative performance of the method using MIAE or MKLD is fairly similar. Dir-CVKLD unsurprisingly achieves the best performance, as it coincides with the reference density, while Dir-AIC performs comparably at $d=3$. Importantly, in both cases and except for Dir-CVISE, the performance of Dirichlet-based estimators is largely unaffected by zero-replacement, indicating that preprocessing of compositions is unnecessary in this framework.

\begin{figure}[!tbh]
    \centering
    \includegraphics[width=0.45\textwidth,
        clip,trim={0.8cm 0.8cm 1cm 0.5cm}]{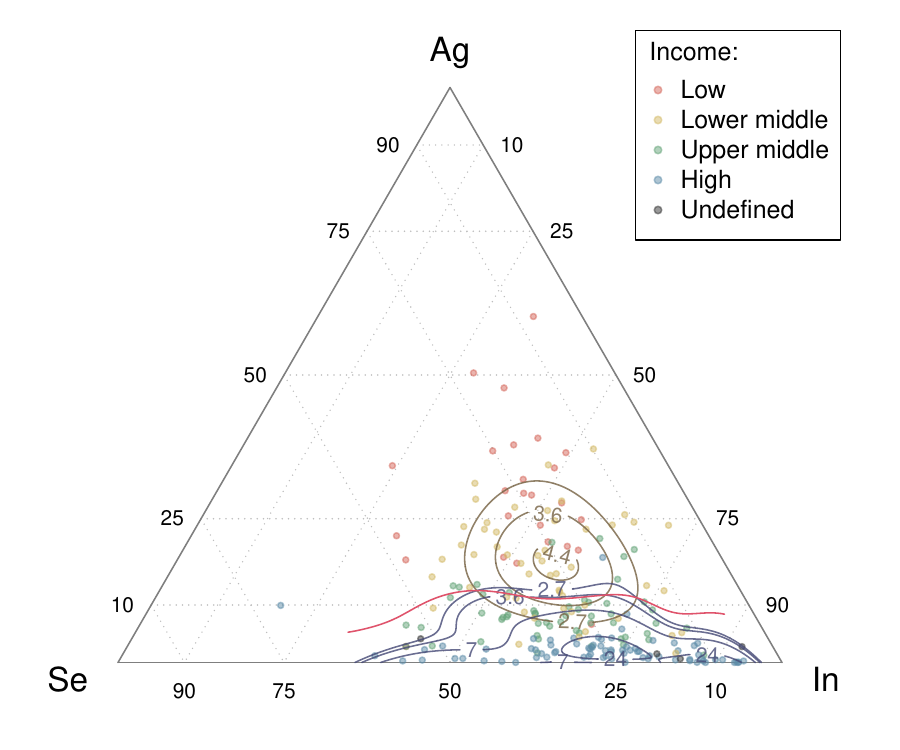}
    \caption{\small National GDP compositions colored by income level, with joint density contours of the two merged income groups (Low + Lower-middle and Upper-middle + High), and the decision boundary (red line).}
    \label{fig:GDPcls}
\end{figure}

We also investigate whether a country's income level can be predicted from the relative distribution of GDP across sectors. Based on Gross National Income (GNI) per capita, \cite{IncomeGroups2023} classifies countries into four levels (excluding four undefined countries): 76 high income, 54 upper middle income, 52 lower middle income, and 26 low income. We merge the upper two and the lower two levels into two groups. Figure~\ref{fig:GDPcls} shows the countries colored by income level, the joint density contours of the two merged groups (each obtained by scaling the Dir-CVKLD density estimates by group proportion), and the red decision boundary separating the two groups. Overall, higher-income countries tend to have a smaller agricultural share in GDP, with a proportion of about $0.1$ roughly delineating the two merged groups. Since the national GDP dataset is relatively small, we did not compare classification results across different estimators here. More detailed classification experiments are presented in Sections~\ref{sec:USPS}--\ref{sec:skin}.

\begin{figure*}[!htb]
    \centering
    \includegraphics[page=3, width=0.325\textwidth,
        clip,trim={1.2cm 0.8cm 1.8cm 0.6cm}]{digit.pdf}
    \includegraphics[page=1, width=0.325\textwidth,
        clip,trim={1.2cm 0.8cm 1.8cm 0.6cm}]{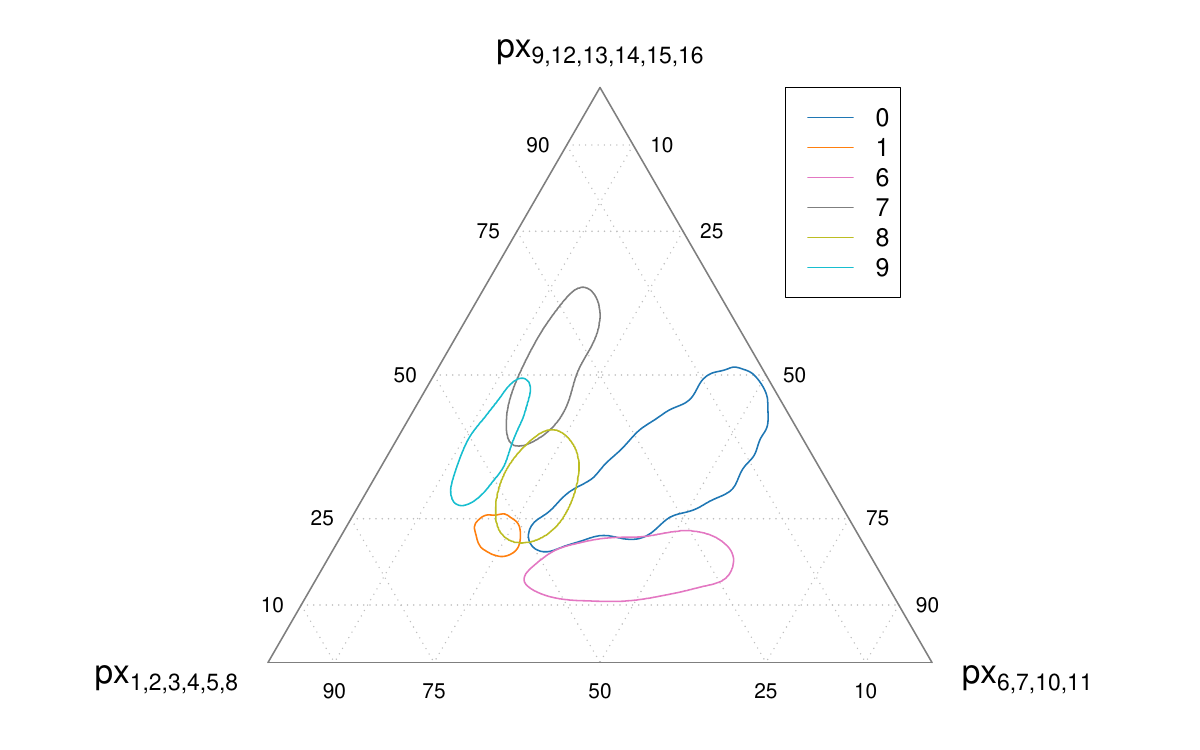}
    \includegraphics[page=2, width=0.325\textwidth,
        clip,trim={1.2cm 0.8cm 1.8cm 0.6cm}]{digit.pdf}
    \caption{\small Marginal probability contour plots on $\Delta^2$ of Dir-CVKLD estimated from USPS dataset.}
    \label{fig:digits}
\end{figure*}

\subsection{Classification for USPS dataset}
\label{sec:USPS}

The USPS dataset is a standard benchmark dataset for handwritten digit classification, originally collected by the \cite{USPS1994} and published by New York University. It contains $9,298$ grayscale images of digits $0$-$9$, each represented as a $16 \times 16$ pixel array, with $7,291$ images for training and $2,007$ for testing. The dataset is transformed to compositional data to remove the influence of ink quantity and focus on digit shape. To reduce dimensionality while retaining local structure, fixed-grid spatial pooling is applied: each $16 \times 16$ image is divided into 16 non-overlapping $4 \times 4$ blocks, and the pixel proportions within each block are aggregated to form its compositional value. This yields a $4 \times 4$ representation, treated as a 16-dimensional compositional vector for classification.

Figure~\ref{fig:digits} presents the $80\%$ contour levels on $\Delta^2$. Each contour corresponds to the marginal Dirichlet mixture density from the Dir-CVKLD method for a digit, with different part aggregations shown in each plot. The digit classes are well separated, and since these $\Delta^2$ plots are marginals of $\Delta^{15}$, they illustrate class separability in the full space.

The density estimators can be directly applied to classification. Let $C$ denote the set of classes. For each class $c \in C$, the class-conditional density $f_c$ is estimated nonparametrically from the training data. A composition $\xb \in \Delta^d$ is then classified by maximum unnormalized posterior:
\begin{align*}
    \hat{c}(\xb) = \argmax_{c \in C} \pi_c f_c(\xb),
\end{align*}
where $\pi_c$ is the empirical class prior, given by the class proportion in the training set, thereby correcting for class imbalance. For evaluating classification performance of different estimators on the USPS dataset, where the digits ``0'' and ``1'' occur more frequently and the data are mildly imbalanced, accuracy tends to favor majority classes and can be misleading. Instead, we adopt the F-score, defined using $\mathrm{Precision} = \mathrm{TP}\,/\,(\mathrm{TP}+\mathrm{FP})$ and $\mathrm{Recall} = \mathrm{TP}\,/\,(\mathrm{TP}+\mathrm{FN})$ as
\begin{align}
    \mathrm{F}_\beta =  \frac{(1 + \beta^2)\,\mathrm{Precision} \cdot \mathrm{Recall}} {\beta^2\,\mathrm{Precision} + \mathrm{Recall}} = \frac{(1 + \beta^2)\,\mathrm{TP}}{(1 + \beta^2)\,\mathrm{TP} + \beta^2\,\mathrm{FN} + \mathrm{FP}},
    \label{eq:F1-score}
\end{align}
where $\beta$ controls the relative weight of recall and precision, we set $\beta=1$ to balance the two, yielding the F1-score. For multi-class data, we adopt the so-called micro-F1 score, defined as 
\begin{align*}
    \text{Micro-F1} = \frac{2\,\sum_c\mathrm{TP}_c}{2\,\sum_c\mathrm{TP}_c + \sum_c\mathrm{FN}_c + \sum_c\mathrm{FP}_c},
\end{align*}
where the subscript $c$ indexes the classes.

Table~\ref{tab:Digitcl} reports the micro-F1 score on the test dataset. Gaussian kernel estimators are excluded due to the high computational cost in $d=15$. All Dirichlet-based estimators are applied to the original data, while ilr-MDE is developed on multivariate zero-replaced data. To assess significant differences in estimator performance, we employed a bootstrap procedure: predicted labels on the test set were resampled $10^4$ times with replacement to obtain bootstrap distributions of micro-F1 scores, and differences between the two methods were summarized by 95\% confidence intervals. The results show that Dir-KDE and Dir-CVKLD achieve the best performance, with no significant difference, while the Gaussian-based ilr-MDE also performs satisfactorily. Dir-AIC remains unreliable in high dimensions.

\begin{table}[!ht]
    \centering
    \caption{\small Micro-F1 score ($\%$) on the USPS test set. Bold highlights the best two estimators with no significant difference.}
    \setlength{\tabcolsep}{3pt}
    \begin{tabular}{ccccc}
        \toprule
        ilr-MDE & Dir-KDE & Dir-AIC & Dir-CVISE & Dir-CVKLD \\ 
        \midrule
        90.43 & $\boldsymbol{91.53}$ & 85.30 & 89.54 & $\boldsymbol{91.43}$ \\
        \bottomrule
    \end{tabular}
    \label{tab:Digitcl}
\end{table}

\subsection{Skin detection for Pratheepan dataset}
\label{sec:skin}

Pratheepan \citep{Tan2012} is an image dataset curated for human skin detection. The images, randomly downloaded from Google, were captured under diverse lighting conditions, cameras, and color enhancement settings. The dataset comprises both portrait images (individual faces with simple backgrounds) and group photos (multiple subjects with complex backgrounds). Pixels are labeled as skin or non-skin. Instead of representing pixels in the RGB format, which encodes both color and luminance, we retain only the relative proportions of three color channels, yielding compositional data on $\Delta^2$, which we refer to as the RGB composition. This mitigates the effect of illumination variation across images and provides a more robust feature space for skin detection.

We focus on skin detection in group photographs where multiple individuals and complex backgrounds pose a greater challenge. Out of a total of 46 group images, 32 were randomly selected as the training set with a total of $1.18\times10^6$ skin pixels and $8.09\times10^6$ non-skin pixels. From these, $10^4$ pixels were randomly sampled from each class, with the number of pixels drawn from each image allocated in proportion to its share of the total training pixels, to construct the skin model $\hat{f}_\text{skin}$ and non-skin model $\hat{f}_\text{non-skin}$. The RGB composition vector $\xb$ of a pixel is classified as skin if both conditions hold:
\begin{align*}
    \begin{cases}
        \hat{f}_\text{skin}(\xb) > \text{Quantile}_p \bigl(\{\hat{f}_\text{skin}(\xb_i)\}_{i=1}^n \bigr), \\[0.2em]
        \gamma \, \hat{f}_\text{skin}(\xb) > \hat{f}_\text{non-skin}(\xb),
    \end{cases}
\end{align*}
where $\{\xb_i\}_{i=1}^n$ is the sample used to train $\hat{f}_\text{skin}$, $\text{Quantile}_p(\cdot)$ represents the $p \times 100\%$ sample quantile, and $\gamma$ is a tunable decision threshold. The first condition ensures that $\xb$ lies in a typical skin region, while the second is a Bayesian likelihood ratio test distinguishing skin from non-skin.

\begin{figure}[!h]
    \centering
    \includegraphics[width=0.45\textwidth,clip,
    trim={0.6cm 1.7cm 0.6cm 1.5cm}]{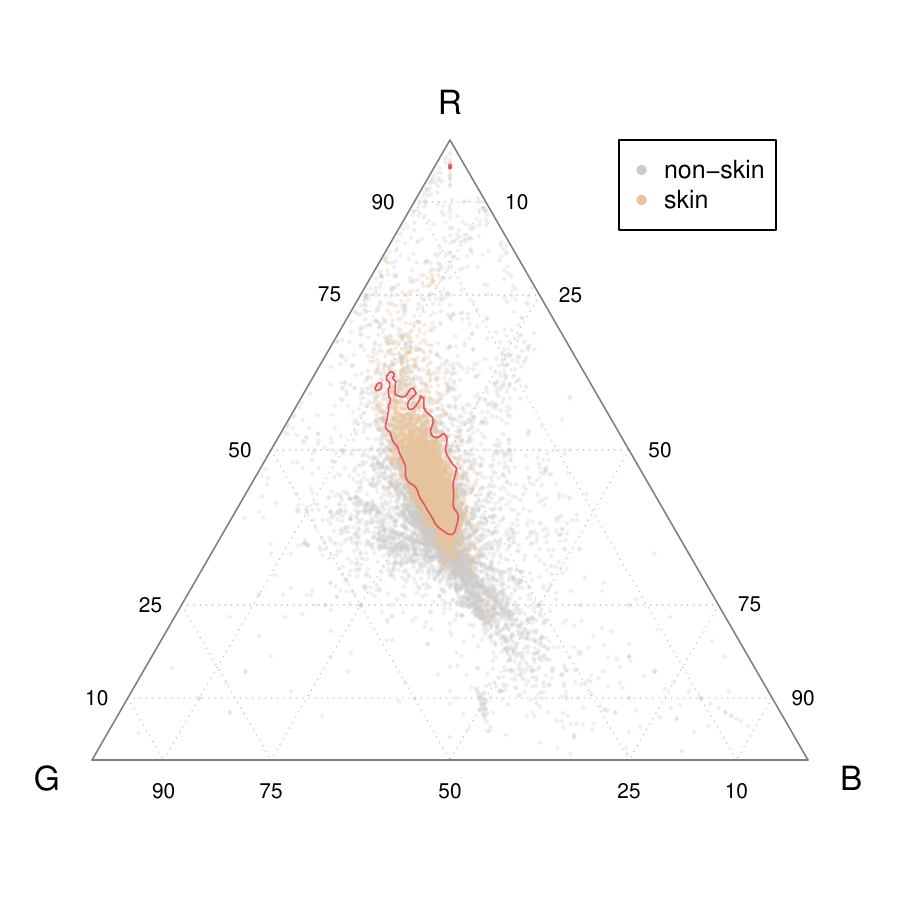}
    \caption{\small Decision boundary under optimal parameters (maximizing F2-score).}
    \label{fig:skinmodel}
\end{figure}

To determine appropriate values for $p$, $\gamma$, and the bandwidth factors $\eta\,_\text{skin}$ and $\eta\,_\text{non-skin}$ of two models, a validation set of $10^5$ pixels was randomly sampled from the training set, excluding those used for model fitting. A grid search was performed over $p \in [0, 0.2]$ (step 0.05), $\gamma \in [0.1, 2.0]$ (step 0.05), and $\eta\,_\text{skin}$, $\eta\,_\text{non-skin} \in [0.1, 1.0]$ (step 0.1). To avoid missing true skin regions, some false positives can be tolerated. Thus, the objective was set to maximize the F2-score by choosing $\beta=2$ in \eqref{eq:F1-score}, thereby emphasizing recall. The results show that $(p, \gamma, \eta\,_\text{skin}, \eta\,_\text{non-skin}) = (0.05, 0.7, 0.2, 0.1)$ yield the highest F2-score. We also selected $\eta\,_\text{skin}$ and $\eta\,_\text{non-skin}$ individually using the bandwidth selector $\hat{h}_\mathrm{CVKLD}$ and obtained the same optimal values. Figure~\ref{fig:skinmodel} shows the pixel samples in RGB composition space, colored by their ground-truth label. The red curve denotes the decision boundary, with pixels inside classified as skin.

\begin{figure*}[!ht]
  \centering
  \includegraphics[width=0.23\textwidth]{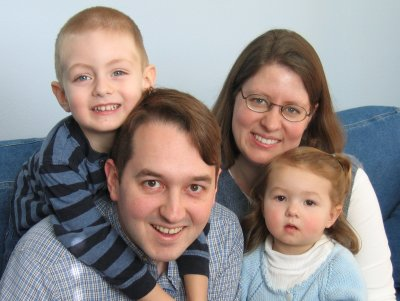}
  \includegraphics[width=0.23\textwidth]{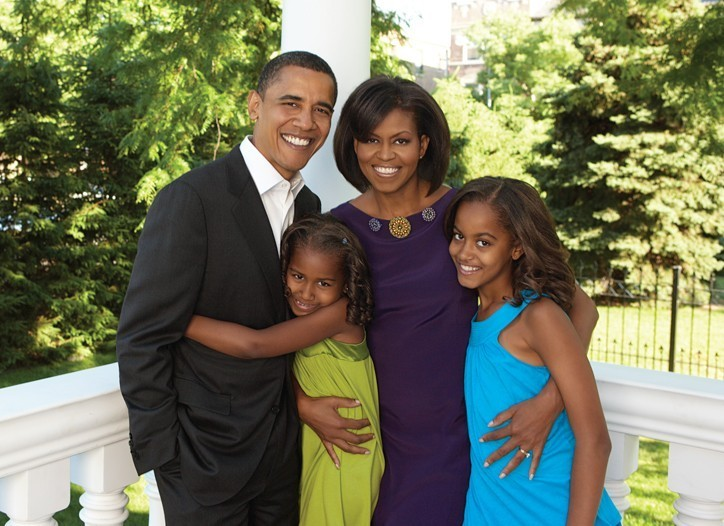}
  \includegraphics[width=0.23\textwidth]{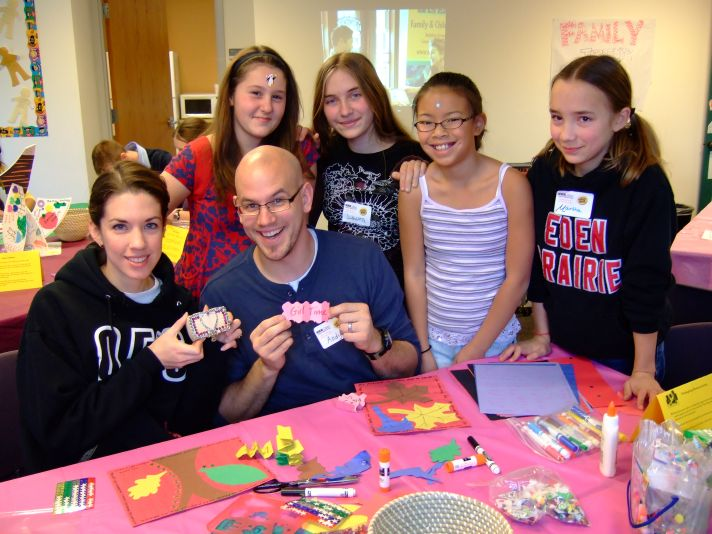}
  \includegraphics[width=0.23\textwidth]{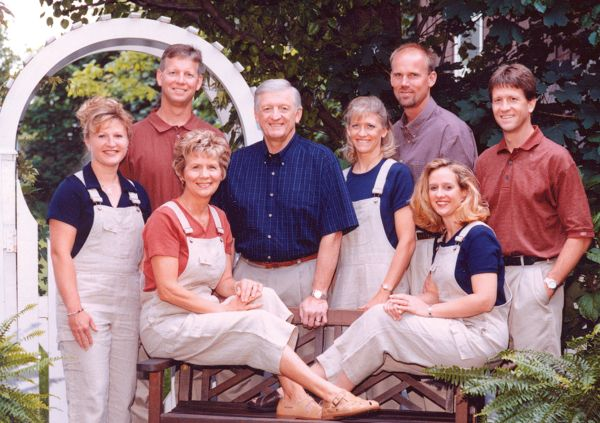}

  \par\medskip

  \includegraphics[width=0.23\textwidth]{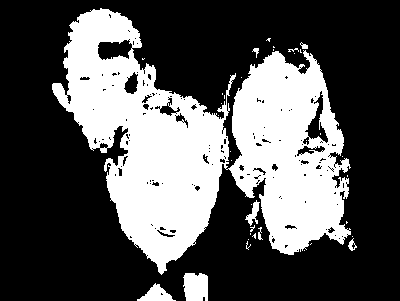}
  \includegraphics[width=0.23\textwidth]{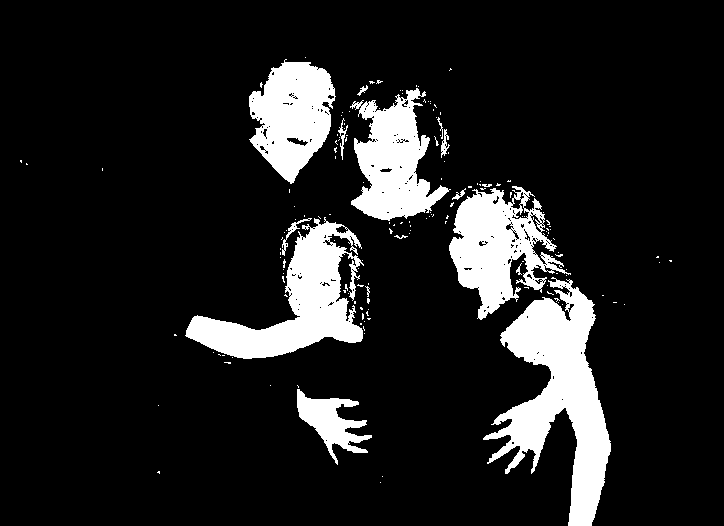}
  \includegraphics[width=0.23\textwidth]{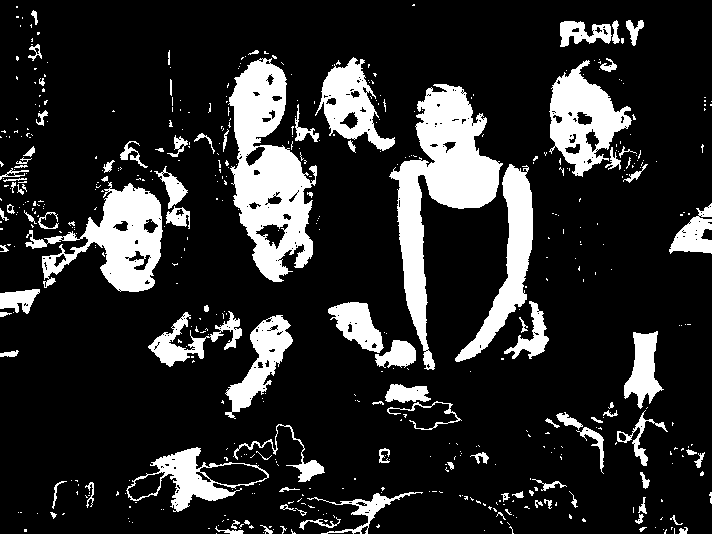}
  \includegraphics[width=0.23\textwidth]{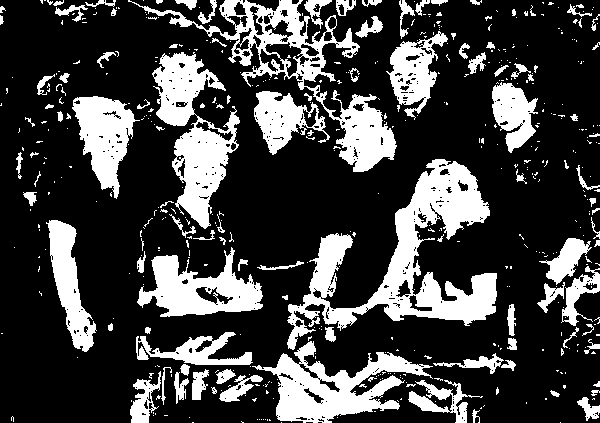}
  
  \caption{\small Original images (top row) and detection results (bottom row), with skin regions shown in white. From left to right, the four cases are denoted as (a) -- (d).}
  \label{fig:pngimg}
\end{figure*}

Figure~\ref{fig:pngimg} shows four test images with detected skin regions marked in white. Table~\ref{tab:skin} reports their precision, recall, and F2-scores. Our method achieved high recall and F2-scores, but relatively low precision. This is reasonable, since the non-skin region is much larger than the skin region in our dataset. Hence, even a small fraction of misclassified non-skin pixels produces many false positives, thereby reducing precision. Thus, precision alone is not an appropriate evaluation criterion in this setting. Since our method only relies on the relative RGB proportions of individual pixels, it is limited in distinguishing skin pixels from non-skin pixels with almost identical RGB compositions in Figure~\ref{fig:skinmodel} (e.g., the painting in (c) or the leaves and desk in (d)), which constitutes the primary source of false positives. Nevertheless, owing to its high recall, the method successfully detects most skin pixels, making it suitable for preliminary screening of skin regions, the results of which can perhaps be further refined with more detailed adjustments.

\begin{table}[!ht]
    \centering
    \caption{\small Precision, recall, and F2-score for test images (a–d).}
    \small
    \begin{tabular}{lcccc}
        \toprule
        & (a) & (b) & (c) & (d) \\
        \midrule
        Precision & 65.60 & 65.84 & 65.37 & 41.37 \\ 
        Recall & 97.65 & 94.50 & 87.28 & 92.63 \\
        F2-score & 88.96 & 86.93 & 81.80 & 74.23 \\
        \bottomrule
    \end{tabular}
    \label{tab:skin}
\end{table}

\section{Discussion}
\label{sec:diss}

We proposed nonparametric mixture density estimation for compositional data. A nonparametric mixture density is expressed as the integral of a one-parameter component density with respect to an unspecified mixing distribution. Two component families are considered: Gaussian distributions applied to log-ratio transformed data, and Dirichlet distributions defined directly on the simplex. We focused on Dirichlet components, as they naturally accommodate zeros without requiring zero-replacement or log-ratio transformations, thereby preserving the original data structure on the simplex. Moreover, Dirichlet mixtures afford components supported on the simplex boundary, yielding smooth transitions of density near the boundary.

In nonparametric Dirichlet mixtures, the parameter vector $\alphab$ serves as the mixing parameter, while the concentration parameter $\alpha_0 = \alphab^\top \mathbf{1}$ controls the smoothness of the estimated density. For each fixed $\alpha_0$, the mixing distribution is obtained by likelihood maximization. The optimal $\alpha_0$ is selected via deterministic criterion or cross-validation approaches. 

Across both simulations and real data, the proposed methods consistently outperform traditional kernel density estimation for compositional data. In tasks relying on density estimates, such as classification and skin detection, they also achieve robust and competitive performance. Among all estimators, Dir-CVKLD is particularly notable for its adaptability and accuracy, showing stable performance across sample sizes and dimensions.

However, the Dirichlet distribution has a limitation as its boundary behavior changes discontinuously with respect to $\alphab$. When $\alpha_j = 1$, the mode lies on the simplex boundary with positive density, but any increase to $\alpha_j > 1$ shifts the mode into the interior and forces the boundary density to zero. This parameter-driven discontinuity affects all Dirichlet-based methods. An important direction for future research is exploring alternative component distributions, still within the framework proposed in this article, that provide smoother boundary behavior as the parameter value varies.

\begin{appendices}

\section{Simulation results for MISE}\label{sec:mise}

Tables \ref{tab:chacon_MISER}--\ref{tab:dirichlet_MISER} report the mean ISE (MISE) in Euclidean space for the Gaussian and Dirichlet mixture simulations, respectively, using the same format as Table \ref{tab:chacon_MIAE}. The performance results are consistent with those observed in Section~\ref{sec:simu} with other two measures. In Dirichlet mixture simulations, Dir-CVISE achieves the best performance overall in heterogeneous scenarios at $d=5$, perhaps due to its bandwidth selector is designed to minimize MISE, giving it an advantage under this measure, even though the selector and the evaluation are based on different spaces.

\begin{sidewaystable*}[p]
    \centering
    \caption{\small MISE with standard errors (subscripts) in Euclidean space, based on Monte Carlo replications in the Gaussian mixture simulations; bold highlights the minimum and those not significantly different from it in each scenario.}
    \setlength{\tabcolsep}{3pt}
    \small
    \begin{tabular}{llcccccccccccc}
    \toprule
    Settings & Estimators & (1) & (2) & (3) & (4) & (5) & (6) & (7) & (8) & (9) & (10) & (11) & (12) \\
    \midrule
    \multirow{7}{*}{\centering\makecell[l]{$d=2$, \\ $n=100$ \\[3pt] ($\times 1000$)}}
    
    & ilr-PI & $9.91_{0.13}$ & $11.03_{0.14}$ & $\boldsymbol{81.88}_{0.82}$ & $\boldsymbol{10.16}_{0.08}$ & $11.18_{0.10}$ & $8.01_{0.08}$ & $14.75_{0.11}$ & $16.32_{0.11}$ & $9.41_{0.09}$ & $29.26_{0.14}$ & $13.69_{0.11}$ & $36.63_{0.13}$ \\
    & ilr-CV & $17.00_{0.40}$ & $17.85_{0.38}$ & $99.08_{1.39}$ & $13.62_{0.20}$ & $12.09_{0.21}$ & $12.66_{0.26}$ & $15.72_{0.30}$ & $15.64_{0.26}$ & $13.61_{0.23}$ & $29.66_{0.32}$ & $16.80_{0.28}$ & $\boldsymbol{31.84}_{0.35}$ \\
    & ilr-MDE & $\boldsymbol{6.31}_{0.19}$ & $\boldsymbol{7.46}_{0.21}$ & $98.60_{1.14}$ & $14.33_{0.11}$ & $10.89_{0.14}$ & $\boldsymbol{7.43}_{0.14}$ & $\boldsymbol{8.17}_{0.25}$ & $\boldsymbol{9.95}_{0.17}$ & $\boldsymbol{8.93}_{0.12}$ & $31.81_{0.27}$ & $\boldsymbol{9.32}_{0.20}$ & $34.83_{0.28}$ \\
    & Dir-KDE & $7.84_{0.12}$ & $17.55_{0.18}$ & $357.89_{0.95}$ & $13.64_{0.10}$ & $\boldsymbol{7.78}_{0.08}$ & $8.42_{0.11}$ & $15.20_{0.13}$ & $13.76_{0.13}$ & $12.39_{0.11}$ & $37.78_{0.32}$ & $16.11_{0.13}$ & $47.25_{0.25}$ \\
    & Dir-MDE & $9.87_{0.16}$ & $20.65_{0.23}$ & $302.79_{0.92}$ & $14.67_{0.13}$ & $9.00_{0.11}$ & $7.79_{0.11}$ & $15.30_{0.20}$ & $12.72_{0.13}$ & $13.61_{0.13}$ & $32.69_{0.30}$ & $17.72_{0.16}$ & $48.21_{0.21}$ \\
    & Dir-CVISE & $10.43_{0.21}$ & $22.64_{0.35}$ & $298.94_{0.76}$ & $15.95_{0.23}$ & $15.89_{0.55}$ & $8.36_{0.16}$ & $15.32_{0.17}$ & $12.93_{0.16}$ & $13.90_{0.15}$ & $\boldsymbol{28.19}_{0.29}$ & $18.48_{0.23}$ & $42.41_{0.24}$ \\
    & Dir-CVKLD & $10.36_{0.19}$ & $21.50_{0.27}$ & $386.80_{0.94}$ & $14.33_{0.11}$ & $8.82_{0.11}$ & $7.97_{0.12}$ & $14.82_{0.15}$ & $12.18_{0.13}$ & $13.19_{0.13}$ & $29.71_{0.25}$ & $16.86_{0.14}$ & $44.32_{0.22}$ \\

    \midrule
    
    \multirow{7}{*}{\centering\makecell[l]{$d=2$, \\ $n=500$ \\[3pt] ($\times 1000$)}}
    & ilr-PI & $3.39_{0.04}$ & $3.94_{0.04}$ & $31.77_{0.32}$ & $\boldsymbol{4.11}_{0.03}$ & $4.67_{0.04}$ & $2.99_{0.03}$ & $4.77_{0.04}$ & $5.39_{0.04}$ & $3.55_{0.03}$ & $12.31_{0.07}$ & $5.43_{0.04}$ & $19.21_{0.07}$ \\
    & ilr-CV & $4.56_{0.07}$ & $5.28_{0.08}$ & $31.21_{0.33}$ & $4.44_{0.04}$ & $3.73_{0.04}$ & $3.66_{0.05}$ & $4.58_{0.05}$ & $4.78_{0.05}$ & $4.12_{0.04}$ & $10.13_{0.07}$ & $5.67_{0.06}$ & $\boldsymbol{11.81}_{0.07}$ \\
    & ilr-MDE & $\boldsymbol{1.35}_{0.04}$ & $\boldsymbol{1.59}_{0.05}$ & $\boldsymbol{29.60}_{0.41}$ & $4.31_{0.04}$ & $\boldsymbol{2.87}_{0.05}$ & $\boldsymbol{1.79}_{0.04}$ & $\boldsymbol{2.09}_{0.11}$ & $\boldsymbol{2.60}_{0.03}$ & $\boldsymbol{2.36}_{0.03}$ & $\boldsymbol{9.12}_{0.09}$ & $\boldsymbol{3.54}_{0.12}$ & $14.55_{0.12}$ \\
    & Dir-KDE & $2.92_{0.03}$ & $7.14_{0.06}$ & $232.83_{0.55}$ & $5.29_{0.03}$ & $3.28_{0.03}$ & $2.96_{0.03}$ & $5.49_{0.04}$ & $4.99_{0.04}$ & $4.51_{0.03}$ & $14.78_{0.30}$ & $6.63_{0.04}$ & $28.65_{0.48}$ \\
    & Dir-AIC & $3.40_{0.04}$ & $8.03_{0.07}$ & $238.99_{0.26}$ & $5.45_{0.04}$ & $3.17_{0.03}$ & $2.62_{0.03}$ & $4.98_{0.04}$ & $4.28_{0.04}$ & $4.51_{0.04}$ & $12.93_{0.09}$ & $7.18_{0.04}$ & $24.61_{0.11}$ \\
    & Dir-CVISE & $3.56_{0.05}$ & $9.57_{0.12}$ & $241.08_{0.26}$ & $5.61_{0.05}$ & $10.76_{0.25}$ & $2.83_{0.04}$ & $4.93_{0.05}$ & $4.69_{0.04}$ & $4.84_{0.04}$ & $10.58_{0.07}$ & $8.24_{0.08}$ & $19.38_{0.09}$ \\
    & Dir-CVKLD & $3.74_{0.05}$ & $8.67_{0.09}$ & $238.86_{0.25}$ & $5.05_{0.04}$ & $3.56_{0.04}$ & $2.83_{0.03}$ & $4.70_{0.04}$ & $4.61_{0.04}$ & $4.80_{0.04}$ & $11.64_{0.08}$ & $6.97_{0.05}$ & $22.61_{0.09}$ \\
    
    \midrule
    
    \multirow{7}{*}{\centering\makecell[l]{$d=5$, \\ $n=100$ \\[3pt] ($\times 1000$)}} 
    & ilr-PI & $10.42_{0.06}$ & $4.68_{0.03}$ & $\boldsymbol{359.67}_{1.43}$ & $5.06_{0.02}$ & $1.76_{0.01}$ & $10.58_{0.06}$ & $8.59_{0.04}$ & $8.43_{0.04}$ & $8.31_{0.04}$ & $62.23_{0.15}$ & $96.68_{0.36}$ & $\boldsymbol{40.66}_{0.08}$ \\
    & ilr-CV & $4.03_{0.04}$ & $2.21_{0.02}$ & $487.81_{2.37}$ & $3.13_{0.02}$ & $1.97_{0.01}$ & $4.19_{0.06}$ & $5.21_{0.06}$ & $5.06_{0.04}$ & $4.30_{0.04}$ & $76.45_{0.22}$ & $92.45_{0.36}$ & $44.31_{0.14}$ \\
    & ilr-MDE & $\boldsymbol{0.77}_{0.01}$ & $\boldsymbol{0.47}_{0.00}$ & $508.49_{1.11}$ & $\boldsymbol{1.62}_{0.01}$ & $2.02_{0.00}$ & $1.33_{0.01}$ & $4.20_{0.05}$ & $3.72_{0.03}$ & $\boldsymbol{2.02}_{0.01}$ & $84.76_{0.08}$ & $90.44_{0.36}$ & $45.08_{0.05}$ \\
    & Dir-KDE & $0.97_{0.00}$ & $1.32_{0.00}$ & $594.25_{0.51}$ & $1.95_{0.00}$ & $\boldsymbol{0.97}_{0.01}$ & $1.39_{0.01}$ & $3.22_{0.01}$ & $2.89_{0.01}$ & $2.27_{0.01}$ & $80.06_{0.10}$ & $90.00_{0.36}$ & $44.36_{0.05}$ \\
    & Dir-AIC & $1.41_{0.01}$ & $2.13_{0.01}$ & $1373.55_{31.90}$ & $2.62_{0.01}$ & $1.49_{0.00}$ & $1.83_{0.01}$ & $4.36_{0.02}$ & $3.77_{0.02}$ & $2.94_{0.01}$ & $87.41_{0.10}$ & $90.98_{0.36}$ & $46.50_{0.05}$ \\
    & Dir-CVISE & $1.26_{0.01}$ & $1.91_{0.08}$ & $916.10_{29.61}$ & $2.41_{0.08}$ & $1.72_{0.10}$ & $1.34_{0.01}$ & $2.93_{0.02}$ & $2.46_{0.01}$ & $2.24_{0.02}$ & $\boldsymbol{54.90}_{0.29}$ & $94.29_{0.49}$ & $41.38_{0.12}$ \\
    & Dir-CVKLD & $1.22_{0.01}$ & $1.66_{0.00}$ & $590.94_{0.53}$ & $2.06_{0.01}$ & $1.11_{0.01}$ & $\boldsymbol{1.31}_{0.01}$ & $\boldsymbol{2.90}_{0.01}$ & $\boldsymbol{2.44}_{0.01}$ & $2.17_{0.01}$ & $78.20_{0.17}$ & $\boldsymbol{89.73}_{0.36}$ & $44.77_{0.06}$ \\
    
    \midrule
    
    \multirow{7}{*}{\centering\makecell[l]{$d=5$, \\ $n=500$ \\[3pt] ($\times 1000$)}}  
    & ilr-PI & $3.08_{0.01}$ & $1.43_{0.00}$ & $\boldsymbol{325.37}_{1.30}$ & $1.90_{0.00}$ & $1.05_{0.00}$ & $3.26_{0.01}$ & $3.28_{0.01}$ & $3.21_{0.01}$ & $2.76_{0.01}$ & $47.72_{0.10}$ & $87.87_{0.41}$ & $33.53_{0.07}$ \\
    & ilr-CV & $1.42_{0.01}$ & $0.92_{0.00}$ & $376.02_{1.49}$ & $1.52_{0.01}$ & $1.15_{0.00}$ & $1.55_{0.01}$ & $2.06_{0.01}$ & $2.02_{0.01}$ & $1.73_{0.01}$ & $55.26_{0.13}$ & $86.86_{0.41}$ & $36.09_{0.09}$ \\
    & ilr-MDE & $\boldsymbol{0.15}_{0.00}$ & $\boldsymbol{0.09}_{0.00}$ & $400.69_{1.46}$ & $\boldsymbol{1.11}_{0.01}$ & $1.68_{0.01}$ & $\boldsymbol{0.41}_{0.01}$ & $\boldsymbol{0.70}_{0.04}$ & $\boldsymbol{0.68}_{0.01}$ & $\boldsymbol{0.56}_{0.01}$ & $77.41_{0.10}$ & $\boldsymbol{86.04}_{0.41}$ & $43.20_{0.07}$ \\
    & Dir-KDE & $0.52_{0.00}$ & $0.82_{0.00}$ & $576.56_{0.71}$ & $1.25_{0.00}$ & $\boldsymbol{0.58}_{0.00}$ & $0.76_{0.00}$ & $1.88_{0.00}$ & $1.64_{0.00}$ & $1.35_{0.00}$ & $67.93_{0.09}$ & $86.53_{0.41}$ & $40.21_{0.07}$ \\
    & Dir-AIC & $0.84_{0.00}$ & $1.51_{0.00}$ & $594.07_{3.68}$ & $1.83_{0.01}$ & $1.10_{0.00}$ & $0.91_{0.01}$ & $2.65_{0.01}$ & $2.17_{0.01}$ & $1.85_{0.01}$ & $85.04_{0.08}$ & $87.27_{0.41}$ & $45.79_{0.07}$ \\
    & Dir-CVISE & $0.59_{0.00}$ & $1.15_{0.02}$ & $598.25_{5.20}$ & $1.59_{0.19}$ & $3.39_{0.74}$ & $0.57_{0.00}$ & $1.60_{0.00}$ & $1.30_{0.00}$ & $1.23_{0.01}$ & $\boldsymbol{22.74}_{0.14}$ & $86.90_{0.36}$ & $\boldsymbol{28.57}_{0.09}$ \\
    & Dir-CVKLD & $0.57_{0.00}$ & $1.04_{0.00}$ & $561.07_{0.68}$ & $1.31_{0.00}$ & $0.60_{0.00}$ & $0.56_{0.00}$ & $1.57_{0.00}$ & $1.31_{0.00}$ & $1.21_{0.00}$ & $57.90_{0.08}$ & $86.29_{0.41}$ & $38.70_{0.10}$ \\
    \bottomrule
    \end{tabular}
    \label{tab:chacon_MISER}
\end{sidewaystable*}

\begin{sidewaystable*}[p]
    \centering
    \caption{\small MISE with standard errors (subscripts) in Euclidean space, based on Monte Carlo replications in the Dirichlet mixture simulations; bold highlights the minimum and those not significantly different from it in each scenario.}
    \setlength{\tabcolsep}{3pt}
    \small
    \begin{tabular}{llcccccccccccc}
    \toprule
    Settings & Estimators & (1) & (2) & (3) & (4) & (5) & (6) & (7) & (8) & (9) & (10) & (11) & (12) \\
    \midrule
    \multirow{7}{*}{\centering\makecell[l]{$d=2$, \\ $n=100$ \\[3pt] ($\times 1000$)}}
    & ilr-PI & $19.92_{0.25}$ & $16.98_{0.21}$ & $5.52_{0.07}$ & $24.21_{0.23}$ & $14.35_{0.08}$ & $7.01_{0.08}$ & $19.45_{0.14}$ & $301.11_{1.25}$ & $22.13_{0.07}$ & $9.88_{0.08}$ & $8.53_{0.10}$ & $\boldsymbol{12.31}_{0.04}$ \\
    & ilr-CV & $33.02_{0.74}$ & $29.99_{0.66}$ & $9.14_{0.20}$ & $36.06_{0.69}$ & $14.52_{0.23}$ & $11.44_{0.24}$ & $20.96_{0.31}$ & $223.54_{2.55}$ & $\boldsymbol{17.20}_{0.13}$ & $12.25_{0.18}$ & $13.90_{0.30}$ & $14.41_{0.12}$ \\
    & ilr-MDE & $14.50_{0.34}$ & $14.09_{0.29}$ & $4.72_{0.09}$ & $24.85_{0.41}$ & $12.12_{0.12}$ & $5.94_{0.11}$ & $16.48_{0.16}$ & $237.92_{1.97}$ & $23.91_{0.09}$ & $9.99_{0.11}$ & $8.87_{0.13}$ & $13.29_{0.04}$ \\
    & Dir-KDE & $12.55_{0.19}$ & $16.97_{0.23}$ & $6.06_{0.08}$ & $19.83_{0.24}$ & $14.02_{0.18}$ & $\boldsymbol{5.48}_{0.08}$ & $16.54_{0.14}$ & $245.08_{1.67}$ & $22.32_{0.10}$ & $10.69_{0.17}$ & $\boldsymbol{7.56}_{0.09}$ & $13.73_{0.10}$ \\
    & Dir-AIC & $7.76_{0.23}$ & $\boldsymbol{5.28}_{0.16}$ & $1.59_{0.07}$ & $\boldsymbol{14.58}_{0.26}$ & $\boldsymbol{4.15}_{0.11}$ & $5.93_{0.12}$ & $\boldsymbol{14.57}_{0.16}$ & $192.32_{2.07}$ & $22.75_{0.40}$ & $8.87_{0.11}$ & $9.36_{0.12}$ & $14.43_{0.07}$ \\
    & Dir-CVISE & $10.58_{0.49}$ & $9.12_{0.40}$ & $2.22_{0.11}$ & $16.17_{0.43}$ & $7.38_{0.21}$ & $7.22_{0.21}$ & $16.61_{0.34}$ & $\boldsymbol{175.00}_{2.00}$ & $18.91_{0.29}$ & $8.94_{0.12}$ & $10.07_{0.16}$ & $16.50_{0.24}$ \\
    & Dir-CVKLD & $\boldsymbol{7.24}_{0.24}$ & $6.78_{0.18}$ & $\boldsymbol{1.47}_{0.07}$ & $\boldsymbol{14.56}_{0.31}$ & $5.92_{0.12}$ & $5.80_{0.10}$ & $\boldsymbol{14.60}_{0.15}$ & $212.06_{1.48}$ & $19.14_{0.12}$ & $\boldsymbol{8.56}_{0.10}$ & $9.56_{0.13}$ & $14.06_{0.07}$ \\
    
    \midrule
    
    \multirow{7}{*}{\centering\makecell[l]{$d=2$, \\ $n=500$ \\[3pt] ($\times 1000$)}}
    & ilr-PI & $7.08_{0.07}$ & $6.08_{0.06}$ & $2.06_{0.02}$ & $9.43_{0.08}$ & $5.95_{0.03}$ & $2.54_{0.03}$ & $7.05_{0.05}$ & $147.26_{0.54}$ & $12.46_{0.04}$ & $3.85_{0.03}$ & $3.07_{0.03}$ & $8.93_{0.02}$ \\
    & ilr-CV & $9.44_{0.14}$ & $7.97_{0.11}$ & $2.74_{0.04}$ & $11.19_{0.13}$ & $4.56_{0.04}$ & $3.32_{0.04}$ & $6.84_{0.06}$ & $80.34_{0.53}$ & $6.53_{0.03}$ & $4.02_{0.04}$ & $3.94_{0.05}$ & $7.06_{0.03}$ \\
    & ilr-MDE & $4.17_{0.10}$ & $4.14_{0.08}$ & $1.40_{0.03}$ & $6.64_{0.11}$ & $4.00_{0.04}$ & $2.10_{0.03}$ & $4.57_{0.05}$ & $80.43_{0.35}$ & $9.18_{0.06}$ & $3.47_{0.03}$ & $2.98_{0.04}$ & $9.62_{0.03}$ \\
    & Dir-KDE & $4.50_{0.06}$ & $5.80_{0.06}$ & $2.17_{0.02}$ & $7.34_{0.07}$ & $4.47_{0.03}$ & $2.01_{0.02}$ & $5.98_{0.04}$ & $123.70_{0.71}$ & $8.49_{0.04}$ & $3.42_{0.03}$ & $\boldsymbol{2.72}_{0.03}$ & $\boldsymbol{6.42}_{0.03}$ \\
    & Dir-AIC & $1.57_{0.04}$ & $\boldsymbol{1.19}_{0.04}$ & $	\boldsymbol{0.30}_{0.01}$ & $3.40_{0.06}$ & $\boldsymbol{0.81}_{0.02}$ & $\boldsymbol{1.50}_{0.02}$ & $4.37_{0.04}$ & $90.55_{0.80}$ & $6.35_{0.06}$ & $2.95_{0.03}$ & $2.84_{0.03}$ & $8.01_{0.04}$ \\
    & Dir-CVISE & $1.91_{0.07}$ & $2.76_{0.05}$ & $0.42_{0.02}$ & $3.54_{0.08}$ & $1.64_{0.05}$ & $1.85_{0.04}$ & $4.41_{0.06}$ & $\boldsymbol{76.61}_{0.49}$ & $\boldsymbol{5.35}_{0.06}$ & $2.98_{0.03}$ & $3.05_{0.04}$ & $7.46_{0.06}$ \\
    & Dir-CVKLD & $\boldsymbol{1.46}_{0.05}$ & $2.45_{0.04}$ & $\boldsymbol{0.29}_{0.01}$ & $\boldsymbol{3.15}_{0.06}$ & $1.17_{0.02}$ & $\boldsymbol{1.51}_{0.03}$ & $\boldsymbol{4.10}_{0.04}$ & $84.28_{1.14}$ & $6.38_{0.08}$ & $\boldsymbol{2.83}_{0.03}$ & $2.93_{0.03}$ & $7.51_{0.03}$ \\
    
    \midrule
    
    \multirow{7}{*}{\centering\makecell[l]{$d=5$, \\ $n=100$ \\[3pt] ($\times 1000$)}}
    & ilr-PI & $17.87_{0.12}$ & $32.61_{0.22}$ & $1.75_{0.01}$ & $44.11_{0.27}$ & $16.30_{0.08}$ & $7.25_{0.07}$ & $12.19_{0.06}$ & $3169.60_{5.48}$ & $26.05_{0.06}$ & $6.85_{0.05}$ & $14.75_{0.10}$ & $22.09_{0.05}$ \\
    & ilr-CV & $8.27_{0.07}$ & $17.07_{0.13}$ & $1.01_{0.01}$ & $20.06_{0.27}$ & $13.89_{0.11}$ & $4.23_{0.06}$ & $13.25_{0.13}$ & $3516.58_{8.47}$ & $30.17_{0.09}$ & $4.90_{0.03}$ & $7.78_{0.07}$ & $22.89_{0.08}$ \\
    & ilr-MDE & $2.21_{0.02}$ & $4.96_{0.04}$ & $0.39_{0.00}$ & $5.28_{0.04}$ & $10.14_{0.05}$ & $1.49_{0.01}$ & $11.86_{0.03}$ & $3690.15_{5.69}$ & $31.53_{0.03}$ & $2.70_{0.01}$ & $2.68_{0.02}$ & $21.34_{0.04}$ \\
    & Dir-KDE & $1.70_{0.01}$ & $6.22_{0.04}$ & $0.30_{0.00}$ & $6.10_{0.04}$ & $7.77_{0.04}$ & $\boldsymbol{0.68}_{0.00}$ & $9.42_{0.03}$ & $3572.10_{5.14}$ & $28.78_{0.04}$ & $1.96_{0.01}$ & $2.13_{0.01}$ & $\boldsymbol{19.43}_{0.04}$ \\
    & Dir-AIC & $\boldsymbol{0.83}_{0.01}$ & $4.28_{0.07}$ & $0.06_{0.00}$ & $\boldsymbol{3.61}_{0.04}$ & $5.19_{0.06}$ & $0.85_{0.01}$ & $11.66_{0.03}$ & $4698.62_{241.14}$ & $50.67_{6.02}$ & $2.34_{0.01}$ & $2.59_{0.02}$ & $21.15_{0.04}$ \\
    & Dir-CVISE & $0.98_{0.02}$ & $2.21_{0.05}$ & $0.06_{0.00}$ & $3.74_{0.05}$ & $3.30_{0.13}$ & $0.80_{0.01}$ & $\boldsymbol{5.60}_{0.08}$ & $\boldsymbol{2587.93}_{9.89}$ & $\boldsymbol{15.36}_{0.28}$ & $\boldsymbol{1.68}_{0.01}$ & $2.00_{0.02}$ & $19.72_{0.08}$ \\
    & Dir-CVKLD & $0.85_{0.01}$ & $\boldsymbol{1.93}_{0.04}$ & $\boldsymbol{0.06}_{0.00}$ & $\boldsymbol{3.64}_{0.04}$ & $\boldsymbol{2.42}_{0.03}$ & $0.74_{0.01}$ & $7.00_{0.06}$ & $3426.38_{6.14}$ & $26.33_{0.08}$ & $1.74_{0.01}$ & $\boldsymbol{1.93}_{0.02}$ & $19.45_{0.04}$ \\
    
    \midrule
    
    \multirow{7}{*}{\centering\makecell[l]{$d=5$, \\ $n=500$ \\[3pt] ($\times 1000$)}}
    & ilr-PI & $5.22_{0.02}$ & $9.33_{0.03}$ & $0.52_{0.00}$ & $12.96_{0.04}$ & $7.20_{0.02}$ & $2.14_{0.01}$ & $6.21_{0.02}$ & $2810.81_{4.79}$ & $22.17_{0.04}$ & $2.38_{0.01}$ & $4.41_{0.02}$ & $17.98_{0.04}$ \\
    & ilr-CV & $3.09_{0.01}$ & $6.86_{0.03}$ & $0.41_{0.00}$ & $7.28_{0.03}$ & $7.53_{0.02}$ & $1.62_{0.01}$ & $7.14_{0.02}$ & $2937.79_{5.54}$ & $24.95_{0.06}$ & $2.24_{0.01}$ & $2.97_{0.01}$ & $18.95_{0.04}$ \\
    & ilr-MDE & $0.76_{0.01}$ & $1.68_{0.02}$ & $0.14_{0.00}$ & $2.07_{0.01}$ & $3.32_{0.02}$ & $0.51_{0.00}$ & $7.31_{0.03}$ & $2988.82_{5.87}$ & $28.58_{0.04}$ & $1.51_{0.01}$ & $1.14_{0.01}$ & $19.58_{0.03}$ \\
    & Dir-KDE & $0.84_{0.01}$ & $3.13_{0.02}$ & $0.28_{0.00}$ & $3.33_{0.02}$ & $4.02_{0.01}$ & $0.31_{0.00}$ & $6.18_{0.02}$ & $3292.69_{4.71}$ & $23.81_{0.04}$ & $1.30_{0.01}$ & $1.17_{0.01}$ & $17.82_{0.04}$ \\
    & Dir-AIC & $\boldsymbol{0.16}_{0.00}$ & $1.44_{0.02}$ & $0.01_{0.00}$ & $1.69_{0.02}$ & $1.81_{0.02}$ & $0.45_{0.00}$ & $9.64_{0.03}$ & $2760.20_{9.43}$ & $29.89_{0.04}$ & $1.86_{0.01}$ & $1.47_{0.01}$ & $19.86_{0.04}$ \\
    & Dir-CVISE & $0.20_{0.00}$ & $0.41_{0.01}$ & $0.01_{0.00}$ & $0.79_{0.01}$ & $0.59_{0.01}$ & $\boldsymbol{0.21}_{0.00}$ & $\boldsymbol{1.87}_{0.02}$ & $\boldsymbol{1439.12}_{12.21}$ & $\boldsymbol{4.17}_{0.07}$ & $\boldsymbol{0.71}_{0.00}$ & $0.63_{0.01}$ & $\boldsymbol{13.98}_{0.07}$ \\
    & Dir-CVKLD & $0.18_{0.00}$ & $\boldsymbol{0.38}_{0.01}$ & $\boldsymbol{0.01}_{0.00}$ & $\boldsymbol{0.73}_{0.01}$ & $\boldsymbol{0.47}_{0.01}$ & $\boldsymbol{0.21}_{0.00}$ & $3.00_{0.02}$ & $3069.62_{4.53}$ & $15.55_{0.09}$ & $0.89_{0.01}$ & $\boldsymbol{0.61}_{0.01}$ & $17.68_{0.04}$ \\
    \bottomrule
    \end{tabular}
    \label{tab:dirichlet_MISER}
\end{sidewaystable*}

\end{appendices}

\bibliographystyle{Chicago}

\end{document}